\documentclass[aps,showpacs,pre,onecolumn,10pt]{revtex4-1}

\usepackage[T1]{fontenc}
\usepackage[latin1]{inputenc}
\usepackage{ae}
\usepackage{amsmath}
\usepackage{amssymb}
\usepackage{amsthm}
\usepackage{graphicx}
\usepackage{epstopdf}
\usepackage{subfigure}
\usepackage{multirow}
\usepackage{booktabs}
\usepackage{color}
\usepackage{extramarks}
\usepackage[colorlinks=true,linkcolor=blue,citecolor=blue]{hyperref}

\newtheorem{lemma}{Lemma}
\newtheorem{remark}{Remark}

\newtheorem{theorem}{Theorem}
\newcommand{\diag}{\mathrm{diag}}
\begin{document}

\title{Multiple double-valley and single-valley dark solitons in the complex modified Korteweg--de Vries equation: shape-preserving collisions and shape-altering collisions }
\author{Yingcan Huang$^{1}$}
\author{Jingsong He$^{1}$ }
\author{T. Kanna$^2$ }
\author{Jiguang Rao$^{3}$}
\affiliation {\vspace{0.1in}
$^{1}$Institute for Advanced Study, Shenzhen University, Shenzhen, Guangdong 518060, P.\ R.\ China\\
$^{2}$Nonlinear Waves Research Lab,  PG and Research Department of Physics, Bishop Heber College, Tiruchirappalli - 620 017, Tamil Nadu, India\\
$^{3}$School of Mathematics and Statistics, Hubei University of Science and Technology, Xianning, Hubei, 437300, P.\ R.\ China\\
}

\thanks{Email of corresponding authors:hejingsong@szu.edu.cn(J.He);\;\;\; kanna\_phy@bhc.edu.in (T. Kanna)}

\begin{abstract}

The shape-preserving and shape-altering collisions of general dark solitons are investigated in the complex modified Korteweg--de Vries equation. The obtained dark soliton solutions are classified into two distinct types, referred to as type-I and type-II dark solitons, which exhibit fundamentally different structural and dynamical characteristics. A single type-I dark soliton is symmetric about its center and admits three distinct valley profiles, namely single-valley, double-valley, and flat-bottom structures, whereas a single type-II dark soliton only supports a single-valley profile. These two types of dark solitons differ not only in their phase behaviors as the spatial variable varies from $-\infty$ to $+\infty$, but also in their velocity--amplitude relations. For multiple pure type-I dark solitons, collisions are shape-preserving; however, an unusual collective effect is revealed in which modifying the parameters of one soliton can induce changes in the profiles and amplitudes of the other solitons, even though all collisions remain elastic in nature. In contrast, multiple pure type-II dark solitons behave independently, undergoing only phase shifts without any modification of their shapes or amplitudes. When type-I and type-II dark solitons coexist, their interactions lead to genuinely shape-altering collisions, where the valley structures and amplitudes of type-I dark solitons are modified, while type-II solitons remain unaffected except for phase shifts. Through asymptotic analysis, it is further shown that the influence of type-II dark solitons on type-I dark solitons is confined to the pre-collision stage and disappears asymptotically after the interaction. These results reveal rich collision-induced dynamical behaviors of dark solitons in the complex modified Korteweg--de Vries equation and provide insights into nontrivial interaction mechanisms of localized nonlinear excitations.

\end{abstract}

\maketitle



\section{Introduction}
Solitons are fundamental to the study of nonlinear physics and are renowned for their remarkable ability to preserve both shape and velocity over long distances without dispersion \cite{Kivshar-2003}. In contrast to conventional waves, which typically broaden and attenuate during propagation, solitons exhibit particle-like behavior, maintaining localized and stable profiles through a precise balance between dispersion and nonlinearity in the medium \cite{Akhmediev-1997}. This distinctive feature has attracted considerable attention because of its wide applicability across diverse technological contexts \cite{Infeld-2000}. In particular, optical solitons have been extensively studied in fiber-optic communication systems, where they enable long-distance information transmission with minimal energy loss, offering a compelling solution to signal degradation \cite{Agrawal-2001}. Beyond telecommunications, solitons play important roles in fields such as fluid dynamics \cite{Whitham-1999}, plasma physics \cite{Petviashvili-2016}, and Bose--Einstein condensates (BECs) \cite{Kevrekidis-2007}.

Bright and dark solitons represent two fundamental classes of solitonic structures, distinguished by their characteristic waveforms and diverse applications in nonlinear science \cite{soliton-book}.  Bright solitons manifest as intense, localized peaks in self-focusing media, where the wave energy is concentrated in a small region, producing a burst-like appearance \cite{Kivshar-2003}. In contrast, dark solitons appear as dips or voids in a continuous wave field, stabilized within self-defocusing media \cite{Kivshar-1998}. The formation of these solitons depends on the nature of the nonlinearity in their environment: bright solitons require self-focusing properties to maintain their integrity, while dark solitons arise from self-defocusing effects.
These distinctions make each type of soliton influential across diverse scientific domains. Bright solitons, for example, are valuable in optical fiber networks, where their robustness supports efficient data transmission. Dark solitons, with their unique low-intensity profiles, contribute to advancements in photonics and atomic physics, where they are studied in phenomena like optical vortices. Apart from the obvious difference in the
density of bright and dark solitons, the latter have distinctive dynamic properties. For instance, multi-component dark vector solitons typically undergo trivial interactions \cite{Biondini-JPA,Ohta-Yang}, whereas bright vector solitons can exchange energy between components, leading to polarization changes \cite{Kanna-PRL,Stalin-PRL}. Their role in fluid dynamics and beyond highlights the solitons' wide-ranging influence, as they continue to inspire innovation and provide insights into fundamental wave behavior across various fields.

Solitons are intriguing mathematical entities that arise as solutions to specific integrable nonlinear evolution equations \cite{Mark-Book}. Their exceptional collision dynamics and various other characteristics have made them valuable in practical applications. One such crucial physical model is the complex modified Korteweg-de Vries (cmKdV) equation \cite{Suhubi-JES, Foursov-JMP-2000, Ablowitz-CHAOS-2006},
\begin{equation}\label{CmKdV-V0}
\begin{aligned}
&u_t + u_{xxx}+6\gamma|u|^2u_x = 0,
\end{aligned}
\end{equation}
which supports both bright and dark soliton solutions.
Notably, the nonlinearity coefficient $\gamma$ can be either positive or negative, depending on the physical context. For instance, in describing light propagation in liquid-crystal waveguides, this coefficient is positive \cite{Leblond-2013}. Conversely, in deriving the cmKdV equation for strongly dispersive waves in a weakly nonlinear medium using a multiple-scale expansion (as shown in Ref. \cite{Leblond-2013}), the nonlinear coefficient takes a negative value. This coefficient can typically be normalized to $\gamma = \pm1$, where the case $\gamma = 1$ corresponds to bright solitons, while $\gamma = -1$ corresponds to dark solitons. For $\gamma = 1$, Eq.~\eqref{CmKdV-V0} reduces to the focusing complex modified Korteweg-de Vries (cmKdV) equation, which has been extensively studied through diverse analytical frameworks. The $N$-bright soliton solutions were systematically constructed via the Darboux transformation \cite{zha2008darboux}, while higher-order soliton solutions were investigated using both the inverse scattering method \cite{zhang2020bound} and refined Darboux transformation techniques \cite{liu2018dynamics, zhang2020soliton}. Beyond fundamental soliton structures, breather solutions were derived in Darboux transformation framework \cite{lv2022breather}, and rogue wave solutions were comprehensively analyzed using Darboux transformation methods \cite{ankiewicz2010rogue, zhaqilao2013nth, he2014few} complemented by bilinear approaches \cite{yangrogue}. Recent studies on the focusing cmKdV investigate the long-time asymptotics of localized solutions \cite{Stewart-SIAM-2025}, the soliton-collision dynamics of the (2+1)-dimensional nonlocal complex mKdV system \cite{Ge-ND-2025}, and the large-order asymptotics of multi-rational solitons \cite{Weng-JDE-2025}.
 When $\gamma = -1$, Eq.~\eqref{CmKdV-V0} reduces to the defocusing cmKdV equation, where progress on dark soliton studies has been relatively limited. The fundamental single dark soliton solution was first established through Painlev\'{e} analysis and truncation methods \cite{Mohammad-JPA-1995}, followed by the systematic derivation of $N$-dark soliton solutions via binary Darboux transformation \cite{zhang2019general}. Subsequent extensions to multi-component systems yielded dark vector soliton solutions through Darboux transformation methodologies \cite{ye2021darboux, ye2022bound}. The Cauchy problem with step-like initial conditions was resolved through the Whitham modulation theory, as demonstrated in works by Kodama \textit{et al.} \cite{kodama2009whitham} and subsequent refinements \cite{zeng2023whitham, wang2022complete}. In addition to classical approaches, improved bilinear formulations have yielded novel exact solutions for nonlocal variants of the cmKdV equation, broadening its solution space \cite{li2019improved}. More recently, gradient-optimized physics-informed neural networks have emerged as efficient solvers, offering a flexible deep-learning framework to approximate nonlinear wave dynamics beyond traditional methods \cite{li2022gradient}. Notably, all reported dark solitons in the cmKdV equation exhibit shape-preserving collisions with single-valley profiles.
 

From a dynamical viewpoint, the robustness of soliton collisions is often regarded as a hallmark of integrable nonlinear wave systems. In this framework, dark solitons are commonly viewed as particle-like localized excitations whose interactions are purely elastic, characterized by the preservation of their shapes and amplitudes up to phase shifts. This elastic-collision paradigm has been widely adopted in the study of dark solitons and underpins the conventional understanding of their stability and interaction properties. However, this picture is largely based on dark solitons with relatively simple internal structures. When more intricate internal degrees of freedom are involved, such as multi-valley profiles, it is not a priori clear whether the standard elastic-collision paradigm remains valid. In particular, it remains an open question whether collisions between such structurally rich dark solitons necessarily preserve their internal profiles, or whether interaction-induced structural transformations can occur even within an integrable setting. Addressing this issue is essential for elucidating how internal structures influence soliton interactions and for advancing the understanding of nonlinear wave dynamics beyond the conventional particle-like description.

This paper conducts an in-depth study of dark solitons in the cmKdV equation with $\gamma = -1$, namely,
\begin{equation}\label{CmKdV}
u_t + u_{xxx} - 6|u|^2 u_x = 0.
\end{equation}
We demonstrate that both shape-preserving and shape-altering collisions of dark solitons occur in Eq.~\eqref{CmKdV}. In particular, the shape-preserving collisions of one specific type of dark soliton reveal novel solitonic behaviors, while shape-altering collisions are rarely reported in previous studies. Motivated by the dynamical issues discussed above, our investigation focuses on three main aspects. First, we systematically construct general dark soliton solutions of the cmKdV equation. This equation admits two types of dark solitons with distinct properties, which are classified as  type-I and type-II dark solitons. The general solutions are obtained via the KP bilinear method \cite{Hirota, Sato-RIMS-1981, Jimbo-RIMS-1983, Date-PhysD-1982}, where the $\tau$-function is expressed as the determinant of a block-structured matrix with four sub-blocks. Notably, the dimension of the first sub-block is twice the number of type-I solitons, whereas the dimension of the last sub-block equals the number of type-II solitons. The detailed derivations are provided in Appendix~\ref{App-1}.
Second, we analyze the dynamics of the fundamental dark soliton solutions. Specifically, there exist precisely two types of fundamental solutions. To clarify their differences, we examine single-soliton solutions of each type in detail, focusing on velocity, amplitude, and profile. A key result is that the type-I dark soliton can exhibit three distinct profiles---double-valley, flat-bottom, and single-valley---depending on its velocity. The precise conditions for each profile are discussed in the paper. In contrast, the type-II soliton always maintains a single-valley profile regardless of velocity. To the best of our knowledge, type-I dark solitons that admit three distinct profiles have not been reported before in  scalar integrable systems.
Third, we investigate the collision dynamics of multi-soliton solutions. We first study collisions between two solitons, covering three cases: two type-I solitons, two type-II solitons, and mixed type-I and type-II solitons. Shape-altering collisions occur only in the mixed case, while the other two cases exhibit shape-preserving collisions. We then extend our analysis to collisions among arbitrary numbers of solitons in the cmKdV equation and derive the asymptotic forms of the solutions before and after collisions. The asymptotic analysis confirms that shape-altering collisions occur exclusively in mixed type-I and type-II cases. Specifically, a type-I soliton is influenced only by solitons of the same type before collision, but after collision, it is affected by both types. This mechanism accounts for the inelastic nature of mixed collisions.

The remainder of the paper is organized as follows. Section~\ref{sec2} presents explicit formulas for the general dark soliton solutions and analyzes the dynamics of the two types of single solitons. Section~\ref{Collision-T} investigates collisions between two solitons in the three basic cases (pure type-I, pure type-II, and mixed). Section~\ref{Collision-N} extends the analysis to collisions among arbitrary numbers of type-I and type-II solitons. Section~\ref{conclusion} concludes the paper.

\section{Dark soliton solutions to cmKdV equation}\label{sec2}
In this section, we first present the general dark soliton solutions of the cmKdV equation \eqref{CmKdV}, which are expressed in terms of block determinants. When the determinant blocks are of specific orders, these general solutions reduce to two distinct types of dark solitons. We then analyze the properties of these two types of single dark solitons. They differ not only in their mathematical structures but also in their velocity-amplitude relations and valley profiles under detailed examination.

\subsection{Explicit formulas for general dark soliton solutions of the cmKdV equation}
We present the general dark soliton solutions to the cmKdV equation \eqref{CmKdV}. A detailed derivation of these solutions is given in Appendix \ref{App-1}.  $\\$

\begin{theorem}\label{definition of solution}
The cmKdV equation \eqref{CmKdV} admits the following dark soliton solutions
\begin{equation}\label{D-SO}
\begin{aligned}
u=\frac{g}{f},
\end{aligned}
\end{equation}
where $f=\overline{\delta}_0,g=\overline{\delta}_1$, and
\begin{equation}\label{so-fg}
\begin{aligned}
\overline{\delta}_n=\det
\left(\begin{matrix}
\mathcal{M}^{[n,1,1]}_{i,j}&\mathcal{M}^{[n,1,2]}_{i,j}\\
\mathcal{M}^{[n,2,1]}_{i,j}&\mathcal{M}^{[n,2,2]}_{i,j}
\end{matrix}
\right),
\end{aligned}
\end{equation}
with
\begin{equation}\label{so-M}
\begin{aligned}
&\mathcal{M}_{i,j}^{[n,1,1]}=\Gamma_1+\left(\begin{matrix}
\overline{\mathrm{m}}_{i,j}^{(n)}
\end{matrix}\right)_{2N_1\times{2N_1}}, \quad
\mathcal{M}_{i,j}^{[n,1,2]}=\left(\begin{matrix}
\overline{\mathrm{m}}_{i,2N_1+j}^{(n)}
\end{matrix}\right)_{2N_1\times{N_2}},\\
&\mathcal{M}_{i,j}^{[n,2,1]}=\left(\begin{matrix}
\overline{\mathrm{m}}_{2N_1+i,j}^{(n)}
\end{matrix}\right)_{N_2\times{2N_1}}, \quad
\mathcal{M}_{i,j}^{[n,2,2]}=\Gamma_2+\left(\begin{matrix}
\overline{\mathrm{m}}_{2N_1+i,2N_1+j}^{(n)}
\end{matrix}\right)_{N_2\times{N_2}},\\
&\Gamma_1=\diag\left(e^{-\zeta_1},e^{-\zeta_2},\cdots,e^{-\zeta_{2N_1}}\right), \quad
\Gamma_2=\diag\left(e^{-\zeta_{2N_1+1}},e^{-\zeta_{2N_1+2}},\cdots,e^{-\zeta_{2N_1+N_2}}\right).
\end{aligned}
\end{equation}
Here,
\begin{equation}\label{mx-sj}
\begin{aligned}
 \overline{\mathrm{m}}_{i,j}^{(n)}=\frac{1}{p_i+p_j^*}\left(-\frac{p_i}{p_j^*}\right)^n, \quad
\zeta_{i}=(p_i+p_i^*)x+\left[3(p_i+p_i^*)-(p_i^3+p_i^{*3})\right]t+\zeta_{i,0}.
\end{aligned}
\end{equation}
The parameters are subject to the following constraints
\begin{equation}\label{pa-1}
\begin{aligned}
|p_i|=1, \quad p_{N_1+s}=-p_s, \quad \zeta_{N_1+s,0}=-\zeta_{s,0}+\mathrm{i}\pi
\end{aligned}
\end{equation}
for $i=1,2,\cdots,2N_1+N_2$ and $s=1,2,\cdots,N_1$, with $p_i\in\mathbb{C}$ and $\zeta_{s,0}\in\mathbb{R}$.

\end{theorem}

\begin{remark}\label{remark-1}
In Eq.~\eqref{pa-1}, the constraint \(|p_i| = 1\) can be equivalently expressed as \(p_i = e^{\mathrm{i}\vartheta_i}\), where \(\vartheta_i \in \left(-\pi, \pi\right)\). Under this form, the relation \(p_{N_1+s} = -p_s\) becomes \(e^{\mathrm{i}\vartheta_{N+s}} = e^{\mathrm{i}(\vartheta_s + \pi)}\). In the subsequent analysis of the properties and collisions of dark solitons,
this representation will also be used to simplify related formulas.
\end{remark}

\begin{remark}\label{remark-2}
Under the parameter restrictions in Eq.~\eqref{pa-1}, the associated elements exhibit the following correlations:
\begin{equation}
\begin{aligned}
\overline{m}_{N_1+s,N_1+\ell}^{(n)} &= -\overline{m}_{s,\ell}^{(n)}, \quad
\overline{m}_{N_1+s,\ell}^{(n)} = -\overline{m}_{s,N_1+\ell}^{(n)}, \quad
e^{\zeta_{N_1+s}}= -e^{-\zeta_s},
\end{aligned}
\end{equation}
for \(1 \leq s, \ell \leq N_1\). Specifically, when \(N_2 = 0\), these correlations imply that the function \(\overline{\delta}_n\), considered as a function of \(\zeta_1, \zeta_2, \cdots, \zeta_N\), satisfies the following symmetry:
\begin{equation}
\begin{aligned}
\overline{\delta}_n(\zeta_1, \zeta_2, \cdots, \zeta_{N_1}) = \overline{\delta}_n(-\zeta_1, -\zeta_2, \cdots, -\zeta_{N_1}).
\end{aligned}
\end{equation}
If we further set \(\zeta_{s,0} = 0\) in the expression for \(\zeta_i\) in Eq.~\eqref{mx-sj}, then \(\overline{\delta}_n\) satisfies the following symmetry relation with respect to \(x\) and \(t\):
\begin{equation}
\begin{aligned}
\overline{\delta}_n(-x,-t) = \overline{\delta}_n(x,t).
\end{aligned}
\end{equation}
Consequently, the dark soliton solutions given by Eq.~\eqref{D-SO} exhibit the symmetry
\begin{equation}
\begin{aligned}
u(-x,-t) = u(x,t),
\end{aligned}
\end{equation}
indicating that they are parity-time symmetric under the specific parameter conditions \(N_2=0\) and \(\zeta_{s,0}=0\).
In this case ($N_2=0$), the solutions of Eq.~\eqref{D-SO} correspond to a particular type of dark solitons, exhibiting double valleys, flat-bottom-shaped valleys, or single valleys, as will be discussed in detail later.
\end{remark}

\subsection{Dynamics of two types of single dark soliton solutions}
As we briefly explained in Remark \ref{remark-2}, when the orders \(N_1\) and \(N_2\) of the sub-blocks in the block determinants take different values, the block determinant solution \eqref{D-SO} will correspond to different forms of solitons. It is necessary to study in detail the properties of the dark soliton solution \eqref{D-SO} corresponding to the most specific values of \(N_1\) and \(N_2\), namely \(N_1 = 1, N_2 = 0\) and \(N_1 = 0, N_2 = 1\), which represent two types of single dark solitons with different characteristics. When \(N_1\) and \(N_2\) in solution \eqref{D-SO} take more general values, they correspond to the collisions of multiple dark soliton solutions of these two cases.  To better distinguish these two types of solitons, we refer to the solutions \eqref{D-SO} with \(N_1 \neq 0, N_2 = 0\) as type-I dark solitons, and the solutions \eqref{D-SO} with \(N_1 = 0, N_2 \neq 0\) as type-II dark solitons. Naturally, the solution \eqref{D-SO} with \(N_1 N_2 \neq 0\) corresponds to a coexistence of type-I dark solitons and type-II dark solitons.

\subsubsection{Dynamics of single type-I dark soliton}
Setting $N_1=1,N_2=0$ in Eqs.~\eqref{so-fg}--\eqref{mx-sj}, the  associated functions $f$ and $g$ of solution \eqref{D-SO} are expressed as
\begin{equation}\label{1fg-cosh}
\begin{aligned}
f=\left|\begin{matrix}
e^{-\zeta_1}+\overline{\mathrm{m}}_{1,1}^{(0)}&\overline{\mathrm{m}}_{1,2}^{(0)}\\
-\overline{\mathrm{m}}_{1,2}^{(0)}&-e^{\zeta_1}-\overline{\mathrm{m}}_{1,1}^{(0)}
\end{matrix}\right|, \quad
g=\left|\begin{matrix}
e^{-\zeta_1}+\overline{\mathrm{m}}_{1,1}^{(1)}&\overline{\mathrm{m}}_{1,2}^{(1)}\\
-\overline{\mathrm{m}}_{1,2}^{(1)}&-e^{\zeta_1}-\overline{\mathrm{m}}_{1,1}^{(1)}
\end{matrix}\right|,
\end{aligned}
\end{equation}
where
\begin{equation}\label{1so-cosh-M}
\begin{aligned}
\overline{\mathrm{m}}_{1,1}^{(n)}=&\frac{1}{p_1+p_1^*}(-\frac{p_1}{p_1^*})^n, \quad
\overline{\mathrm{m}}_{1,2}^{(n)}=\frac{1}{p_1-p_1^*}(\frac{p_1}{p_1^*})^n, \quad
\zeta_{1}=(p_1+p_1^*)x+\left[3(p_1+p_1^*)-(p_1^3+p_1^{*3})\right]t+\zeta_{1,0},
\end{aligned}
\end{equation}
and $|p_1|=1$. From Eq.~\eqref{D-SO}, a single soliton solution can be expressed in terms of hyperbolic functions as
\begin{equation}\label{1D-so}
\begin{aligned}
u=\frac{\chi_0^{(1)}+\Theta_0^{(1)}\cosh(\zeta_1)}{\chi_0^{(0)}+\Theta_0^{(0)}\cosh(\zeta_1)},
\end{aligned}
\end{equation}
with $\chi_0^{(n)}=1-\frac{4}{(p_1^2-p_1^{*2})^2}(\frac{p_1}{p_1^*})^{2n}$ and $\Theta_0^{(n)}=\frac{2}{p_1+p_1^*}(-\frac{p_1}{p_1^*})^{n}$.
As stated in Remark \ref{remark-2}, this solution satisfies the symmetry $u(\zeta_1)=u(-\zeta_1)$.
By setting $p_1=e^{\mathrm{i}\vartheta_1}$, Eq.~\eqref{1D-so} can be expressed in the equivalent form:
\begin{equation}\label{1so-cosh}
\begin{aligned}
u=&\frac{\left(\cot^22\vartheta_1-\cos2\vartheta_1\sec\vartheta_1\cosh\zeta_1\right)
+2\mathrm{i}\left(\cot2\vartheta_1-\sin\vartheta_1\cosh\zeta_1\right)}{1+\csc^2(2\vartheta_1)+\sec\vartheta_1\cosh\zeta_1},
\end{aligned}
\end{equation}
with $\zeta_1=2\cos\vartheta_1\left[x+(4-2\cos2\vartheta_1)t\right]+\zeta_{1,0}$, and $\vartheta_1\in(-\pi,\pi),\zeta_{1,0}\in\mathbb{R}$. When one takes $\sec\vartheta_1>0$, i.e., $\vartheta_1\in(-\frac{1}{2}\pi,\frac{1}{2}\pi)$, the denominator of this soliton solution remains strictly positive, so the soliton solution is smooth. The velocity of the soliton is given by
\begin{equation}\label{1-v}
\begin{aligned}
\nu=2\cos(2\vartheta_1)-4.
\end{aligned}
\end{equation}
In addition, $|u|$ approaches constant background amplitude $1$ as $x\rightarrow\pm\infty$. As $x$ varies from
$-\infty$ to $+\infty$, the phase of $u$ acquires no shift.

This dark soliton solution is characterized by the two parameters $\vartheta_1$ (i.e., $p_1$) and $\zeta_{1,0}$. The parameter \(\zeta_{1,0}\), which is embedded in \(\zeta_1\), determines the position of the dark soliton. The parameter \(\vartheta_1\) not only influences the smoothness of the solution but also determines the waveform of the dark soliton, which in turn is governed by the sign of \(\Delta_0\). Here, \(\Delta_0=\left.\left(\partial_{\zeta_1\zeta_1}^2\left|u\right|\right)\right|_{\zeta_1=0}\). Based on the sign of \(\Delta_0\), the soliton can be classified into three types: a single-valley dark soliton when  \(\Delta_0>0\), a flat-bottom-shaped dark soliton when \(\Delta_0=0\), and a double-valley dark soliton when \(\Delta_0<0\). It is easy to verify that $\left.\left(\partial_{\zeta_1}|u|\right)\right|_{\zeta_1=0}=0$, which indicates that if $\left|u(\zeta_1)\right|$ is regarded as a function of $\zeta_1$, then $\zeta_1=0$ is a critical point of $|u(\zeta_1)|$. When $\Delta_0 > 0$ (i.e., $\left.\frac{\partial^2 |u|}{\partial \zeta_1^2}\right|_{\zeta_1=0} > 0$), the point $\zeta_1 = 0$ is a local minimum of $|u(\zeta_1)|$. Thus, $|u|$ exhibits only one extremal line in the $(x, t)$ plane, manifesting as a dark soliton with a single-valley profile.
When $\Delta_0 = 0$, the point $\zeta_1 = 0$ is a inflection point of $|u(\zeta_1)|$, indicating that in the vicinity of $\zeta_1 = 0$ within the $(x, t)$ plane, $|u|$ attains no extremum. Consequently, $u$ in the $(x, t)$ plane presents a dark soliton with flat-bottom-shaped profile.
When $\Delta_0 < 0$ (i.e., $\left.\frac{\partial^2 |u|}{\partial \zeta_1^2}\right|_{\zeta_1=0} < 0$), the point $\zeta_1 = 0$ is a local maximum of $|u(\zeta_1)|$ in the vicinity of $\zeta_1 = 0$. In the detailed analysis below, this case will be shown to correspond to a dark soliton with a double-valley profile.

\begin{figure}[!htbp]
\centering
\subfigure{\includegraphics[height=5cm,width=5.5cm]{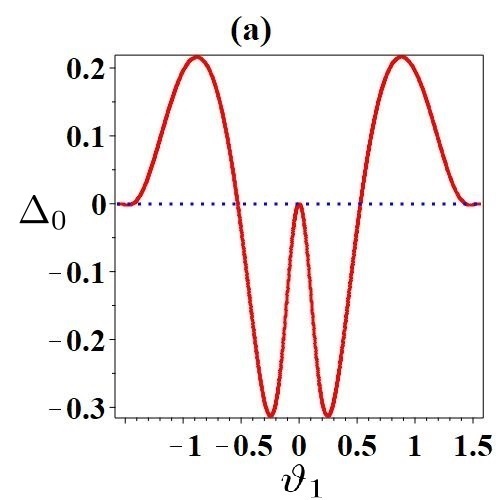}}
\subfigure{\includegraphics[height=5cm,width=5.5cm]{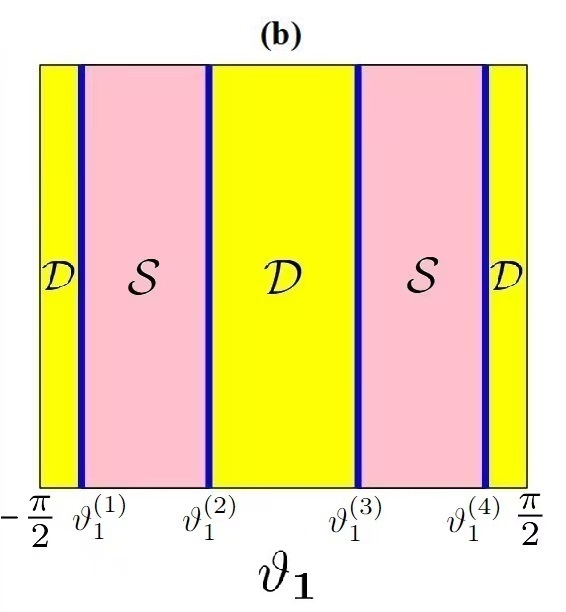}}
\subfigure{\includegraphics[height=5cm,width=5.5cm]{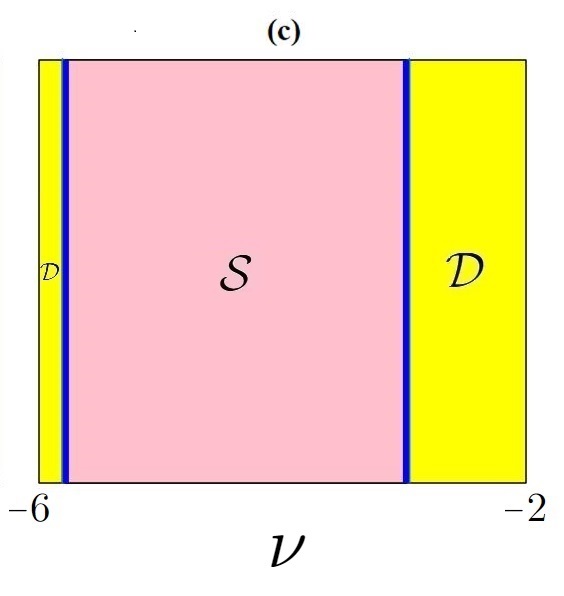}}\label{Fig1-c}
\caption{(Color online) The valley profile alterations of the dark soliton solutions \eqref{1so-cosh} as $\vartheta_1$ varies: `$\mathcal{\mathcal{D}}$' represents double-valley (yellow region); `$\mathcal{\mathcal{S}}$' represents single-valley (pink region); and the blue solid line corresponds to the flat-bottom.
~}\label{Fig1}
\end{figure}

Next, we will specifically discuss the valley properties of these three types of solitons based on the sign of $\Delta_0$, where $\Delta_0$ is explicitly formulated as
\begin{equation}
\begin{aligned}
\Delta_0=\frac{16\sin^2\vartheta_1\cos^3\vartheta_1\left(8\cos\vartheta_1\sin^2\vartheta_1-4\cos^2\vartheta_1\sin^2\vartheta_1-1\right)}
{\left(4\cos\vartheta_1\sin^2\vartheta_1+4\cos^2\vartheta_1\sin^2\vartheta_1-1\right)^2}.
\end{aligned}
\end{equation}
The variation of the sign of $\Delta_0$ with respect to $\vartheta_1$ is shown in Fig.~\ref{Fig1}(a). This confirms that for suitable values of $\vartheta_1$, the conditions $\Delta_0>0$, $\Delta_0=0$, and $\Delta_0<0$ can all occur. When \(\Delta_0 > 0\), \(|u|\) has only one minimum at \(\zeta_1 = 0\), and corresponds to a dark soliton with single-valley profile.  Along the extremum $\zeta_1=0$, the intensity of this soliton is
\begin{equation}\label{1-AM}
\begin{aligned}
\mathcal{I}=\sqrt{1-\frac{16 \cos^3{\vartheta_1}\sin^2{\vartheta_1}}{1+4\cos{\vartheta_1}\sin^2{\vartheta_1}(\cos{\vartheta_1}+1)}}.
\end{aligned}
\end{equation}
When \(\Delta_0 = 0\), namely, $\left(\partial^2_{\zeta_1\zeta_1}|u|\right)\Big|_{\zeta_1=0}=0$,  it also follows that $\left(\partial_{\zeta_1} |u|\right)\big|_{\zeta_1=0}=0$. This indicates that the center of the soliton at \(\zeta_1 = 0\) is not an extremum of \(|u|\). Therefore, $|u|$ exhibits a dark soliton with flat-bottom--shaped profile. The intensity of the flattop-shaped soliton  is also given by Eq.~\eqref{1-AM}. When \(\Delta_0 < 0\), \(|u|\) exhibits three extrema: one local maximum at \(\zeta_1 = 0\) and two local minima at \(\zeta_1 = \Lambda^{\pm}\), where
\begin{equation}
\begin{aligned}
\Lambda^{\pm}=\ln\left[\frac{1}{\gamma_0}\left(\Lambda_0 \pm\sqrt{\Lambda_1}\right)\right],
\end{aligned}
\end{equation}
with
\begin{equation}
\begin{aligned}
\Lambda_0&={2\sin^2(2\vartheta_1)(\cos 2\vartheta_1 - 2\sin^4 \vartheta_1) + 1},\quad
\gamma_0={2\sin \vartheta_1 \sin 2\vartheta_1(\sin^2 2\vartheta_1 + 1)},\\
\Lambda_1&=\left[\left(\sin^2(2\vartheta_1)+1\right)^2-16\sin^2(2\vartheta_1)\cos^2\vartheta_1\right]
\left[\left(\sin^2(2\vartheta_1)+1\right)^2-4\sin^2(2\vartheta_1)\cos^2\vartheta_1\right].
\end{aligned}
\end{equation}
These extrema are identified by solving \(\partial_{\zeta_1} |u| = 0\), and they satisfy the conditions \(\partial^2_{\zeta_1 \zeta_1} |u| \Big|_{\zeta_1=0}<0\) and \(\partial^2_{\zeta_1 \zeta_1} |u| \Big|_{\zeta_1=\Lambda^{\pm}}> 0\).
The extremum of \(|u|\) at \(\zeta_1 = 0\) is still expressed by Eq.~\eqref{1-AM}. Along the other two extremum lines, the intensities of the dark soliton are equal and are given by
\begin{equation}\label{1-AMs}
\begin{aligned}
\overline{\mathcal{I}}=\frac{(2\cos{\vartheta_1^2-1})^2 |\sin{\vartheta_1}|}{\sqrt{16\cos^8{\vartheta_1}-48\cos^6{\vartheta_1}+40\cos^4{\vartheta_1}-8\cos^2{\vartheta_1}+1}}.
\end{aligned}
\end{equation}
Since $\overline{\mathcal{I}} < \mathcal{I} < 1$, the valleys of the dark solitons, i.e., their minimum-intensity positions, are located at the two extremum lines $\zeta_1 = \Lambda^{\pm}$.
Furthermore, since the two extrema are symmetric about $\zeta_1=0$ and have equal amplitudes,
the soliton exhibits a symmetric double-valley profile.

\begin{figure}[!htbp]
\centering
\subfigure{\includegraphics[height=5cm,width=5.5cm]{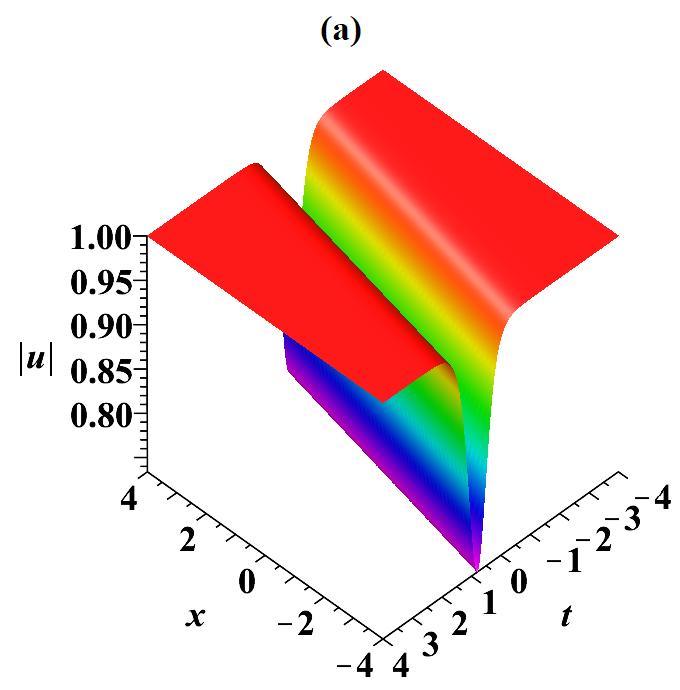}}
\subfigure{\includegraphics[height=5cm,width=5.5cm]{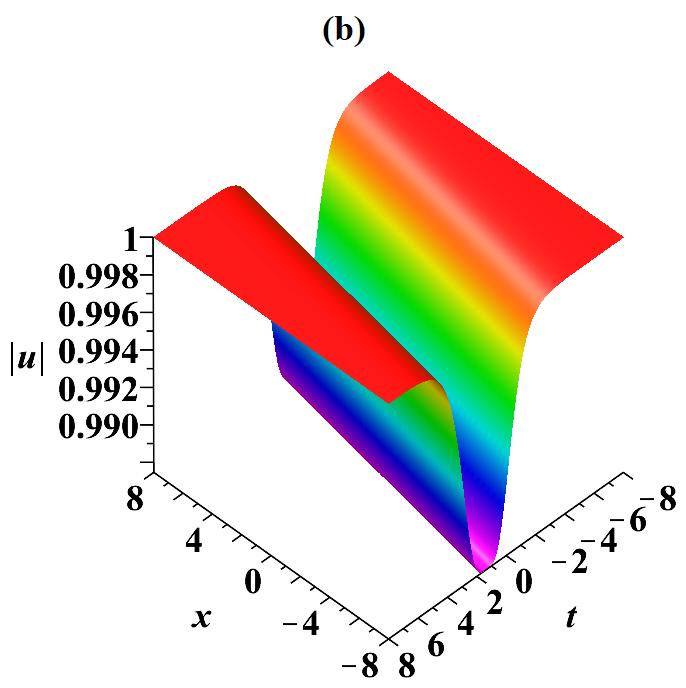}}
\subfigure{\includegraphics[height=5cm,width=5.5cm]{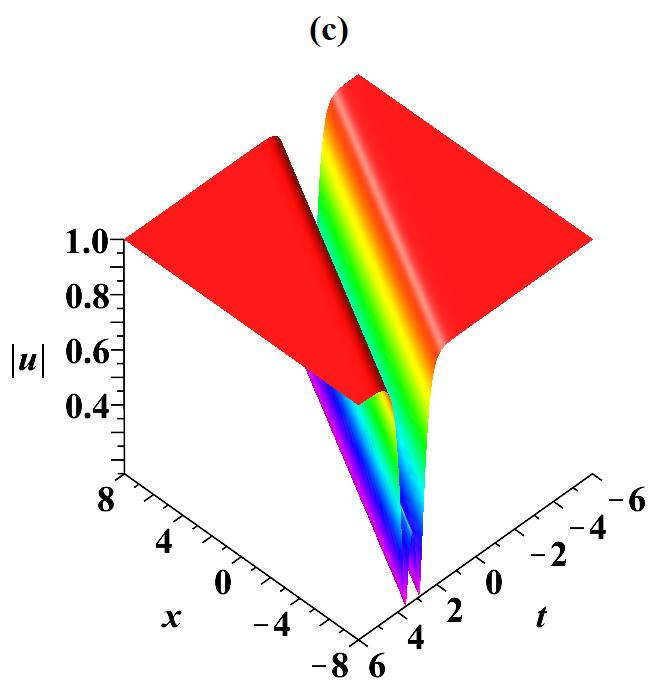}}
\subfigure{\includegraphics[height=4cm,width=5.5cm]{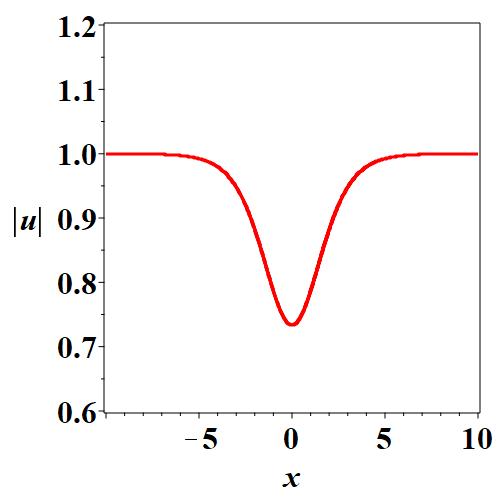}}
\subfigure{\includegraphics[height=4cm,width=5.5cm]{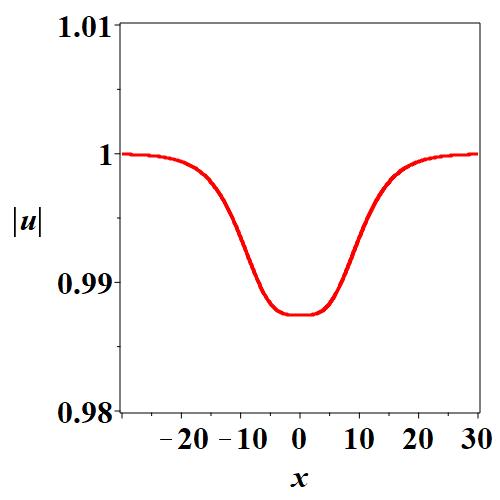}}
\subfigure{\includegraphics[height=4cm,width=5.5cm]{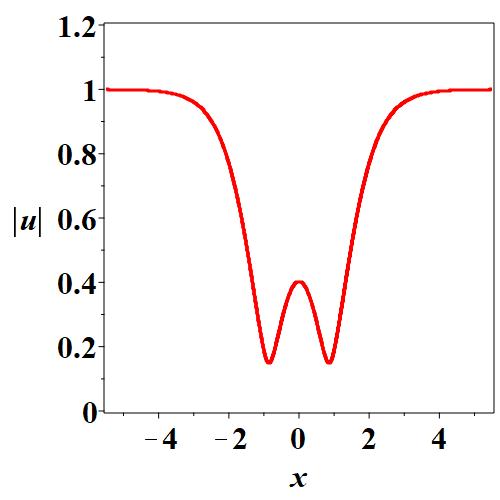}}
\caption{(Color online) The dark soliton solutions \eqref{1so-cosh} with three distinct valley profiles:  (a) the single-valley-shaped profiles with parameters $\vartheta_1=\frac{\pi}{3},\overline{\zeta}_{1,0}=0$; (b) the flat-bottom-shaped profile with parameters $\vartheta_1=\arctan\left(\frac{\sqrt {-2+2\,\sqrt {5-2\,\sqrt {5}}+2\,\sqrt {5}}}{1-\sqrt {5-2\,\sqrt {5}}}\right),\overline{\zeta}_{1,0}=0$; (c) the double-valley-shaped profiles with parameters $\vartheta_1=\frac{\pi}{9},\overline{\zeta}_{1,0}=0$. The bottom panels are intensity plots at $t=0$.
~}\label{Fig2}
\end{figure}

To better illustrate the correspondence between the soliton profiles and the range of values for \(\vartheta_1\), we will revisit the relationship between the sign of \(\Delta_0\) and \(\vartheta_1\). The condition $\Delta_0=0$ holds when $\vartheta_1$ takes the values
$\vartheta_1^{(1)}, \vartheta_1^{(2)}, \vartheta_1^{(3)}, \vartheta_1^{(4)}$,
which correspond to the flat-bottom soliton. For \(\vartheta_1 \in \left(-\frac{1}{2}\pi, \vartheta_1^{(1)}\right) \cup \left(\vartheta_1^{(2)}, \vartheta_1^{(3)}\right) \cup \left(\vartheta_1^{(4)}, \frac{1}{2}\pi\right)\), we have \(\Delta_0  <0\), leading to a double-valley-shaped dark soliton. Conversely, for \(\vartheta_1 \in \left(\vartheta_1^{(1)}, \vartheta_1^{(2)}\right) \cup \left(\vartheta_1^{(3)}, \vartheta_1^{(4)}\right)\), the condition \(\Delta_0 >  0\) holds, resulting in a single-valley-shaped dark soliton. Here $\vartheta_{1}^{(1,2,3,4)}$ are given as
\begin{equation}
\begin{aligned}
\vartheta_1^{(1)}=&\arctan\left(-\frac{\sqrt {-2+2\,\sqrt {5-2\,\sqrt {5}}+2\,\sqrt {5}}}{1-\sqrt {5-2\,\sqrt {5}}}\right), \quad
\vartheta_1^{(2)}=\arctan\left(-\frac{\sqrt {-2-2\,\sqrt {5-2\,\sqrt {5}}+2\,\sqrt {5}}}{1+\sqrt {5-2\,\sqrt {5}}}\right),\\
\vartheta_1^{(3)}=&\arctan\left(\frac{\sqrt {-2-2\,\sqrt {5-2\,\sqrt {5}}+2\,\sqrt {5}}}{1+\sqrt {5-2\,\sqrt {5}}}\right), \quad
\vartheta_1^{(4)}=\arctan\left(\frac{\sqrt {-2+2\,\sqrt {5-2\,\sqrt {5}}+2\,\sqrt {5}}}{1-\sqrt {5-2\,\sqrt {5}}}\right).
\end{aligned}
\end{equation}
The variations in the valley profiles of the dark soliton solutions \eqref{1so-cosh} as $\vartheta_1$ changes are shown in Fig.~\ref{Fig1}(b).
It is clear that the transition from a double-valley to a single-valley profile necessarily passes through an intermediate flat-bottom profile.

The single dark soliton solution \eqref{1so-cosh} with its three distinct valley profiles is illustrated in Fig.~\ref{Fig2}, corresponding to the conditions $\Delta_0>0$, $\Delta_0=0$, and $\Delta_0<0$, respectively.


Finally, we establish the relationship between the velocity $\nu$ and the valley profiles of the single type-I soliton.
Since the relation between $\vartheta_1$ and $\nu$ is given in Eq.~\eqref{1-v},
the dependence of the valley profiles on $\vartheta_1$ in Fig.~\ref{Fig1}(b)
can be equivalently expressed in terms of the velocity $\nu$,
as illustrated in Fig.~\ref{Fig1}(c).
\begin{figure}[!htbp]
\centering
\subfigure{\includegraphics[height=5cm,width=5.5cm]{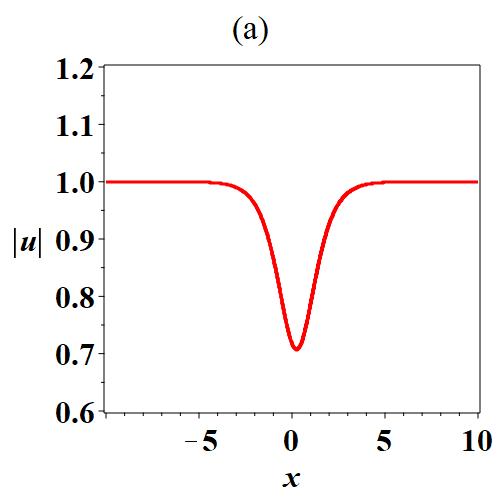}}
\subfigure{\includegraphics[height=5cm,width=5.5cm]{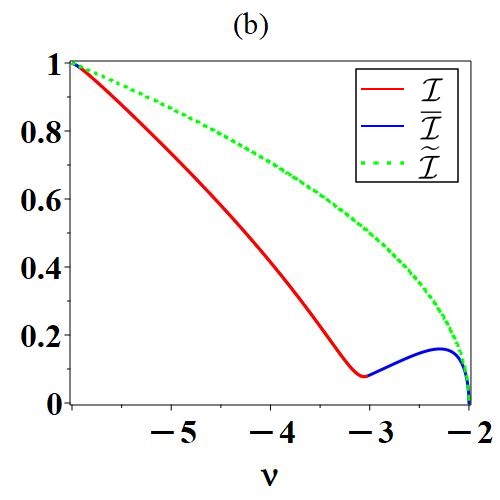}}
\caption{(Color online) (a) The intensity plot of type-II dark soliton solutions \eqref{1so-tanh} at $t=0$ with parameters $\vartheta_1=\tfrac{\pi}{3}$ and $\zeta_{1,0}=0$.
(b) The amplitude-velocity relations of the two types of single dark soliton solutions in Eqs.~\eqref{1so-cosh} and \eqref{1so-tanh}.
~}\label{Fig3}
\end{figure}

\subsubsection{Dynamics of single type-II dark solitons}
By setting $N_1=0$ and $N_2=1$ in Eqs.~\eqref{so-fg}--\eqref{mx-sj}, the single type-II dark soliton solution can be derived as
\begin{equation}\label{1so-tanh-0}
\begin{aligned}
u=\frac{e^{-\zeta_1}-\frac{1}{2}e^{2\mathrm{i}\vartheta_1}\sec\vartheta_1}{e^{-\zeta_1}+\frac{1}{2}\sec\vartheta_1}.
\end{aligned}
\end{equation}
To make the difference between this type-II soliton solution and the type-I dark soliton in Eq.~\eqref{1so-cosh}
more evident at the level of mathematical expressions, it can be rewritten as
\begin{equation}\label{1so-tanh}
\begin{aligned}
u=&-\frac{1}{2}\left[e^{2\mathrm{i}\vartheta_1}-1+(e^{2\mathrm{i}\vartheta_1}+1)\tanh\frac{\zeta_1+\zeta_0}{2}\right],
\end{aligned}
\end{equation}
with $\zeta_0=-\ln(2\cos\vartheta_1)$.
As $x$ varies from $-\infty$ to $+\infty$, the phase of $u$ changes by $(\pi+2\vartheta_1)$.
The magnitude $|u|$ has a single extremum at $\zeta_1+\zeta_0=0$, where it attains its minimum value,
\begin{equation}\label{1tan-AM-1}
\begin{aligned}
\widetilde{\mathcal{I}}=|\sin\vartheta_1|.
\end{aligned}
\end{equation}
Hence, this dark soliton always has a single-valley profile. Its velocity coincides with that of the type-I dark soliton in Eq.~\eqref{1so-cosh},
and is also given by Eq.~\eqref{1-v}. Therefore, the relationship between the amplitude and velocity of this single valley soliton can be expressed by the following function:
\begin{equation}\label{1tan-AM-v}
\begin{aligned}
 \nu=-4\widetilde{\mathcal{I}}^2-2.
\end{aligned}
\end{equation}
An example of this profile is displayed in Fig.~\ref{Fig3}(a).

Finally, we summarize the main differences between these two types of solitons:
\begin{itemize}
    \item[(i)] The type-I dark solitons in Eq.~\eqref{1so-cosh} acquire no phase shift as \(x\) varies from \(-\infty\) to \(+\infty\), whereas the phase of the type-II dark soliton shifts by an amount of \((\pi + 2\vartheta_1)\) as \(x\) varies from \(-\infty\) to \(+\infty\);
    \item[(ii)] The type-I dark solitons exhibit three distinct valley profiles: double-valley, single-valley, and flat-bottom, while the type-II dark soliton only exhibits a single-valley profile;
    \item[(iii)] The two types of dark solitons obey different amplitude-velocity relations. To further visually emphasize the differences in velocity and amplitude between these two types of solitons, we present the evolution of amplitude with velocity in Fig. \ref{Fig3}(b). Additionally, it can be directly inferred from the figure that the velocity is closely related to the valley profile of the type-I dark soliton. There are exactly two velocity intervals in which the type-I dark soliton exhibits a double-valley profile (corresponding to the velocity intervals of the blue line), while in a broader velocity interval, the type-I dark soliton displays a single-valley profile.
\end{itemize}

\section{Collisions between two dark soliton solutions}\label{Collision-T}
The solution \eqref{D-SO} represents two dark solitons when the condition $N_1 + N_2 = 2$ is satisfied in Eqs. \eqref{so-fg}--\eqref{pa-1}. These two solitons can exhibit three different configurations depending on the values of $N_1$ and $N_2$: (i) two type-I dark solitons for $N_1 = 2$ and $N_2 = 0$; (ii) two type-II dark solitons for $N_1 = 0$ and $N_2 = 2$; (iii) a mixture of one type-I dark soliton and one type-II dark soliton for $N_1 = 1$ and $N_2 = 1$. In what follows, we analytically examine the dynamics of these three kinds of two-soliton collisions.

For convenience, we denote the type-I soliton along the trajectory
$\zeta_{k_1}\simeq \mathcal{O}(1)$ as ${\bf S}_{k_1}$,
and the type-II soliton along the trajectory
$\zeta_{2N_1+k_2}\simeq \mathcal{O}(1)$ as ${\bf \overline{S}}_{k_2}$,
where $k_1=1,2,\dots,N_1$ and $k_2=1,2,\dots,N_2$.
The asymptotic expressions of ${\bf S}_{k_1}$ as $t\to\pm\infty$ are denoted by $u_{k_1}^{\pm}$,
while those of ${\bf \overline{S}}_{k_2}$ are denoted by $\overline{u}_{k_2}^{\pm}$.
Our asymptotic analysis will be performed under the following velocity ordering:
\begin{equation}\label{velo-DN}
\nu_1<\cdots<\nu_{k_1}<\cdots<\nu_{N_1}
<\overline{\nu}_{1}<\cdots<\overline{\nu}_{k_2}
<\cdots<\overline{\nu}_{N_2}<0,
\end{equation}
where $\nu_{k_1}$ and $\overline{\nu}_{k_2}$, the velocities of ${\bf S}_{k_1}$ and ${\bf \overline{S}}_{k_2}$,
are defined by
\begin{equation}
\begin{aligned}
\nu_{k_1} &= 2\cos(2\vartheta_{k_1})-4, \\
\overline{\nu}_{k_2} &= 2\cos(2\vartheta_{2N_1+k_2})-4.
\end{aligned}
\end{equation}

\subsection{Collisions between two type-I dark solitons}
By setting $N_1=2$ and $N_2=0$ in Eqs.~\eqref{so-fg}--\eqref{pa-1}, the solution \eqref{D-SO} represents two type-I dark solitons, with associated functions $f$ and $g$ given by
\begin{equation}\label{2fg-cosh}
\begin{aligned}
f=&\left|\begin{matrix}
e^{-\zeta_1}+\overline{\mathrm{m}}_{1,1}^{(0)}&\overline{\mathrm{m}}_{1,2}^{(0)}&\overline{\mathrm{m}}_{1,3}^{(0)}&\overline{\mathrm{m}}_{1,4}^{(0)}\\
\overline{\mathrm{m}}_{2,1}^{(0)}&e^{-\zeta_2}+\overline{\mathrm{m}}_{2,2}^{(0)}&\overline{\mathrm{m}}_{2,3}^{(0)}&\overline{\mathrm{m}}_{2,4}^{(0)}\\
-\overline{\mathrm{m}}_{1,3}^{(0)}&-\overline{\mathrm{m}}_{1,4}^{(0)}&-e^{\zeta_1}-\overline{\mathrm{m}}_{1,1}^{(0)}&-\overline{\mathrm{m}}_{1,2}^{(0)}\\
-\overline{\mathrm{m}}_{2,3}^{(0)}&-\overline{\mathrm{m}}_{2,4}^{(0)}&-\overline{\mathrm{m}}_{2,1}^{(0)}&-e^{\zeta_2}-\overline{\mathrm{m}}_{2,2}^{(0)}
\end{matrix}\right|,\\
g=&\left|\begin{matrix}
e^{-\zeta_1}+\overline{\mathrm{m}}_{1,1}^{(1)}&\overline{\mathrm{m}}_{1,2}^{(1)}&\overline{\mathrm{m}}_{1,3}^{(1)}&\overline{\mathrm{m}}_{1,4}^{(1)}\\
\overline{\mathrm{m}}_{2,1}^{(1)}&e^{-\zeta_2}+\overline{\mathrm{m}}_{2,2}^{(1)}&\overline{\mathrm{m}}_{2,3}^{(1)}&\overline{\mathrm{m}}_{2,4}^{(1)}\\
-\overline{\mathrm{m}}_{1,3}^{(1)}&-\overline{\mathrm{m}}_{1,4}^{(1)}&-e^{\zeta_1}-\overline{\mathrm{m}}_{1,1}^{(1)}&-\overline{\mathrm{m}}_{1,2}^{(1)}\\
-\overline{\mathrm{m}}_{2,3}^{(1)}&-\overline{\mathrm{m}}_{2,4}^{(1)}&-\overline{\mathrm{m}}_{2,1}^{(1)}&-e^{\zeta_2}-\overline{\mathrm{m}}_{2,2}^{(1)}
\end{matrix}\right|,
\end{aligned}
\end{equation}
where $\overline{\mathrm{m}}_{i,j}^{(n)}$ and $\zeta_i$ are defined in Eq.~\eqref{mx-sj}, subject to $|p_i|=1$ with $p_3=-p_1$ and $p_4=-p_2$.
The two type-I dark solitons are located along $\zeta_1\simeq\mathcal{O}(1)$ and $\zeta_2\simeq\mathcal{O}(1)$,
with velocities $\nu_1=2\cos(2\vartheta_1)-4$ and $\nu_2=2\cos(2\vartheta_2)-4$. Here $|p_i|=1$ is equivalently written as $p_i=e^{\mathrm{i}\vartheta_i}$. We denote the solitons by ${\bf S}_1$ and ${\bf S}_2$,  with velocity ordering $\nu_1<\nu_2<0$, to distinguish them in the subsequent asymptotic analysis. As before, we impose $\sec\vartheta_i>0$ to ensure smoothness of the solution. Along the trajectories \(\zeta_1 \simeq \mathcal{O}(1)\) and \(\zeta_2 \simeq \mathcal{O}(1)\), as well as in the limits \(t \rightarrow \pm\infty\), the asymptotic expressions for these two solitons are organized as follows:

(a)\underline{ Before collision $t\rightarrow-\infty$}\\

Soliton ${\bf S}_1 \left(\zeta_1\simeq \mathcal{O}(1),\zeta_2\rightarrow+\infty\right)$:
\begin{equation}\label{2so-Asy-1-}
\begin{aligned}
u_1^{-}\simeq\frac{\chi_1^{(1)}+\Theta_1^{(1)}\cosh(\zeta_1+\Omega_1)}
{\chi_1^{(0)}+\Theta_1^{(0)}\cosh(\zeta_1+\Omega_1)},
\end{aligned}
\end{equation}

Soliton ${\bf S}_2 \left(\zeta_2\simeq\mathcal{O}(1),\zeta_1\rightarrow-\infty\right)$:
\begin{equation}\label{2so-Asy-2-}
\begin{aligned}
u_2^{-}\simeq\frac{\chi_2^{(1)}+\Theta_2^{(1)}\cosh(\zeta_2+\Omega_2)}
{\chi_2^{(0)}+\Theta_2^{(0)}\cosh(\zeta_2+\Omega_2)},
\end{aligned}
\end{equation}
where
\begin{equation}\label{Chi}
\begin{aligned}
\chi_1^{(n)}=&(-\frac{p_2}{p_2^*})^n\frac{1}{p_2+p_2^*}\left[1-(\frac{p_1}{p_1^*})^{2n}\frac{4|p_1^2-p_2^2|^2}{(p_1^2-p_1^{*2})^2|p_1^2-p_2^{*2}|^2}\right],\\
\chi_2^{(n)}=&(-\frac{p_1}{p_1^*})^n\frac{1}{p_1+p_1^*}\left[1-(\frac{p_2}{p_2^*})^{2n}\frac{4|p_1^2-p_2^2|^2}{(p_2^2-p_2^{*2})^2|p_1^2-p_2^{*2}|^2}\right],\\
\Theta_1^{(n)}=&\frac{2}{(p_1+p_1^*)(p_2+p_2^*)}\left|\frac{p_1^2-p_2^2}{p_1^2-p_2^{*2}}\right|(\frac{p_1p_2}{p_1^*p_2^*})^n, \quad n=0,1,\\
\Omega_1=&\ln\left|\frac{(p_1-p_2)(p_1-p_2^*)}{(p_1+p_2)(p_1+p_2^*)}\right|, \quad \Theta_2^{(n)}=\Theta_1^{(n)}, \quad \Omega_2=-\Omega_1.
\end{aligned}
\end{equation}

(b)\underline{ After collision $t\rightarrow+\infty$}\\

Soliton ${\bf S}_1 \left(\zeta_1\simeq\mathcal{O}(1),\zeta_2\rightarrow-\infty\right)$:
\begin{equation}\label{2so-Asy-1+}
\begin{aligned}
u_1^{+}\simeq\frac{\chi_1^{(1)}+\Theta_1^{(1)}\cosh(\zeta_1-\Omega_1)}
{\chi_1^{(0)}+\Theta_1^{(0)}\cosh(\zeta_1-\Omega_1)},
\end{aligned}
\end{equation}

\begin{figure}[!htbp]
\centering
\subfigure{\includegraphics[height=5cm,width=5.5cm]{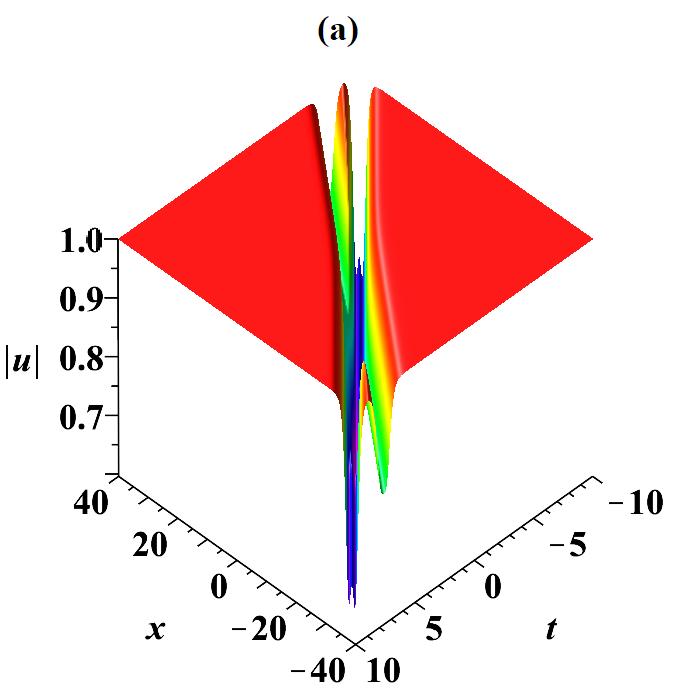}}
\subfigure{\includegraphics[height=5cm,width=5.5cm]{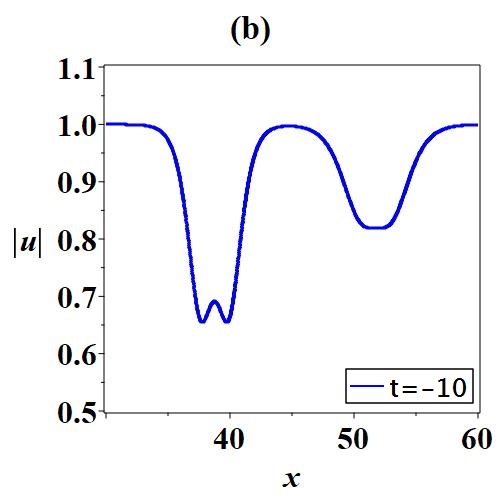}}
\subfigure{\includegraphics[height=5cm,width=5.5cm]{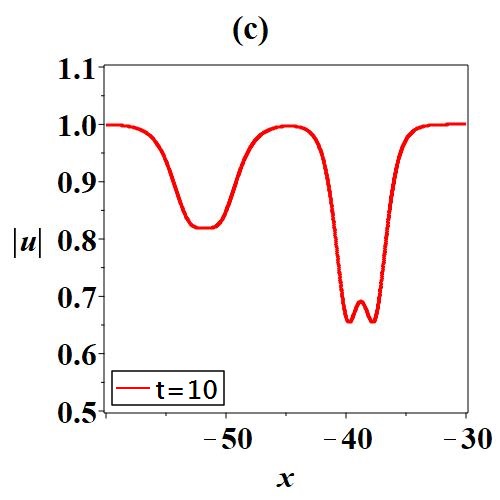}}
\caption{(Color online)(a)The shape-preserving collisions of two type-I dark soliton solutions with $(p_1,p_2)=(e^{\frac{1}{3}\mathrm{i}\pi},e^{\frac{1}{4}\mathrm{i}\pi})$ and
$(\zeta_{1,0},\zeta_{2,0})=(0,0)$. (b) and (c) are the intensity plots of the two type-I dark soliton solutions displayed in panel (a) at $t=-10$ and $t=10$, respectively.
~}\label{Fig4}
\end{figure}

Soliton ${\bf S}_2 \left(\zeta_2\simeq\mathcal{O}(1),\zeta_1\rightarrow+\infty\right)$:
\begin{equation}\label{2so-Asy-2+}
\begin{aligned}
u_2^{+}\simeq\frac{\chi_2^{(1)}+\Theta_2^{(1)}\cosh(\zeta_2-\Omega_2)}
{\chi_2^{(0)}+\Theta_2^{(0)}\cosh(\zeta_2-\Omega_2)}.
\end{aligned}
\end{equation}
The asymptotic analysis above directly implies that
\begin{equation}
\begin{aligned}
u_1^{+}(\zeta_1)=u_1^{-}(\zeta_1+\Delta\Phi_1),\quad u_2^{+}(\zeta_2)=u_2^{-}(\zeta_2+\Delta\Phi_2),
\end{aligned}
\end{equation}
where $\Delta\Phi_1,\Delta\Phi_2$ represent the phase shifts of solitons ${\bf S}_1$ and ${\bf S}_2$ during their collisions, respectively. They are given explicitly by
\begin{equation}
\begin{aligned}
\Delta\Phi_1=-\Delta\Phi_2=-2\Omega_1.
\end{aligned}
\end{equation}
This indicates two properties: (i) the collisions between the two type-I dark solitons do not alter their shapes, resulting in elastic or shape-preserving collisions. (ii) the total phase shifts of the two solitons sum to zero, i.e., $\Delta\Phi_1 + \Delta\Phi_2 = 0$. Fig.~\ref{Fig4} illustrates an example of a shape-preserving collision between a type-I dark soliton with a double-valley profile and a type-I dark soliton with a flat-bottom-shaped profile, demonstrating that both solitons retain their shape after the collision.

Although the collisions between these two solitons do not alter their shapes, one soliton can still influence the other. Specifically, altering the amplitude or velocity of one soliton will induce changes in the shape or amplitude of the other soliton, even if its velocity remains constant. Below, we will discuss the unique collision properties from two perspectives. Before that, we emphasize again that the velocity of soliton ${\bf S}_{k_1}$ ($k_1=1,2$) is determined solely by $p_{k_1}$ ($p_{k_1} = e^{\mathrm{i}\vartheta_{k_1}}$), with the magnitude of the velocity given by $\nu_{k_1} = 2\cos(\vartheta_{k_1}) - 4$. Changing the value of $p_{k_1}$ (or $\vartheta_{k_1}$) only affects the velocity of soliton ${\bf S}_{k_1}$ without altering the velocity of the other soliton ${\bf S}_{3-k_1}$.

\begin{figure}[!htbp]
\centering
\subfigure{\includegraphics[height=5cm,width=6.5cm]{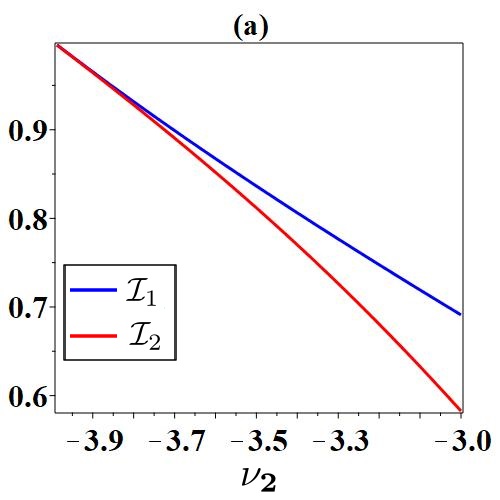}}
\subfigure{\includegraphics[height=5cm,width=6.5cm]{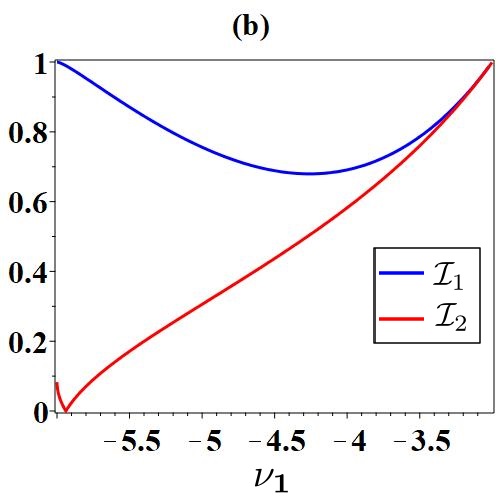}}
\caption{(Color online) (a) The numerical variations of $\mathcal{I}_1$ and $\mathcal{I}_2$ with respect to the velocity $\nu_2$ of soliton ${\bf S}_2$, while the velocity of soliton ${\bf S}_1$ is fixed at $\nu_1=-4$ (i.e., $p_1=e^{\frac{1}{4}\mathrm{i} \pi}$).
(b)The numerical variations of $\mathcal{I}_1$ and $\mathcal{I}_2$ with respect to the velocity $\nu_1$ of soliton ${\bf S}_1$, while the velocity of soliton ${\bf S}_2$ is fixed at $\nu_2 = -3$ (i.e., $p_2=e^{\frac{1}{6}\mathrm{i}\pi}$).
~}\label{Fig5}
\end{figure}

The first aspect concerns the relationship between the amplitudes of the two type-I solitons at their centers and their velocities. Since the amplitudes and shapes of the two type-I solitons remain unchanged before and after their collisions, we will only consider the amplitudes before the collisions and disregard those afterward. From the asymptotic expressions in Eqs. \eqref{2so-Asy-1-} and \eqref{2so-Asy-2-} for solitons \( {\bf S}_1 \) and \( {\bf S}_2 \), it can be deduced that their amplitudes at their centers, where \(\zeta_1 + \Omega_1^- = 0\) and \(\zeta_2 + \Omega_2^- = 0\), are given by:
\begin{equation}
\begin{aligned}
\mathcal{I}_1=&\left.|u_1^{-}|\right|_{\zeta_1=- \Omega_1} = \left|\frac{\chi_1^{(1)} + \Theta_1^{(1)}}{\chi_1^{(0)} + \Theta_1^{(0)}}\right|, \\
\mathcal{I}_2=&\left.|u_2^{-}|\right|_{\zeta_2=- \Omega_2} =\left|\frac{\chi_2^{(1)} + \Theta_2^{(1)}}{\chi_2^{(0)} + \Theta_2^{(0)}}\right|.
\end{aligned}
\end{equation}
Using the explicit expressions for $\chi_{k_1}^{(n)}$ and $\Theta_{k_1}^{(n)}$ in Eq.~\eqref{Chi},
we find that both $\mathcal{I}_1$ and $\mathcal{I}_2$ depend on $p_1$ and $p_2$,
which are related to the velocities of solitons ${\bf S}_1$ and ${\bf S}_2$.
Thus, the amplitudes of the two solitons are jointly determined by both velocities.
To illustrate this correlation, we fix the velocity of one soliton, say $\nu_s$ ($s=1$ or $2$),
and examine how $\mathcal{I}_1$ and $\mathcal{I}_2$ vary with the velocity $\nu_{3-s}$ of the other. For instance, if we take \(p_1 = e^{\frac{1}{4}\mathrm{i}\pi}\), then the velocity of soliton \({\bf S}_1\) is fixed at \(\nu_1 = -4\). To satisfy the velocity constraint \(\nu_1 < \nu_2 \leq 0\) required by our asymptotic analysis, we set the range of \(\vartheta_2\) (\(p_2 = e^{\mathrm{i}\vartheta_2}\)) to \(\left[\frac{\pi}{6}, \frac{\pi}{4}\right)\), which corresponds to the velocity range of soliton \({\bf S}_2\) being \(\nu_2 \in [-3, -4)\). Fig.~\ref{Fig5}(a) shows that while the amplitude $\mathcal{I}_2$ of soliton ${\bf S}_2$ varies with $\nu_2$,
the amplitude $\mathcal{I}_1$ of soliton ${\bf S}_1$ also changes with $\nu_2$,
even though $\nu_1$ remains fixed. Similarly, by setting \(p_2 = e^{\frac{1}{6} \mathrm{i} \pi}\), we have \(\nu_2 = -3\), fixing the velocity of soliton \({\bf S}_2\). The numerical variations of \(\mathcal{I}_1\) and \(\mathcal{I}_2\) with respect to the velocity \(\nu_1\) of soliton \({\bf S}_1\) are shown in Fig.~\ref{Fig5}(b), further demonstrating this property.

The second aspect concerns the impact of the velocities of the two type-I solitons on their valley profiles(or soliton shapes).
As in the single type-I dark soliton case, the valley profile of each soliton can be determined from the sign of $\Delta_{k_1}$,
where
\begin{equation}
\begin{aligned}
\Delta_{k_1} = \left.\left(\partial_{\zeta_{k_1}\zeta_{k_1}}^2 \left|u_{k_1}^{-}\right|\right)\right|_{\zeta_{k_1}=-\Omega_{k_1}^- },\,\,\,\,k_1=1,2.
\end{aligned}
\end{equation}
The valley profiles of soliton ${\bf S}_{k_1}$ can be classified into three categories according to the sign of $\Delta_{k_1}$:
double-valley when $\Delta_{k_1}<0$, flat-bottom when $\Delta_{k_1}=0$, and single-valley when $\Delta_{k_1}>0$.
It follows that $\Delta_{k_1}$ depends only on $p_1$ and $p_2$, and hence on the velocities $\nu_1,\nu_2$ of the two type-I solitons. Due to the complexity of the specific algebraic expression of \( \Delta_{k_1} \) in terms of \( \nu_1 \) and \( \nu_2 \), the explicit form is not provided here. Specifically, the signs of $\Delta_{1}$ and $\Delta_{2}$ are determined simultaneously by the velocities $\nu_1$ and $\nu_2$ of the two solitons.

\begin{figure}[!htbp]
\centering
\subfigure{\includegraphics[height=5cm,width=6.5cm]{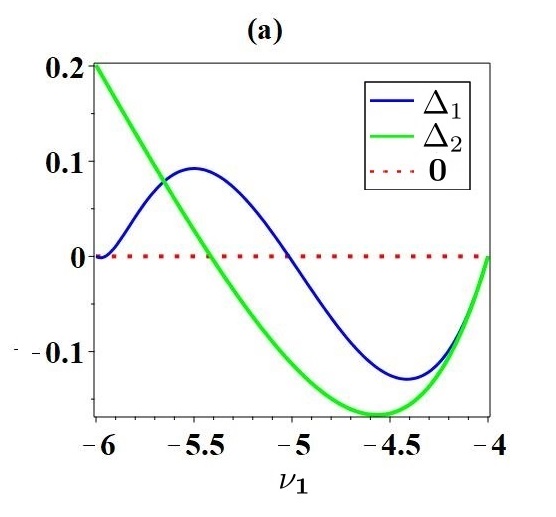}}
\subfigure{\includegraphics[height=5cm,width=6.5cm]{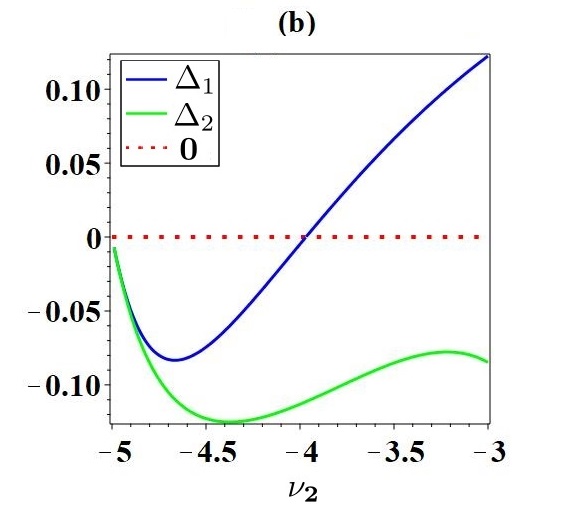}}
\caption{(Color online) (a) The numerical variations of $\mathcal{I}_1$ and $\mathcal{I}_2$ with respect to the velocity $\nu_2$ of soliton ${\bf S}_2$, while the velocity of soliton ${\bf S}_1$ is fixed at $\nu_1=-4$ (i.e., $p_1=e^{\frac{1}{4}\mathrm{i} \pi}$).
(b)The numerical variations of $\mathcal{I}_1$ and $\mathcal{I}_2$ with respect to the velocity $\nu_1$ of soliton ${\bf S}_1$, while the velocity of soliton ${\bf S}_2$ is fixed at $\nu_2 = -3$ (i.e., $p_2=e^{\frac{1}{6}\mathrm{i}\pi}$).
~}\label{Fig6}
\end{figure}

\begin{figure}[!htbp]
\centering
\subfigure{\includegraphics[height=5cm,width=6.5cm]{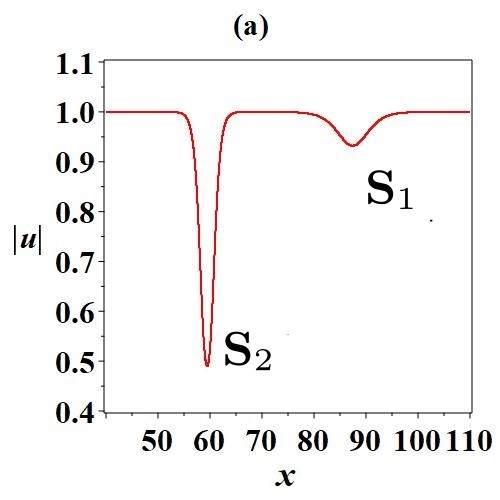}}
\subfigure{\includegraphics[height=5cm,width=6.5cm]{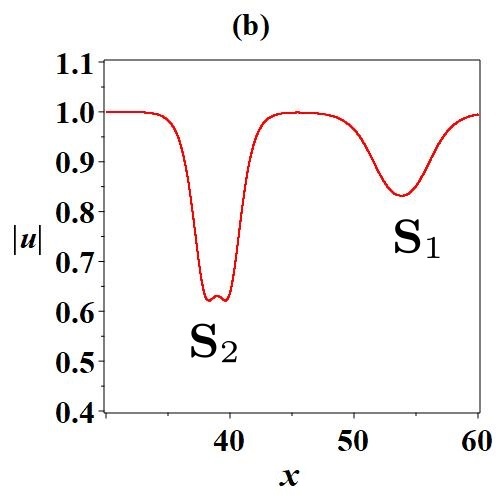}}\\
\subfigure{\includegraphics[height=5cm,width=6.5cm]{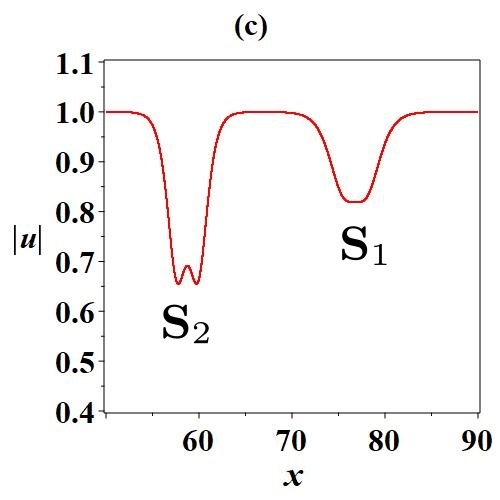}}
\subfigure{\includegraphics[height=5cm,width=6.5cm]{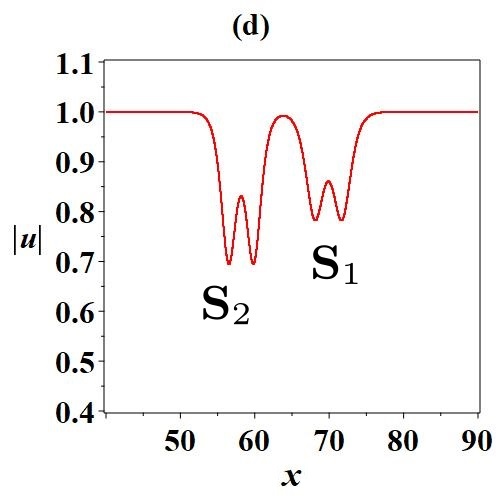}}
\caption{(Color online)The shape alterations of two type-I dark solitons as the velocity $\nu_1$ of ${\bf S}_1$ varies,
with ${\bf S}_2$ fixed at $\nu_2=-4$ ($p_2=e^{\tfrac{1}{4}\mathrm{i}\pi}$).
The solutions \eqref{2fg-cosh} are plotted at $t=-15$ with $\zeta_{1,0}=\zeta_{2,0}=0$:
 (a) two single-valley-shaped solitons with $\nu_1=2\cos(\frac{5}{6}\pi)-4$ (i.e., $p_1=e^{\frac{5}{12}\mathrm{i}\pi}$);
 (b)  a double-valley-shaped soliton  coexisting  with a  single-valley-shaped soliton with $\nu_1=2\cos(\frac{17}{24}\pi)-4$ (i.e., $p_1=e^{\frac{17}{48}\mathrm{i}\pi}$);
 (c) a double-valley-shaped soliton  coexisting  with a  flat-bottom-shaped soliton with $\nu_1=-5$ (i.e., $p_1=e^{\frac{1}{3}\mathrm{i}\pi}$);
 (d) two double-valley-shaped solitons with $\nu_1=2\cos(\frac{7}{12}\pi)-4$ (i.e., $p_1=e^{\frac{7}{24}\mathrm{i}\pi}$).
~}\label{Fig7}
\end{figure}

To illustrate this, we fix the velocity of soliton ${\bf S}_2$ at $\nu_2=-4$ ($p_2=e^{\tfrac{\pi}{4}\mathrm{i}}$).
In this case, both $\Delta_1$ and $\Delta_2$ can be regarded as functions of $\nu_1$.
The numerical evolution of $\Delta_1$ and $\Delta_2$ with respect to $\nu_1$ is shown in Fig.~\ref{Fig6}(a).
We observe that $\Delta_1,\Delta_2>0$ for $\nu_1\in(-6,-5.5)$, indicating that both solitons are single-valley.
By contrast, $\Delta_1,\Delta_2<0$ for $\nu_1\in(-5,-4)$, so both solitons become double-valley. The detailed evolution of the shapes of solitons ${\bf S}_1$ and ${\bf S}_2$
as the velocity $\nu_1$ of ${\bf S}_1$ varies is shown in Fig.~\ref{Fig7}. It clearly illustrates that while the velocity of soliton ${\bf S}_2$ remains fixed at $\nu_2 = -4$, the shapes of the solitons vary with changes in $\nu_1$.
In Fig.~\ref{Fig7}(a), when the velocity of soliton ${\bf S}_1$ is $\nu_1 = 2\cos\left(\frac{5}{6}\pi\right)-4$, both solitons ${\bf S}_1$ and ${\bf S}_2$ are single-valley-shaped. However, when the velocity $\nu_1$ changes to $\nu_1 = -5$ in Fig.~\ref{Fig7}(b), soliton ${\bf S}_1$ remains single-valley-shaped, while the shape of soliton ${\bf S}_2$ transforms to a double-valley shape.
In particular, when the velocity $\nu_1$ of soliton ${\bf S}_1$ changes to $\nu_1 = -4$ in Fig.~\ref{Fig7}(c), the shape of soliton ${\bf S}_1$ becomes flat-bottom-shaped, whereas the shape of soliton ${\bf S}_2$ remains double-valley-shaped. Finally, when the velocity of soliton ${\bf S}_1$ changes to $\nu_1 = 2\cos\left(\frac{7}{12}\pi\right) - 4$ in Fig.~\ref{Fig7}(d), both solitons ${\bf S}_1$ and ${\bf S}_2$ are double-valley-shaped.

We further confirm that the valley profiles of solitons ${\bf S}_1$ and ${\bf S}_2$
vary with the velocity $\nu_2$ of ${\bf S}_2$,
when ${\bf S}_1$ is fixed at $\nu_1=-5$.
This dependence is evident from the evolution of the sign of $\Delta_2$ with respect to $\nu_2$,
as shown in Fig.~\ref{Fig6}(b).

\subsection{Collisions between two type-II dark solitons}
Setting $N_1=0,N_2=1$ in Eqs.\eqref{so-fg}--\eqref{pa-1}, the solution \eqref{D-SO} corresponds to two type-II dark solitons,  with associated functions $f$ and $g$ given by
\begin{equation}\label{1fg-TII}
\begin{aligned}
f=&\left|\begin{matrix}
e^{-\zeta_1}+\overline{\mathrm{m}}_{1,1}^{(0)}&\overline{\mathrm{m}}_{1,2}^{(0)}\\
\overline{\mathrm{m}}_{2,1}^{(0)}&e^{-\zeta_2}+\overline{\mathrm{m}}_{2,2}^{(0)}
\end{matrix}\right|,\\
g=&\left|\begin{matrix}
e^{-\zeta_1}+\overline{\mathrm{m}}_{1,1}^{(1)}&\overline{\mathrm{m}}_{1,2}^{(1)}\\
\overline{\mathrm{m}}_{2,1}^{(1)}&e^{-\zeta_2}+\overline{\mathrm{m}}_{2,2}^{(1)}
\end{matrix}\right|,
\end{aligned}
\end{equation}
 where $\zeta_i$ and $\mathrm{m}_{i,j}^{(n)}$ are explicitly given by Eq.~\eqref{mx-sj}. To reveal the collisions of the two type-II dark solitons, we also  perform the asymptotic analysis for the two type-II dark solitons under the velocity constraint $\overline{\nu}_1<\overline{\nu}_2<0$. After algebraic simplification, the asymptotic expression is concisely written in the following form:

\begin{equation}\label{1AS-TII}
\begin{aligned}
\overline{u}_{k_2}^{\pm}\simeq&-\frac{1}{2}\mu_{k_2}^{\pm}\left[e^{2\mathrm{i}\vartheta_{k_2}}-1+(e^{2\mathrm{i}\vartheta_{k_2}}+1)\tanh\frac{\zeta_{k_2}+\overline{\Omega}^{\pm}_{k_2}}{2}\right],\quad \,{k_2}=1,2,
\end{aligned}
\end{equation}
where
\begin{equation}
\begin{aligned}
&\mu_1^{-}=\mu_2^{+}=1, \quad \mu_1^{+}=e^{\mathrm{i}(2\vartheta_2+\pi)}, \quad \mu_2^{-}=e^{\mathrm{i}(2\vartheta_1+\pi)},\\
&\overline{\Omega}_1^{-}=\ln\left(\frac{1}{2\cos\vartheta_1}\right), \quad \overline{\Omega}_1^{+}=\overline{\Omega}_1^{-}+\Delta\overline{\Phi}, \quad
\overline{\Omega}_2^{-}=\overline{\Omega}_2^{+}+\Delta\overline{\Phi},\\
&\overline{\Omega}_2^{+}=\ln\left(\frac{1}{2\cos\vartheta_2}\right), \quad
\Delta\overline{\Phi}=2\ln\left|\frac{e^{\mathrm{i}\vartheta_1}-e^{\mathrm{i}\vartheta_2}}{e^{\mathrm{i}\vartheta_1}+e^{-\mathrm{i}\vartheta_2}}\right|.
\end{aligned}
\end{equation}
Here we take $p_1=e^{\mathrm{i}\vartheta_1}$ and $p_2=e^{\mathrm{i}\vartheta_2}$.
The asymptotic expressions yield
$|\overline{u}_1^{+}(\zeta_1)|=|\overline{u}_1^{-}(\zeta_1+\Delta\overline{\Phi})|$
and
$|\overline{u}_2^{+}(\zeta_2)|=|\overline{u}_2^{-}(\zeta_2-\Delta\overline{\Phi})|$,
showing that collisions between two type-II solitons are shape-preserving.
One soliton acquires a phase shift $\Delta\overline{\Phi}$, the other $-\Delta\overline{\Phi}$,
so the total phase shift vanishes. The amplitude of a soliton centered around $\zeta_k\simeq\mathcal{O}(1)$ is
$\widetilde{\mathcal{I}}_k=|\sin\vartheta_k|$,
depends only on its own velocity and is independent of the other soliton.
This characteristic is completely different from that of type-I solitons, where the amplitude of each soliton is dependent on the velocities of both solitons.

\subsection{Collisions between a type-I dark soliton and a type-II dark soliton}
Setting $N_1=N_2=1$ in Eqs.~\eqref{so-fg}--\eqref{pa-1}, the solution \eqref{D-SO} corresponds to a type-I dark soliton coexisting with a type-II dark soliton,
and associated functions $f$ and $g$ given by
\begin{equation}\label{1fg-TI-TII}
\begin{aligned}
f=&\left|\begin{matrix}
e^{-\zeta_1}+\overline{\mathrm{m}}_{1,1}^{(0)}&\overline{\mathrm{m}}_{1,2}^{(0)}&\overline{\mathrm{m}}_{1,3}^{(0)}\\
-\overline{\mathrm{m}}_{1,2}^{(0)}&-e^{-\zeta_1}-\overline{\mathrm{m}}_{1,1}^{(0)}&\overline{\mathrm{m}}_{2,3}^{(0)}\\
\overline{\mathrm{m}}_{1,3}^{(0)*}&\overline{\mathrm{m}}_{2,3}^{(0)*}&e^{-\zeta_3}+\overline{\mathrm{m}}_{3,3}^{(0)}
\end{matrix}\right|,\\
g=&\left|\begin{matrix}
e^{-\zeta_1}+\overline{\mathrm{m}}_{1,1}^{(1)}&\overline{\mathrm{m}}_{1,2}^{(1)}&\overline{\mathrm{m}}_{1,3}^{(1)}\\
-\overline{\mathrm{m}}_{1,2}^{(1)}&-e^{-\zeta_1}-\overline{\mathrm{m}}_{1,1}^{(1)}&\overline{\mathrm{m}}_{2,3}^{(1)}\\
\overline{\mathrm{m}}_{1,3}^{(-1)*}&\overline{\mathrm{m}}_{2,3}^{(-1)*}&e^{-\zeta_3}+\overline{\mathrm{m}}_{3,3}^{(1)}
\end{matrix}\right|,
\end{aligned}
\end{equation}
 where $\zeta_k$ and $\overline{m}_{i,j}^{(n)}$ are explicitly defined in Eq.~\eqref{mx-sj}. To reveal the collisions between the two mixed dark solitons, asymptotic analysis is performed for them under the velocity correlation $\nu_1 < \overline{\nu}_1 < 0$. Here $\nu_1$($=2\cos2\vartheta_1-4$) is the velocity of the type-I dark soliton and $\overline{\nu}_1$($=2\cos2\vartheta_3-4$) is the velocity of the type-II dark soliton. We denote the type-I dark soliton by ${\bf S}_1$ and the type-II dark soliton by ${\bf \overline{S}}_1$. After algebraic simplification, the asymptotic expression is concisely written in the following form:

(a)\underline{ Before collision $t\rightarrow-\infty$}\\

Soliton ${\bf S}_1 \left(\zeta_1\simeq \mathcal{O}(1),\zeta_3\rightarrow+\infty\right)$:
\begin{equation}\label{Mso-Asy-1-}
\begin{aligned}
u_1^{-}\simeq\frac{\widetilde{\chi}_1^{(1)-}+\widetilde{\Theta}_1^{(1)-}\cosh(\zeta_1+\widetilde{\Omega}_1^{-})}
{\widetilde{\chi}_1^{(0)-}+\widetilde{\Theta}_1^{(0)-}\cosh(\zeta_1+\widetilde{\Omega}_1^{-})},
\end{aligned}
\end{equation}

Soliton ${\bf \overline{S}}_1 \left(\zeta_3\simeq \mathcal{O}(1),\zeta_1\rightarrow-\infty\right)$:
\begin{equation}\label{Mso-Asy-2-}
\begin{aligned}
\overline{u}_1^{-}\simeq-\frac{1}{2}\widehat{\mu}_1^{-}\left[e^{2\mathrm{i}\vartheta_3}-1+(e^{2\mathrm{i}\vartheta_3}+1)\tanh\frac{\zeta_3+\widehat{\Omega}^{-}_1}{2}\right],
\end{aligned}
\end{equation}
where
\begin{equation}\label{M-Chi-}
\begin{aligned}
\widetilde{\chi}_1^{(n)-}=&-(-\frac{p_3}{p_3^*})^n\frac{1}{p_3+p_3^*}\left[1-(\frac{p_1}{p_1^*})^{2n}\frac{4|p_1^2-p_3^2|^2}{(p_1^2-p_1^{*2})^2|p_1^2-p_3^{*2}|^2}\right],\\
\widetilde{\Theta}_1^{(n)-}=&-\frac{2}{(p_1+p_1^*)(p_3+p_3^*)}\left|\frac{p_1^2-p_3^2}{p_1^2-p_3^{*2}}\right|(\frac{p_1p_3}{p_1^*p_3^*})^n, \quad n =0,1,\\
\widetilde{\Omega}_1^{-}=&\ln\left|\frac{(p_1-p_3)(p_1-p_3^*)}{(p_1+p_3)(p_1+p_3^*)}\right|, \quad \widehat{\Omega}_1^{-}=\ln\left(\frac{1}{p_3+p_3^*}\left|\frac{p_1+p_3}{p_1-p_3^*}\right|^2\right), \quad
\widehat{\mu}_1^{-}=\widehat{\mu}_1=-\frac{p_1}{p_1^*}.
\end{aligned}
\end{equation}

(b)\underline{ After collision $t\rightarrow+\infty$}\\

Soliton ${\bf S}_1 \left(\zeta_1\simeq \mathcal{O}(1),\zeta_3\rightarrow-\infty\right)$:
\begin{equation}\label{Mso-Asy-1+}
\begin{aligned}
u_1^{+}\simeq\frac{\widetilde{\chi}_1^{(1)+}+\widetilde{\Theta}_1^{(1)+}\cosh(\zeta_1)}
{\widetilde{\chi}_1^{(0)+}+\widetilde{\Theta}_1^{(0)+}\cosh(\zeta_1)},
\end{aligned}
\end{equation}

\begin{figure}[!htbp]
\centering
\subfigure{\includegraphics[height=5cm,width=7.5cm]{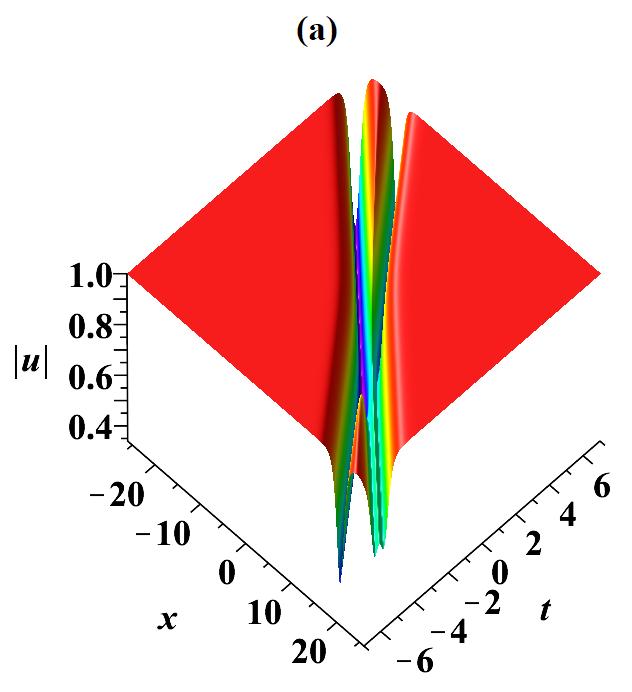}}
\subfigure{\includegraphics[height=5cm,width=6.5cm]{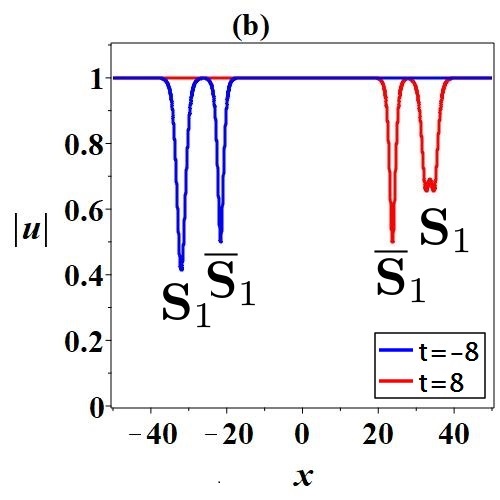}}
\caption{(Color online)(a) The shape-altering collision between a type-I dark soliton and a type-II dark soliton with $(p_1,p_3)=(e^{\frac{1}{4}\mathrm{i}\pi},e^{\frac{1}{6}\mathrm{i}\pi})$ and $(\zeta_{1,0},\zeta_{3,0})=(0,0)$.
(b) The intensity plot of the two solitons in panel (a) at $t=\pm8$.~}\label{Fig8}
\end{figure}

Soliton ${\bf \overline{S}}_1 \left(\zeta_3\simeq \mathcal{O}(1),\zeta_1\rightarrow+\infty\right)$:
\begin{equation}\label{Mso-Asy-2+}
\begin{aligned}
\overline{u}_1^{+}\simeq-\frac{1}{2}\widehat{\mu}_1^{+}\left[e^{2\mathrm{i}\vartheta_3}-1+(e^{2\mathrm{i}\vartheta_3}+1)\tanh\frac{\zeta_3+\hat{\Omega}^{+}_1}{2}\right],
\end{aligned}
\end{equation}
where
\begin{equation}\label{M-Chi+}
\begin{aligned}
\widetilde{\chi}_1^{(n)+}=&1-\frac{4}{(p_1^2-p_1^{*2})^2}(\frac{p_1}{p_1^*})^{2n}, \quad
\widetilde{\Theta}_1^{(n)+}=\frac{2}{p_1+p_1^*}(-\frac{p_1}{p_1^*})^{n}, \quad n=0,1,\\
\hat{\Omega}_1^{+}=&\ln\left(\frac{1}{p_3+p_3^*}\left|\frac{p_1-p_3}{p_1+p_3^*}\right|^2\right), \quad \widehat{\mu}_1^{+}=\widehat{\mu}_1.
\end{aligned}
\end{equation}
Based on the above asymptotic expressions,
we now analyze the collision dynamics of the two mixed-type solitons. First, the algebraic expressions of $\widetilde{\chi}_1^{(n)\pm}$ and $\widetilde{\Theta}_1^{(n)\pm}$ ($n=0,1$) imply that
\(
|u_1^{+}(\zeta_1)| \neq |u_1^{-}(\zeta_1+\Delta\widetilde{\Phi}_1)|,
\)
where $\Delta\widetilde{\Phi}_1=-\widetilde{\Omega}_1$ denotes the phase shift of the type-I dark soliton. This implies that the type-I dark soliton \({\bf S}_1\) undergoes a shape change after the collision. In contrast, since $|\overline{u}_1^{+}(\zeta_3)| = |\overline{u}_1^{-}(\zeta_3 + \Delta \widehat{\Phi}_1)|$, where \(\Delta \widehat{\Phi}_1 = \widehat{\Omega}_1^{+} - \widehat{\Omega}_1^{-}\) represents the phase shift experienced by the type-II dark soliton, this indicates that the type-II dark soliton maintains its shape after the collision. Therefore, since shape changes occur in the type-I dark soliton, the collision between these mixed-type solitons is classified as a shape-altering collision. A typical example of such a shape-altering collision is displayed in Fig. \ref{Fig8}. In this example, the type-I soliton is double-valley-shaped at \( t = -10 \) but transforms into a single-valley shape at \( t = 10 \). In contrast, the type-II soliton ${\bf \overline{S}}_1$ maintains its shape throughout the process.

Second, we consider the asymptotic expression \( u_1^{-} \) of the type-I soliton \({\bf S}_1\) in Eq.~\eqref{Mso-Asy-1-} before the collision. The phase shift factor \(\widetilde{\Omega}_1^{-}\) and the shape-related factors \(\widetilde{\chi}_1^{(n)-}\) and \(\widetilde{\Theta}_1^{(n)-}\), are all expressed in terms of the parameters \( p_1(= e^{\mathrm{i}\vartheta_1}) \) and \( p_3(= e^{\mathrm{i}\vartheta_3}) \). Specifically, \( p_1 \) and \( p_3 \) correspond to the velocities of solitons \({\bf S}_1\) and \({\bf \overline{S}}_1\), respectively, where \(\nu_1 = 2\cos(2\vartheta_1) - 4\) and \(\overline{\nu}_1 = 2\cos(2\vartheta_3) - 4\). This implies that, prior to the collision, the shape of soliton \({\bf S}_1\) is influenced not only by its own velocity \(\nu_1\), but also by the velocity \(\overline{\nu}_1\) of the soliton \({\bf \overline{S}}_1\).
In contrast, for the asymptotic expression \( u_1^{+} \) of the type-I soliton \({\bf S}_1\) in Eq.~\eqref{Mso-Asy-1+} after the collision, only the phase shift factor \(\widetilde{\Omega}_1^{+}\) remains dependent on the parameter \( p_3 \), but the factors determining the soliton's shape, \(\widetilde{\chi}_1^{(n)+}\) and \(\widetilde{\Theta}_1^{(n)+}\), rely solely on \( p_1 \) and are independent of \( p_3 \). Thus, after the collision, the shape of soliton \({\bf S}_1\) depends only on its own velocity \(\nu_1\) and is unaffected by the velocity $\overline{\nu}_1$ of soliton \({\bf \overline{S}}_1\).
This observation can also be confirmed by examining the signs of factors $\widetilde{\Delta}_1^{\pm}$, where $\widetilde{\Delta}_1^{-}=\left.\left(\partial_{\zeta_1\zeta_1}|u_1^{-}|\right)\right|_{\zeta_1=-\widetilde{\Omega}_1^{-}}$ and $\widetilde{\Delta}_1^{+}=\left.\left(\partial_{\zeta_1\zeta_1}|u_1^{+}|\right)\right|_{\zeta_1=0}$. The similar analysis for the shapes of the type-I soliton based on the this factor
has already been conducted in detail earlier, as shown in Fig. \ref{Fig2} and Fig. \ref{Fig6}, along with the related discussions. Therefore, we do not repeat it here.

\begin{figure}[!htbp]
\centering
\subfigure{\includegraphics[height=5cm,width=5.85cm]{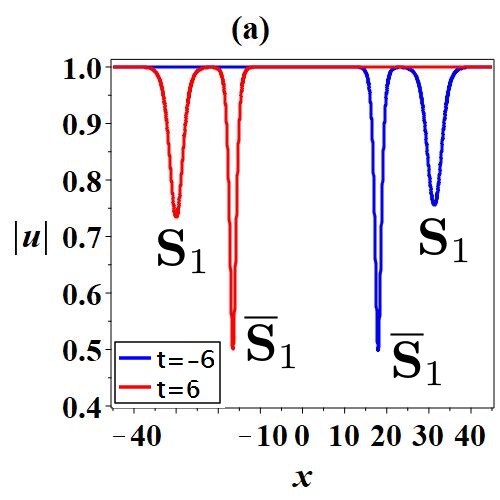}}
\subfigure{\includegraphics[height=5cm,width=5.85cm]{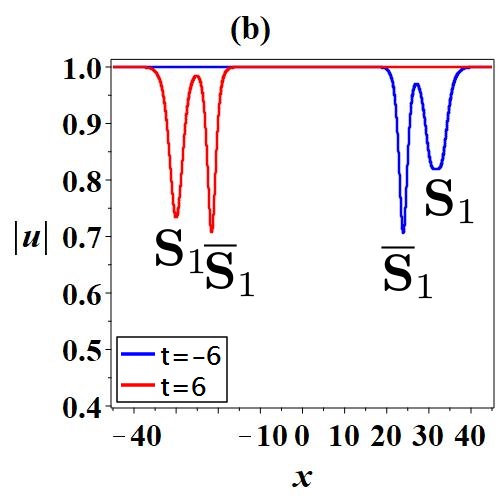}}
\subfigure{\includegraphics[height=5cm,width=5.85cm]{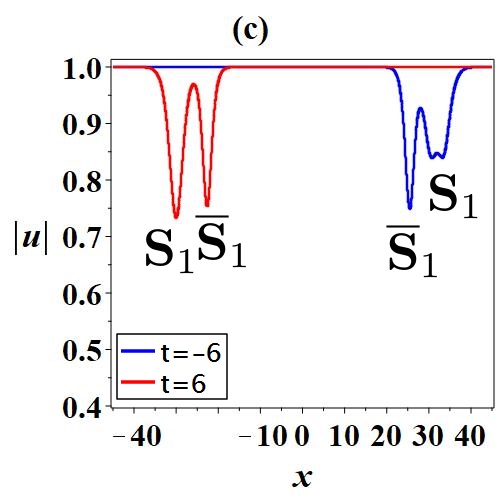}}
\caption{(Color online) The shape-altering collisions of a type-I dark soliton with fixed velocity $\nu_1 = -5$ (i.e.,$p_1 = e^{\frac{1}{3}\mathrm{i}\pi}$) and a type-II dark soliton whose  velocity $\overline{\nu}_1$ varies: (a) $\overline{\nu}_1=-3$ (i.e., $p_3=e^{\frac{1}{6}\mathrm{i}\pi}$); (b) $\overline{\nu}_1=-4$ (i.e., $p_3=e^{\frac{1}{4}\mathrm{i}\pi}$); (c) $\overline{\nu}_1=2\cos(\frac{13}{24}\pi)$ (i.e., $p_3=e^{\frac{13}{48}\mathrm{i}\pi}$). The parameters $\zeta_{1,0}$ and $\zeta_{2,0}$ are set to $0$.~}\label{Fig9}
\end{figure}

To provide a more intuitive illustration of this property,
we present intensity plots in Fig.~\ref{Fig9},
showing a collision between a type-I dark soliton $\mathbf{S}_1$ with fixed velocity $\nu_1=-5$
(i.e., $p_1=e^{\frac{1}{3}\mathrm{i}\pi}$) and a type-II dark soliton $\mathbf{\overline{S}}_1$
with varying velocity $\overline{\nu}_1$.
The plots clearly demonstrate that, before the collision,
the shape of soliton $\mathbf{S}_1$ evolves from a single-valley profile (panel (a)),
to a flat-bottom profile (panel (b)),
and finally to a double-valley profile (panel (c)) as $\overline{\nu}_1$ changes.
After the collision, however, soliton $\mathbf{S}_1$ consistently retains a single-valley shape,
showing no further alteration.

Finally, we examine the shape changes of the type-I soliton ${\bf S}_1$ before and after the collision
by analyzing its intensity profile along the soliton center.
Before the collision, the center of ${\bf S}_1$ is located at $\zeta_1+\widetilde{\Omega}_1=0$,
whereas after the collision it shifts to $\zeta_1=0$.
Accordingly, the explicit expressions for the intensity of ${\bf S}_1$ along its center are
\begin{equation}\label{Intensity-AX}
\begin{aligned}
\mathcal{I}_1^{-}=&\left.\left(\left|u_1^{-}\right|\right)\right|_{\zeta_1=-\widetilde{\Omega}_1^{-}}=\left|\frac{\widetilde{\chi}_1^{(1)-}+\widetilde{\Theta}_1^{(1)-}}
{\widetilde{\chi}_1^{(0)-}+\widetilde{\Theta}_1^{(0)-}}\right|,\\
\mathcal{I}_1^{+}=&\left.\left(\left|u_1^{+}\right|\right)\right|_{\zeta_1=0}=\left|\frac{\widetilde{\chi}_1^{(1)+}+\widetilde{\Theta}_1^{(1)+}}
{\widetilde{\chi}_1^{(0)+}+\widetilde{\Theta}_1^{(0)+}}\right|.
\end{aligned}
\end{equation}
From the explicit forms of $\widetilde{\chi}_1^{(n)\pm}$ and $\widetilde{\Theta}_1^{(n)\pm}$ ($n=0,1$) in
Eqs.~\eqref{M-Chi-} and \eqref{M-Chi+}, we deduce that
$\mathcal{I}_1^{-}=\mathcal{I}_1^{-}(p_1,p_3)$ an $\mathcal{I}_1^{+}=\mathcal{I}_1^{+}(p_1)$. This indicates that $\mathcal{I}_1^{-}$ is determined by the velocities $\nu_1$ and $\overline{\nu}_1$ of both solitons ${\bf S}_1$ and ${\bf \overline{S}}_1$, while $\mathcal{I}_1^{+}$ is determined solely by the velocity $\nu_1$ of soliton ${\bf S}_1$.
In other words, changing the velocity $\overline{\nu}_1$ of the type-II soliton ${\bf \overline{S}}_1$ affects  amplitude $\mathcal{I}_1^{-}$ of soliton ${\bf S}_1$ before collision, but the after collision amplitude $\mathcal{I}_1^{+}$   remains unchanged.
This is different from the type-I and type-I soliton collision discussed earlier, in which changing the velocity of either soliton modifies its amplitude both before and after the collision,
with the amplitudes remaining equal across the collision.

To provide a clear comparison of the three types of two-soliton interactions discussed above, we summarize their key dynamical characteristics in Table~\ref{tab:two_soliton_collision}.

\begin{table}[htbp]
\centering
\small
\caption{Summary of two-soliton collisions.}
\begin{tabular}{@{}p{3.2cm} p{3.2cm} p{4.0cm} p{5.2cm}@{}}
\toprule
\textbf{Collision type} & \textbf{Shape change} & \textbf{Phase shift} & \textbf{Center amplitude} \\
\midrule
Type-I + Type-I & Shape-preserving 
& $\Delta\Phi_1 = -\,\Delta\Phi_2$ 
& $\mathcal{I}_{1,2}$ depend on $(p_1,p_2)$; amplitudes jointly depend on $(\nu_1,\nu_2)$ \\[4pt]

Type-II + Type-II & Shape-preserving 
& $\Delta\overline{\Phi}_1 = -\,\Delta\overline{\Phi}_2$ 
& $\widetilde{\mathcal{I}}_k = |\sin\vartheta_k|$; each soliton depends only on its own $\vartheta_k$ \\[4pt]

Type-I + Type-II & Shape-altering (Type-I changes; Type-II preserves) 
& Type-I: $\Delta\widetilde{\Phi}_1 = -\widetilde{\Omega}_1 ,\quad \quad$ \quad Type-II: $\Delta\widehat{\Phi}_1 = \widehat{\Omega}_1^{+} - \widehat{\Omega}_1^{-}$ 
& Type-I: before collision depends on $(p_1,p_3)$, after only on $p_1$; Type-II depends solely on its own parameter \\ 
\bottomrule
\end{tabular}
\label{tab:two_soliton_collision}
\end{table}

\section{Collisions between $N_1$ Type-I and $N_2$ Type-II Dark Solitons}\label{Collision-N}
In this section, we investigate collisions between arbitrary $N_1$ Type-I and $N_2$ Type-II dark solitons,
with $\max\{N_1, N_2\} \geq 2$.
This extends the two-soliton collisions discussed in Section~\ref{Collision-T}.
Our analysis is divided into three cases according to the values of $N_1$ and $N_2$:
\begin{itemize}
    \item[i)] collisions among $N_1$ Type-I dark solitons, with $N_1 > 2$ and $N_2 = 0$;
    \item[ii)] collisions among $N_2$ Type-II dark solitons, with $N_1 = 0$ and $N_2 > 2$;
    \item[iii)] mixed collisions between $N_1$ Type-I and $N_2$ Type-II dark solitons,
    with $N_1 N_2 \neq 0$ and $\max\{N_1, N_2\} \geq 2$.
\end{itemize}

To investigate these soliton collisions in detail,
we perform an asymptotic analysis as $t\rightarrow \pm \infty$
for both Type-I and Type-II dark solitons in the three cases outlined above.

\subsection{Collisions among pure $N_1$ type-I dark solitons }
This subsection considers collisions among pure $N_1$ Type-I dark solitons
${\bf S}_1, {\bf S}_2, \ldots, {\bf S}_{N_1}$,
described by the solutions of Eq.~\eqref{D-SO} with $N_1 > 2$ and $N_2 = 0$.
To this end, we first present the asymptotic behavior of the functions $\overline{\delta}_n$
in Eq.~\eqref{so-fg} as $t \to \pm \infty$, which is summarized in the following lemma.
The detailed derivations are given in Appendix~\ref{App-2}.

\begin{lemma}\label{asy-ntype1-delta}
Assume that the velocities of the solitons ${\bf S}_1, {\bf S}_2, \ldots, {\bf S}_{N_1}$
satisfy the ordering in Eq.~\eqref{velo-DN} with $N_2=0$, i.e.,
\begin{equation}\label{velo-N1}
\nu_1 < \nu_2 < \cdots < \nu_{k_1} < \cdots < \nu_{N_1} < 0.
\end{equation}
Along the trajectories $\zeta_{k_1}\simeq\mathcal{O}(1)$, the asymptotic form of
$\overline{\delta}_n$ in Eq.~\eqref{so-fg} as $t\rightarrow\pm\infty$ is
    \begin{equation}\label{delta-T1}
        \begin{aligned}
    \overline{\delta}_n^{(k_1)\pm} \simeq &\rho_{k_1}^{\pm}\left(\chi_{k_1}^{(n)}+\Theta_{k_1}^{(n)}\cosh(\zeta_{k_1}\mp \Omega_{k_1})\right)
        \end{aligned}
    \end{equation}
where
     \begin{equation}\label{delta-T1-1}
        \begin{aligned}
\rho_{k_1}^{-}=&(-1)^{N_1-{k_1}+1}e^{\left(-\sum_{s=1}^{{k_1}-1} \zeta_{s}+\sum_{s={k_1}+1}^{N_1} \zeta_{s}\right)} \mathcal{D}_{[{k_1}+1,N_1+{k_1}-1]},\\
\rho_{k_1}^{+}=&(-1)^{{k_1}}e^{\left(\sum_{s=1}^{{k_1}-1} \zeta_{s}- \sum_{s={k_1}+1}^{N_1} \zeta_{s}\right)} \mathcal{D}_{[1, {k_1}-1]\cup[N_1+{k_1}+1, 2N_1]},\\
\chi_{k_1}^{(n)}=&\left[\left(\prod_{s=1}^{k_1-1}  \prod_{s=k_1+1}^{N_1}\right)\left(-\frac{p_s}{p_s^*}\right)^n\right] \left[1-\left(-\frac{p_{{k_1}}}{p_{{k_1}}^*}\right)^{2n}\frac{4}{(p_{{k_1}}^2-p_{{k_1}}^{*2})^2}\left(\prod_{s=1}^{k_1-1} \prod_{s=k_1+1}^{N_1}\right)\frac{|p_{{k_1}}^2-p_{s}^2|^2}{|p_{{k_1}}^2-p_{s}^{*2}|^2}\right],\\
\Theta_{k_1}^{(n)}=&\frac{2}{p_{{k_1}}+p_{{k_1}}^*}\left[\prod_{s=1}^{N_1}\left(-\frac{p_s}{p_s^*}\right)^n\right]\left[\left(\prod_{s=1}^{k_1-1}\prod_{s=k_1+1}^{N_1}\right)\frac{|p_{{k_1}}^2-p_{s}^2|}{|p_{{k_1}}^2-p_{s}^{*2}|}\right],\\
\Omega_{k_1}=&\ln\left[ \left(\prod_{s=1}^{k_1-1}\frac{|p_{s}+p_{{k_1}}^*||p_{{k_1}}+p_{s}|}{|p_{{k_1}}-p_{s}||p_{s}-p_{{k_1}}^*|}\right)\left(\prod_{s=k_1+1}^{N_1}\frac{|p_{{k_1}}-p_{s}||p_{s}-p_{{k_1}}^*|}{|p_{s}+p_{{k_1}}^*||p_{{k_1}}+p_{s}|}\right)\right],
        \end{aligned}
    \end{equation}
for $k_1=1,2,\cdots,N_1$.
Here, $\mathcal{D}_{[a,b]}$ denotes the Cauchy determinant with entries
$\frac{1}{p_s+p_j^*}$, where the integer indices $s$ and $j$ range from $a$ to $b$.
Similarly, $\mathcal{D}_{[a,b]\cup[c,d]}$ represents the Cauchy determinant with entries
$\frac{1}{p_s+p_j^*}$, where the indices $s$ and $j$ span the combined intervals
$[a,b]$ and $[c,d]$. 
 Since \(\mathcal{D}_{[k_1+1, N_1+k_1-1]}\) and \(\mathcal{D}_{[1, k_1-1] \cup [N_1+k_1+1, 2N_1]}\) will be eliminated in our final asymptotic expressions of type-I dark solitons, we do not provide their explicit algebraic expressions. The notation \(\left(\prod_{s=a}^{b} \prod_{s=c}^{d}\right) f(s)\) represents two separate products over the same variable \(s\): the first product is taken from \(s = a\) to \(s = b\), and the second from \(s = c\) to \(s = d\). This results in
\(
\prod_{s=a}^{b} f(s) \cdot \prod_{s=c}^{d} f(s),
\)
which is the product of all \(f(s)\) values in both ranges.
\end{lemma}

Using the explicit expression of $\overline{\delta}_n^{(k_1)\pm}$ in Eq.~\eqref{delta-T1} of this lemma, the asymptotic expressions for each type-I dark soliton ${\bf S}_{k_1}$ and the general solution \eqref{D-SO} with $N_1>0,N_2=0$ can be yielded, as outlined by the following theorem.
\begin{theorem}\label{asy-ntype1}
As $t\rightarrow\pm\infty$ and under the velocity correlation assumption in Eq.~\eqref{velo-N1}, the asymptotic expression for type-I dark soliton ${\bf S}_{k_1}$ along \(\zeta_{k_1} \simeq \mathcal{O}(1)\) in solution \eqref{D-SO} with $N_1>0,N_2=0$ is
    \begin{equation}\label{type1-Asy-k}
    \begin{aligned}
    u_{k_1}^{\pm}\simeq \frac{\chi_{k_1}^{(1)}+\Theta_{k_1}^{(1)}\cosh(\zeta_{k_1} \mp \Omega_{k_1})}
    {\chi_{k_1}^{(0)}+\Theta_{k_1}^{(0)}\cosh(\zeta_{k_1}\mp \Omega_{k_1})},
    \end{aligned}
    \end{equation}
with $ {k_1}=1,2,\cdots,N_1$. The asymptotics of solution \eqref{D-SO} with $N_1>0,N_2=0$ is
    \begin{equation}\label{type1-Asy-sum}
    \begin{aligned}
    |u^{\pm}|\simeq \sum_{{k_1}=1}^{N_1}|u_{k_1}^{\pm}|-(N_1-1).
    \end{aligned}
    \end{equation}
\end{theorem}

We analyze the properties of the type-I dark solitons and their collisions based on the asymptotic expression \eqref{type1-Asy-k}.
First, the center of type-I dark soliton ${\bf S}_{k_1}$ shifts from
\begin{equation}
\begin{aligned}
\zeta_{k_1}+\Omega_{k_1}=0,
\end{aligned}
\end{equation}
to
\begin{equation}
\begin{aligned}
\zeta_{k_1}-\Omega_{k_1}=0.
\end{aligned}
\end{equation}
Thus the phase shift of this soliton undergoing during this collision is
\begin{equation}
\begin{aligned}
\Delta\Phi_{k_1}=-2\Omega_{k_1}.
\end{aligned}
\end{equation}
Specifically, since $\sum\limits_{k_1=1}^{N_1} \Omega_{k_1} = 0$, it follows that
\begin{equation}
 \sum\limits_{k_1=1}^{N_1} \Delta\Phi_{k_1} = 0,
\end{equation}
which implies that the total phase shifts of the $N_1$ solitons
${\bf S}_1, {\bf S}_2, \dots, {\bf S}_{N_1}$ also sum to zero.

\begin{figure}[!htbp]
\centering
\subfigure{\includegraphics[height=5cm,width=5.5cm]{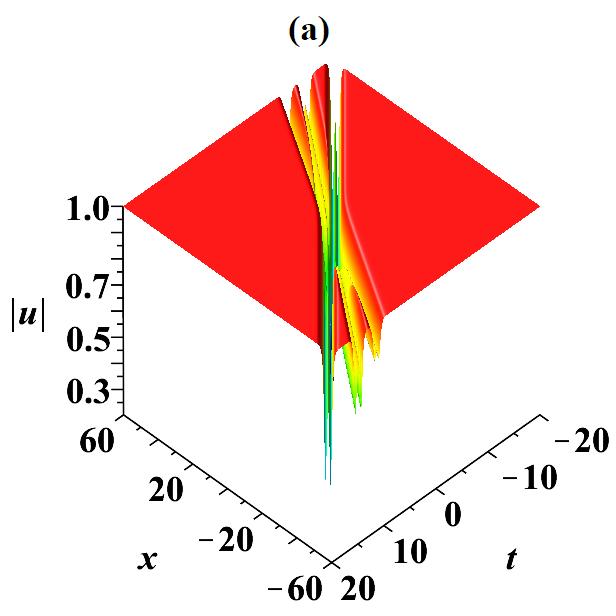}}
\subfigure{\includegraphics[height=5cm,width=5.5cm]{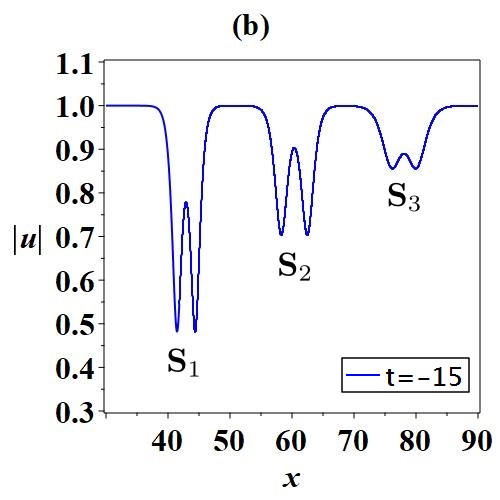}}
\subfigure{\includegraphics[height=5cm,width=5.5cm]{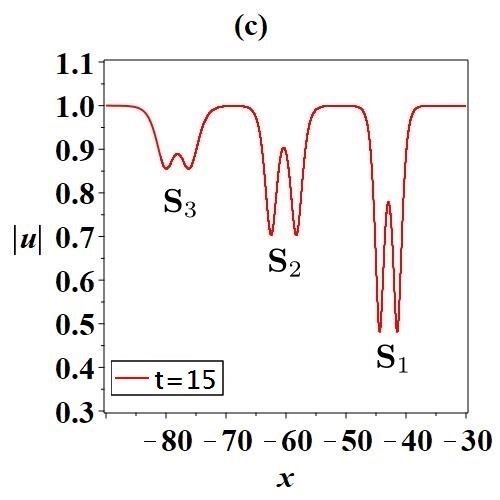}}
\caption{(a) The shape-preserving collisions of three type-I dark soliton solutions with $(p_1,p_2,p_3)=(e^{\frac{1}{3}\mathrm{i}\pi},e^{\frac{1}{4}\mathrm{i}\pi},e^{\frac{1}{6}\mathrm{i}\pi})$ and $(\zeta_{1,0},\zeta_{2,0},\zeta_{3,0})=(0,0,0)$. (b) and (c) are the intensity plots of the three type-I dark soliton solutions displayed in panel (a) at $t=-15$ and $t=15$, respectively.~}\label{Fig10}
\end{figure}

Second, as the asymptotic expressions $u_{k_1}^{-}$ and $u_{k_1}^{+}$ admit the correlation:
\begin{equation}
\begin{aligned}
u_{k_1}^{+}(\zeta_{k_1})=u_{k_1}^{-}(\zeta_{k_1}+\Delta\Phi_{k_1}),
\end{aligned}
\end{equation}
the collision between these $N_1$ pure type-I dark solitons are shape-preserving.
Fig. \ref{Fig10} provides an example of such shape-preserving collision of three type-I dark solitons, illustrating that the type-I dark solitons ${\bf S}_1,{\bf S}_2, {\bf S}_3$ retain their double-valley profiles unaltered before and after collision. Here, we emphasize a comparison with the two type-I dark soliton collisions shown in Fig.~\ref{Fig4}. The parameters $p_1$, $p_2$, and $p_3$ determine the velocities of the type-I solitons ${\bf S}_1,{\bf S}_2, {\bf S}_3$, respectively. In Fig.~\ref{Fig10}, the parameters $p_1$ and $p_2$ are taken the same values as those in Fig.~\ref{Fig4}, yet the velocities of solitons ${\bf S}_1,{\bf S}_2$ are identical in these two figures. However, the shapes of the solitons ${\bf S}_1,{\bf S}_2$ in Fig.~\ref{Fig4} and Fig.~\ref{Fig10} are completely different. This indicates that the participation of the type-I soliton ${\bf S}_3$ in the collision alters the shapes of the type-I dark soliton ${\bf S}_1,{\bf S}_2$.
This phenomenon is related to the third property of pure type-I soliton collisions discussed below.

Our third aspect concerns the shapes of type-I dark solitons and the corresponding amplitudes of the solitons at their centers.
Since type-I dark solitons only undergo shape-preserving collisions, we will focus solely on analyzing their shapes and amplitudes before their collisions, without considering the after collision outcomes. As we discussed earlier, the shape of the type-I dark soliton by ${\bf S}_{k_1}$ determined by the sign of the factor $\Delta_{k_1}$, where the factor $\Delta_{k_1}$ is given by:
\begin{equation}\label{Delta-TI}
\begin{aligned}
\Delta_{k_1} = \left.\left[\partial_{\zeta_{k_1}\zeta_{k_1}}(\left|u_{k_1}^{-}(\zeta_{k_1})\right|)\right]\right|_{\zeta_{k_1}=\Omega_{k_1}},\qquad\qquad\, {k_1}=1,2,\cdots,N_1.
\end{aligned}
\end{equation}
The soliton ${\bf S}_{k_1}$ is single-valley-shaped for $\Delta_{k_1}>0$,  flat-bottom-shaped for $\Delta_{k_1}$ and double-valley-shaped for $\Delta_{k_1}<0$.
One can easily verified that $\Delta_{k_1}$ is not only a function of $p_k$ but also a function of $p_1, p_2, \cdots, p_N$. Here, the parameter $p_k$ determines
the velocity of the soliton ${\bf S}_{k_1}$ as $\nu_{k_1} = 2\cos(2\vartheta_{k_1})-4$ if the parameter $p_{k_1}$ is taken in its equivalent form $p_{k_1} = e^{\mathrm{i}\vartheta_{k_1}}$.
This implies that changing the velocity $\nu_{k_1}$ of a soliton ${\bf S}_{k_1}$ will alter the shapes of all $N_1$ type-I solitons ${\bf S}_1,{\bf S}_2,\cdots,{\bf S}_{N_1}$.
In other words, these $N_1$ type-I dark solitons mutually influence the shapes of each other. This effect is also reflected in their amplitudes at their centers. The algebraic expression for the amplitude of the soliton ${\bf S}_{k_1}$ at its center $\zeta_{k_1}-\Omega_{k_1} = 0$ is given by:
\begin{equation}\label{AM-Ik}
\begin{aligned}
\mathcal{I}_{k_1}&= \left. |u_{k_1}^{-}| \right|_{\zeta_{k_1}=\Omega_{k_1}}=\left| \frac{\chi_{k_1}^{(1)} + \Theta_{k_1}^{(1)}}{\chi_{k_1}^{(0)} + \Theta_{k_1}^{(0)}} \right|,\qquad\qquad\, {k_1}=1,2,\cdots,N_1.
\end{aligned}
\end{equation}
The value of $\mathcal{I}_{k_1}$ is also determined by the parameters $p_1, p_2, \cdots, p_{N_1}$ collectively. This means that changing the amplitude of one type soliton will result in the amplitudes of all other type-I solitons being affected as well.  This can be regarded as a generalization of the pure two dark solitons shown in Fig. \ref{Fig7}. By changing the parameter equivalent to the velocity of one type-I soliton, the shapes and amplitudes of all type-I dark solitons will change.

\subsection{Collisions among pure $N_2$ type-II dark solitons }
This subsection focuses on the collisions among pure $N_2$ type-II dark solitons, described by the solutions \eqref{D-SO} with $N_1=0$ and $N_2>2$.
For this purpose, we first present the asymptotics of the functions $\overline{\delta}_n$ in Eq.~\eqref{so-fg} as $t \rightarrow \pm\infty$, summarized in the following lemma. The detailed derivations are given in Appendix \ref{App-2}.
\begin{lemma}\label{asy-ntype2-delta}
Assume that the velocities of the solitons ${\bf \overline{S}}_1, {\bf \overline{S}}_2, \cdots, {\bf \overline{S}}_{N_2}$ satisfy the correlation in Eq.~\eqref{velo-DN} with $N_1=0$, i.e.,
\begin{equation}\label{velo-N2}
\overline{\nu}_1 < \overline{\nu}_2 < \cdots < \overline{\nu}_{k_2} < \cdots < \overline{\nu}_{N_2} < 0,
\end{equation}
then, along the soliton trajectories $\zeta_{k_2}\simeq\mathcal{O}(1)$ for $k_2=1,2,\cdots,N_2$,
the asymptotic expression of the function $\overline{\delta}_n$ in Eq.~\eqref{so-fg} as $t \rightarrow \pm\infty$ is given by
    \begin{equation}\label{delta-T2}
        \begin{aligned}
     \overline{\delta}_n^{(k_2)\pm}\simeq&\overline{\rho}_{k_2}^{\pm}e^{-\zeta_{k_2}}\left(\mu_{k_2}^{\pm}\right)^n\left[1+\left(y_{k_2}\right)^n e^{\zeta_{k_2}+\overline{\Omega}_{k_2}^{\pm}}\right],
        \end{aligned}
    \end{equation}
where
     \begin{equation}
        \begin{aligned}
\overline{\rho}_{k_2}^{-}=&e^{\left(-\sum\limits_{s=1}^{k_2-1}\zeta_s\right)}\left[ \frac{\prod\limits_{k_2+1\leq s<j \leq N_2}\left(p_j-p_s\right)\left(p_j^*-p_s^*\right)}{\prod\limits_{s, j=k_2+1}^{N_2}\left(p_s+p_j^*\right)}\right],\\
\overline{\rho}_{k_2}^{+}=&e^{\left(-\sum\limits_{s=k_2+1}^{N_2}\zeta_s\right)}\left[\frac{\prod\limits_{1 \leq s<j \leq k_2-1}\left(p_j-p_s\right)\left(p_j^*-p_s^*\right)}{\prod\limits_{s, j=1}^{k_2-1}\left(p_s+p_j^*\right)}\right],\\
\mu_{k_2}^{-}=& \prod_{s=k_2+1}^{N_2}\left(-\frac{p_s}{p_s^*}\right),
\quad \mu_{k_2}^{+}=\prod_{s=1}^{k_2-1}\left(-\frac{p_s}{p_s^*}\right), \quad y_{k_2}=-\frac{p_{k_2}}{p_{k_2}^*},\\
\overline{\Omega}_{k_2}^{-} =& \ln\left[\frac{1}{p_{k_2}+p_{k_2}^*}\prod_{s=k_2+1}^{N_2}\frac{\left|p_s-p_{k_2}\right|^2}{\left|p_{k_2}+p_s^*\right|^2}\right],
\quad \overline{\Omega}_{k_2}^{+} = \ln\left[\frac{1}{p_{k_2}+p_{k_2}^*}\prod_{s=1}^{k_2-1}\frac{\left|p_s-p_{k_2}\right|^2}{\left|p_{k_2}+p_s^*\right|^2}\right].
        \end{aligned}
    \end{equation}
for $k_2=1,2,\cdots,N_2$.
\end{lemma}

Using the explicit expression of $\overline{\delta}_n^{(k_2)\pm}$ in Eq.~\eqref{delta-T2} of this lemma, the asymptotic expressions for each type-II dark soliton ${\bf \overline{S}}_{k_2}$ and the general solution \eqref{D-SO} with $N_1=0$ and $N_2>0$ can be derived, as outlined in the following theorem.
\begin{theorem}\label{asy-ntype2}
As $t\rightarrow\pm\infty$ and under the velocity correlation assumption in Eq.~\eqref{velo-N2}, the asymptotic expression for type-II dark soliton \({\bf \overline{S}}_{k_2}\) in solution \eqref{D-SO} with $N_1=0,N_2>0$  is
    \begin{equation}\label{type2-Asy-k}
    \begin{aligned}
    \overline{u}_{k_2}^{\pm}\simeq  \frac{1}{2}\mu_{k_2}^{\pm}\left[(1+y_{k_2})+(y_{k_2}-1)\tanh(\frac{\zeta_{k_2}+\overline{\Omega}_{k_2}^{\pm}}{2})\right]
    \end{aligned}
    \end{equation}
with $ k_2=1,2,\cdots,N_2$. The asymptotics of solution \eqref{D-SO} with $N_1=0,N_2>0$ is
    \begin{equation}\label{type2-Asy-N1}
    \begin{aligned}
    |u^{\pm}|\simeq \sum_{k_2=1}^{N_2}|\overline{u}_{k_2}^{\pm}|-(N_2-1).
    \end{aligned}
    \end{equation}
\end{theorem}

We briefly summarize the collision properties of \(N_2\) type-II dark solitons, as indicated by the asymptotic expressions \eqref{type2-Asy-k} and \eqref{type2-Asy-N1}. For the type-II dark soliton \({\bf \overline{S}}_{k_2}\), its center shifts from
\begin{equation}
\zeta_{k_2}+\overline{\Omega}_{k_2}^{-}=0,
\end{equation}
to
\begin{equation}
\zeta_{k_2}+\overline{\Omega}_{k_2}^{+}=0,
\end{equation}
for \(k_2=1,2,\cdots N_2\). Thus, the phase shift of this soliton during the collision is
\begin{equation}
\Delta \overline{\Phi}_{k_2}=\overline{\Omega}_{k_2}^{+}-\overline{\Omega}_{k_2}^{-}.
\end{equation}
Specifically,
\begin{equation}
\sum\limits_{k_2=1}^{N}\Delta\overline{\Phi}_{k_2} = 0,
\end{equation}
indicating that the total phase shifts of all the \(N_2\) solitons \({\bf \overline{S}}_{1},{\bf \overline{S}}_{2},\cdots,{\bf \overline{S}}_{N_2}\) also sum to zero.

Since the modulus of the factors \(\mu_{k_2}^{\pm}\) in the asymptotic expression \(u_{k_2}^{\pm}\) of Eq.~\eqref{type2-Asy-k} is always unity, i.e., \(|\mu_{k_2}^{\pm}|=1\), the relationship between \(|\overline{u}_{k_2}^{-}|\) and \(|\overline{u}_{k_2}^{+}|\) can be expressed in the following form of phase-shift transformation:
\begin{equation}
\left|\overline{u}_{k_2}^{+}(\zeta_{k_2})\right|=\left|\overline{u}_{k_2}^{-}(\zeta_{k_2}+\Delta \overline{\Phi}_{k_2})\right|.
\end{equation}
Thus, all \(N_2\) type-II dark solitons generally maintain their shape and integrity, even after colliding with one another. Therefore, the collisions of pure \(N_2\) type-II dark solitons are also shape-preserving.

Finally, the intensity of the type-II dark soliton ${\bf \overline{S}}_{k_2}$ at its center is
\begin{equation}\label{1tan-AM-N2}
\widetilde{\mathcal{I}}_{k_2}=|\sin\vartheta_{k_2}|,
\end{equation}
where $p_{k_2}$ is taken in the equivalent form $p_{k_2}=e^{\mathrm{i}\vartheta_{k_2}}$.
Accordingly, the velocity--amplitude correlation is expressed as
\begin{equation}\label{1tan-AM-vN}
\overline{\nu}_{k_2}=-4\widetilde{\mathcal{I}}_{k_2}^2-2.
\end{equation}
This relation is consistent with that of a single type-II dark soliton given in Eq.~\eqref{1tan-AM-v}, which confirms that the presence of other solitons does not affect the amplitude of ${\bf \overline{S}}_{k_2}$.
In summary, collisions among pure $N_2$ type-II dark solitons are shape-preserving: the solitons undergo only phase shifts, while their individual profiles and amplitudes remain unchanged.

\subsection{Collisions between mixed $N_1$ type-I and $N_2$ type-II dark solitons }
This subsection considers the collisions between mixed type solitons consisting of $N_1$ type-I dark solitons and $N_2$ type-II dark soltions, depicted by the solutions \eqref{D-SO} with $N_1N_2\neq0$ and $\max\{N_1,N_2\}\geq2$.
For this purpose, we first derive the asymptotics of the functions $\overline{\delta}_n$ in Eq.~\eqref{so-fg} along the soliton trajectories as $t \rightarrow \pm \infty$, summarized in the following lemma.

\begin{lemma}\label{asy-ntype1-2-delta}
Under the velocity conditions described in Eq.~\eqref{velo-DN}, the asymptotics of the function
  \(\overline{\delta}_n\) along the trajectories of \( N_1 \) type-I dark solitons and \( N_2 \) type-II dark solitons are characterized as follows.

    \begin{itemize}
        \item Along $\zeta_{k_1}\simeq\mathcal{O}(1)$, i.e., the trajectory of type-I dark soliton ${\bf S}_{k_1}$, the asymptotic expression of function \(\overline{\delta}_n\) as $t\rightarrow\pm\infty$ is given by:
 \begin{equation}\label{M-As-k}
    \begin{aligned}
 \overline{\delta}_n^{(k_1)\pm} \simeq  &\widetilde{\rho}_{k_1}^{\pm} \left(\widetilde{\chi}_{k_1}^{(n)\pm}+\widetilde{\Theta}_{k_1}^{(n)\pm}\cosh(\zeta_{k_1}\mp\widetilde{\Omega}_{k_1}^{\pm})\right),
    \end{aligned}
    \end{equation}
where
    \begin{equation}\label{delta-T1-2-1}
        \begin{aligned}
     \widetilde{\rho}_{k_1}^{-} = &(-1)^{N_1-k_1+1}e^{-\left(\sum\limits_{s=1}^{k_1-1} \zeta_{s}\right)+\left(\sum\limits_{s=k_1+1}^{N_1} \zeta_{s}\right)}\times \mathcal{D}_{[k_1+1,N_1+k_1-1]\cup[2N_1+1,2N_1+N_2]},\\
     \widetilde{\rho}_{k_1}^{+} = &(-1)^{k_1}e^{\left(\sum\limits_{s=1}^{k_1-1} \zeta_{s}\right)-\left(\sum\limits_{s=k_1+1}^{N_1}\zeta_{s}\right)-\left(\sum\limits_{s=2N_1+1}^{2N_1+N_2} \zeta_{s}\right)}\times \mathcal{D}_{[1, k_1-1]\cup[N_1+k_1+1, 2N_1]}, \\
    \widetilde{\chi}_{k_1}^{(n)-}=&\left[\left(\prod_{s=1}^{k_1-1}\prod_{s=k_1+1}^{N_1}\prod_{s=2N_1+1}^{2N_1+N_2}\right)\left(-\frac{p_s}{p_s^*}\right)^n\right]\\
     & \times \left[1-\frac{4}{(p_{k_1}^2-p_{k_1}^{*2})^2}\left(-\frac{p_{k_1}}{p_{k_1}^*}\right)^{2n}\left(\prod_{s=1}^{k_1-1}\prod_{s=k_1+1}^{N_1}\prod_{s=2N_1+1}^{2N_1+N_2}\right)\frac{|p_{k_1}^2-p_{s}^2|^2}{|p_{k_1}^2-p_{s}^{*2}|^2}\right],\\
    \widetilde{\Theta}_{k_1}^{(n)-} =&\frac{2}{p_{k_1}+p_{k_1}^*}\left[\left(\prod_{s=1}^{N_1}\prod_{s=2N_1+1}^{2N_1+N_2}\right)\left(-\frac{p_s}{p_s^*}\right)^n\right]\left[\left(\prod_{s=1}^{k_1-1}\prod_{s=k_1+1}^{N_1}\prod_{s=2N_1+1}^{2N_1+N_2}\right)\frac{|p_{k_1}^2-p_{s}^2|}{|p_{k_1}^2-p_{s}^{*2}|}\right],\\
    \widetilde{\Omega}_{k_1}^{-}=&\ln \left[\left(\prod_{s=1}^{k_1-1}\frac{|p_{s}+p_{k_1}^*||p_{k_1}+p_{s}|}{|p_{k_1}-p_{s}||p_{s}-p_{k_1}^*|}\right)\left(\prod_{s=k_1+1}^{N_1}\prod_{s=2N_1+1}^{2N_1+N_2}\right)\frac{|p_{k_1}-p_{s}||p_{s}-p_{k_1}^*|}{|p_{s}+p_{k_1}^*||p_{k_1}+p_{s}|}\right],
        \end{aligned}
    \end{equation}
    and  $\widetilde{\chi}_{k_1}^{(n)+} = \chi_{k_1}^{(n)}, \widetilde{\Theta}_{k_1}^{(n)+} = \Theta_{k_1}^{(n)}, \widetilde{\Omega}_{k_1}^{+} = -\Omega_{k_1}$, with $\chi_{k_1}^{(n)}$, $\Theta_{k_1}^{(n)}$, and $\Omega_{k_1}$ being defined by Eq.~\eqref{delta-T1-1}.

        \item Along $\zeta_{2N_1+k_2}\simeq\mathcal{O}(1)$, i.e., the trajectory of type-II dark soliton ${\bf \overline{S}}_{k_2}$, the asymptotic expression of function \(\overline{\delta}_n\) as $t\rightarrow\pm\infty$ is given by:
\begin{equation}\label{M-As-Nk}
\begin{aligned}
 \overline{\delta}_n^{(2N_1+k_2)\pm} \simeq&
\widehat{\rho}_{k_2}^{\pm}e^{-\zeta_{2N_1+k_2}} \left(\widehat{\mu}_{k_2}^{\pm}\right)^n\left[1+\widehat{y}_{k_2}^ne^{\zeta_{2N_1+k_2}+\widehat{\Omega}_{k_2}^{\pm}}\right]
\end{aligned}
\end{equation}
where
\begin{equation}\label{delta-T1-2-2}
\begin{aligned}
     \widehat{\rho}_{k_2}^- = &e^{-(\sum_{s=1}^{N_1}\zeta_s+ \sum_{s=2N_1+1}^{2N_1+k_2-1}\zeta_s)}  \mathcal{D}_{[N_1+1,2N_1]\cup[2N_1+k_2+1,2N_1+N_2]},\\
     \widehat{\rho}_{k_2}^+ = &(-1)^{N_1} e^{\sum_{s=1}^{N_1}\zeta_s-\sum_{s=2N_1+k_2+1}^{2N_1+N_2}\zeta_s}  \mathcal{D}_{[1,N_1]\cup[2N_1+1,2N_1+k_2-1]}, \quad \widehat{y}_{k_2}=-\frac{p_{2N_1+k_2}}{p_{2N_1+k_2}^*},\\
      \widehat{\mu}_{k_2}^- = & \left(\prod_{s= 1}^{N_1}\prod_{s= 2N_1+k_2+1}^{2N_1+N_2}\right)\left(-\frac{p_s}{p_s^*}\right), \quad \widehat{\mu}_{k_2}^+ = \left(\prod_{s= 1}^{N_1}\prod_{s= 2N_1+1}^{2N_1+k_2-1}\right)\left(-\frac{p_s}{p_s^*}\right),  \\
     \widehat{\Omega}_{k_2}^{-} = &\ln \left(\frac{1}{p_{2N_1+k_2}+p_{2N_1+k_2}^*}\prod_{s=1}^{N_1}\frac{\left|p_s+p_{2N_1+k_2}\right|^2}{\left|p_{2N_1+k_2}-p_s^*\right|^2}\prod_{s=2N_1+k_2+1}^{2N_1+N_2}\frac{\left|p_s-p_{2N_1+k_2}\right|^2}{\left|p_{2N_1+k_2}+p_s^*\right|^2} \right),\\
    \widehat{\Omega}_{k_2}^{+} = &\ln \left[\frac{1}{p_{2N_1+k_2}+p_{2N_1+k_2}^*}\left(\prod_{s= 1}^{N_1}\prod_{s= 2N_1+1}^{2N_1+k_2-1}\right)\frac{\left|p_s-p_{2N_1+k_2}\right|^2}{\left|p_{2N_1+k_2}+p_s^*\right|^2} \right].
\end{aligned}
\end{equation}
    \end{itemize}
\end{lemma}

From the explicit expressions of $\overline{\delta}_n^{(k_1)\pm}$ and $\overline{\delta}_n^{(2N_1+k_2)\pm}$
given in Eqs.~\eqref{M-As-k} and \eqref{M-As-Nk},
we obtain the asymptotic forms of the two distinct types of dark solitons in solution \eqref{D-SO}
for $N_1N_2\neq0$ and $\max\{N_1,N_2\}\geq2$, as stated in the following theorem.
\begin{theorem}\label{asy-ntype1-2}
As $t\rightarrow\pm\infty$ and under the velocity condition assumed in Eq.~\eqref{velo-N2}, the asymptotic expressions for the two types of dark soliton in mixed soliton solution \eqref{D-SO}  with  $N_1N_2\neq0$ and $\max\{N_1,N_2\}\geq2$ are:
    \begin{itemize}
        \item Along  $\zeta_k\simeq\mathcal{O}(1)$, the asymptotic expression of the type-I dark soliton ${\bf S}_{k_1}$ is
        \begin{equation}\label{type12-Asy-K}
        \begin{aligned}
        u_{k_1}^{\pm} \simeq \frac{\widetilde{\chi}_{k_1}^{(1)\pm}+\widetilde{\Theta}_{k_1}^{(1){\pm}}\cosh(\zeta_{k_1} + \widetilde{\Omega}_{k_1}^{\pm})}
        {\widetilde{\chi}_{k_1}^{(0)\pm}+\widetilde{\Theta}_{k_1}^{(0)\pm}\cosh(\zeta_{k_1}+\widetilde{\Omega}_{k_1}^{\pm})},
        \end{aligned}
        \end{equation}
        Here, $k_1 = 1,2,\cdots,N_1$.
        \item Along  $\zeta_{2N_1+k_2}\simeq\mathcal{O}(1)$, the asymptotic expression of the type-II dark soliton ${\bf \overline{S}}_{k_2}$ is
    \begin{equation}\label{type12-Asy-NK}
    \begin{aligned}
    \overline{u}_{k_2}^{\pm} \simeq&  \frac{1}{2} \widehat{\mu}^{\pm}_{k_2}\left[(1+\widehat{y}_{k_2})+(\widehat{y}_{k_2}-1)\tanh(\frac{\zeta_{2N_1+k_2}+\widehat{\Omega}_{k_2}^{\pm}}{2})\right],
    \end{aligned}
    \end{equation}
    \end{itemize}
    Here, $k_2 = 1,2,\cdots,N_2$.
Therefore, using the asymptotics for both types of dark solitons in Eqs.\eqref{type12-Asy-K} and \eqref{type12-Asy-NK}, the asymptotics of the general mixed dark soliton solution \eqref{D-SO} with $N_1N_2 \neq 0$ and $\max\{N_1, N_2\} \geq 2$ is given by
    \begin{equation}\label{type12-Asy-N}
    \begin{aligned}
    |u^{\pm}|\simeq \sum_{k_1=1}^{N_1}| u_{k_1}^{\pm}|+\sum_{k_2=1}^{N_2}| \overline{u}_{k_2}^{\pm} |-(N_1+N_2-1).
    \end{aligned}
    \end{equation}
\end{theorem}

Next, we provide some comments on these mixed-type soliton collisions based on the asymptotic expressions of Eqs. \eqref{type12-Asy-K} and \eqref{type12-Asy-NK}.
First, the phase shifts experienced by the type-I dark soliton ${\bf S}_{k_1}$ and the type-II dark soliton ${\bf \overline{S}}_{k_2}$, denoted as $\Delta\widetilde{\Phi}_{k_1}$ and $ \Delta\widehat{\Phi}_{k_2}$ respectively, are given by:
    \begin{equation}\label{P-S-M}
    \begin{aligned}
    \Delta\widetilde{\Phi}_{k_1}=&\widetilde{\Omega}_{k_1}^{+}-\widetilde{\Omega}_{k_1}^{-},\\
    \Delta\widehat{\Phi}_{k_2}=&\widehat{\Omega}_{k_2}^{+}-\widehat{\Omega}_{k_2}^{-}.
    \end{aligned}
    \end{equation}
The sum of phase shifts of all the type-I dark solitons and type-II dark solitons admits the relations
    \begin{equation}
    \begin{aligned}
    2\left(\sum\limits_{k_1=1}^{N_1}\Delta\widetilde{\Phi}_{k_1}\right)+\left(\sum\limits_{k_2=1}^{N_2}\Delta\widehat{\Phi}_{k_2}\right)=0.
    \end{aligned}
    \end{equation}

\begin{figure}[!htbp]
\centering
\subfigure{\includegraphics[height=6.0cm,width=7.5cm]{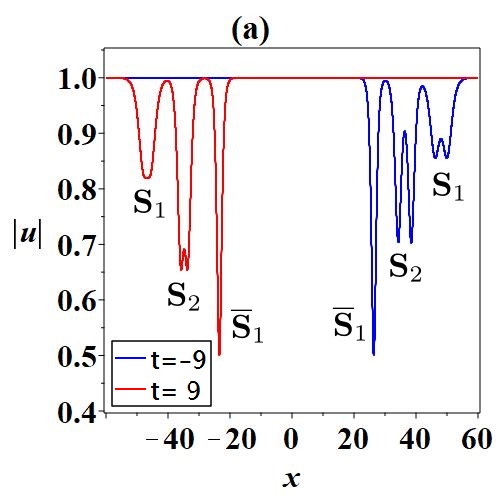}}
\subfigure{\includegraphics[height=6.0cm,width=7.5cm]{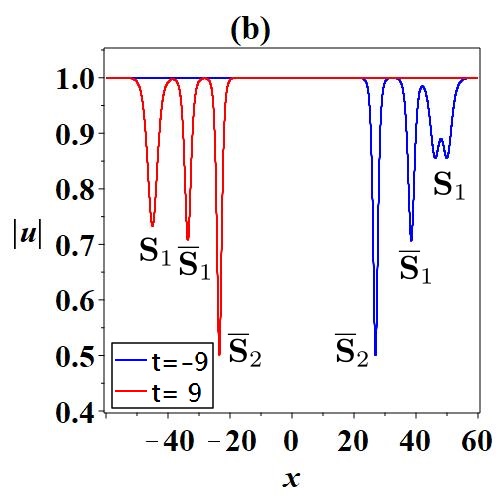}}
\caption{(a) The shape-altering collisions of two type-I dark solitons and a type-II dark soliton with $(N_1,N_2)=(2,1)$, $(p_1,p_2,p_5)=(e^{\frac{1}{3}\mathrm{i}\pi},e^{\frac{1}{4}\mathrm{i}\pi},e^{\frac{1}{6}\mathrm{i}\pi})$ and $(\zeta_{1,0},\zeta_{2,0},\zeta_{5,0})=(0,0,0)$.
(b)  The shape-altering collisions of one type-I dark soliton and two type-II dark solitons with $(N_1,N_2)=(1,2)$, $(p_1,p_3,p_4)=(e^{\frac{1}{3}\mathrm{i}\pi},e^{\frac{1}{4}\mathrm{i}\pi},e^{\frac{1}{6}\mathrm{i}\pi})$ and $(\zeta_{1,0},\zeta_{3,0},\zeta_{4,0})=(0,0,0)$.
~}\label{Fig11}
\end{figure}

Second, it is confirmed that
$\left|u_{k_1}^{+}(\zeta_{k_1})\right|\neq\left|u_{k_1}^{-}(\zeta_{k_1}+\Delta\widetilde{\Phi}_{k_1})\right|$,
while
$\left|\overline{u}_{k_2}^{+}(\zeta_{2N_1+k_2})\right|=\left|\overline{u}_{k_2}^{-}(\zeta_{2N_1+k_2}+\Delta\widehat{\Phi}_{k_2})\right|$.
Thus, the shapes of all type-I dark solitons
${\bf S}_{1},\dots,{\bf S}_{N_1}$ change after the collisions,
whereas the shapes of all type-II dark solitons
${\bf \overline{S}}_{1},\dots,{\bf \overline{S}}_{N_2}$ remain unchanged.
Consequently, collisions between mixed-type solitons are classified as shape-altering.
Two typical examples are shown in Fig.~\ref{Fig11}: one involving two type-I and one type-II dark soliton,
and the other involving one type-I and two type-II dark solitons.
In both cases, we can directly observe that the shapes of the type-I dark solitons undergo significant changes, while the shapes of the type-II dark solitons remain unaffected.

Finally, we comment on the mutual influence between the two distinct types of solitons during collisions. Before proceeding, we need to emphasize that the pure type-I dark soliton ${\bf S}_{k_1}$ is determined by the parameter $p_{k_1}$, while the pure type-II dark soliton $\overline{{\bf S}}_{k_2}$ is controlled by the parameter $p_{2N_1+k_2}$.

We first consider the effect of type-I dark solitons on type-II dark solitons. Since the asymptotic expressions of the type-II soliton $\overline{{\bf S}}_{k_2}$ before and after the collision satisfy
\[
\left|\overline{u}_{k_2}^{+}(\zeta_{2N_1+k_2})\right|=\left|\overline{u}_{k_2}^{-}(\zeta_{2N_1+k_2}+\Delta\widehat{\Phi}_{k_2})\right|,
\]
the parameter $p_{k_1}$ associated with the type-I soliton affects only the phase shift $\Delta\widehat{\Phi}_{k_2}$.
This demonstrates that collisions with type-I solitons induce merely a phase shift in type-II solitons,
while leaving their shapes and amplitudes unchanged.

Next, we consider the effects of type-II dark solitons on type-I dark solitons.
This indicates that the inclusion of type-II dark solitons in the collisions indeed alters both the shapes and amplitudes of the type-I dark solitons.
It is worth noting a key feature of the expressions $\widetilde{\chi}_{k_1}^{(n)\pm}$ and $\widetilde{\Theta}_{k_1}^{(n)\pm}$.
As shown in Eq.~\eqref{delta-T1-2-1} and the subsequent discussion,
$\widetilde{\chi}_{k_1}^{(n)-}$ and $\widetilde{\Theta}_{k_1}^{(n)-}$ depend not only on the parameters $p_1, p_2, \dots, p_{N_1}$,
but also on $p_{2N_1+1}, p_{2N_1+2}, \dots, p_{2N_1+N_2}$.
In contrast, $\widetilde{\chi}_{k_1}^{(n)+}$ and $\widetilde{\Theta}_{k_1}^{(n)+}$ depend only on $p_1, p_2, \dots, p_{N_1}$, and are independent of $p_{2N_1+1}, p_{2N_1+2}, \dots, p_{2N_1+N_2}$.
In other words,
\begin{equation}\label{CH-P}
\begin{aligned}
\widetilde{\chi}_{k_1}^{(n)-} &= \widetilde{\chi}_{k_1}^{(n)-}(p_1, p_2, \dots, p_{N_1}, p_{2N_1+1}, p_{2N_1+2}, \dots, p_{2N_1+N_2}), \\
\widetilde{\chi}_{k_1}^{(n)+} &= \widetilde{\chi}_{k_1}^{(n)+}(p_1, p_2, \dots, p_{N_1}), \\
\widetilde{\Theta}_{k_1}^{(n)-} &= \widetilde{\Theta}_{k_1}^{(n)-}(p_1, p_2, \dots, p_{N_1}, p_{2N_1+1}, p_{2N_1+2}, \dots, p_{2N_1+N_2}), \\
\widetilde{\Theta}_{k_1}^{(n)+} &= \widetilde{\Theta}_{k_1}^{(n)+}(p_1, p_2, \dots, p_{N_1}).
\end{aligned}
\end{equation}

The specific expressions for the amplitude $\mathcal{I}_{k_1}^{\pm}$ of the type-I dark soliton ${\bf S}_{k_1}$ before and after the collisions are given by:
\begin{equation}
\begin{aligned}
\mathcal{I}_{k_1}^{\pm}=&\left.|u_{k_1}^{\pm}|\right|_{\zeta_{k_1}=-\widetilde{\Omega}_{k_1}^{\pm}} = \left|\frac{\widetilde{\chi}_{k_1}^{(1)\pm} + \widetilde{\Theta}_{k_1}^{(1)\pm}}{\widetilde{\chi}_{k_1}^{(0)\pm} + \widetilde{\Theta}_{k_1}^{(0)\pm}}\right|.
\end{aligned}
\end{equation}
Referring to Eq.~\eqref{CH-P}, we conclude that $\mathcal{I}_{k_1}^{-}$ depends on the parameters $p_1, p_2, \dots, p_{N_1}, p_{2N_1+1}, p_{2N_1+2}, \dots, p_{2N_1+N_2}$, while $\mathcal{I}_{k_1}^{+}$ depends only on $p_1, p_2, \dots, p_{N_1}$. Specifically,
\begin{equation}\label{TY-I}
\begin{aligned}
\mathcal{I}_{k_1}^{-}=&\mathcal{I}_{k_1}^{-}(p_1,p_2,\cdots,p_{N_1},p_{2N_1+1},p_{2N_1+2},\cdots,p_{2N_1+N_2}),\\
\mathcal{I}_{k_1}^{+}=&\mathcal{I}_{k_1}^{+}(p_1,p_2,\cdots,p_{N_1}).
\end{aligned}
\end{equation}
Since the type-II dark solitons are characterized by the parameters $p_{2N_1+k_2}$, it follows from Eq.~\eqref{TY-I} that the type-II dark solitons only influence the amplitudes of the type-I soliton before the collision, while they have no effect on the amplitudes of the type-I soliton after the collision. As noted specifically below Eq.~\eqref{delta-T1-2-1},  $\widetilde{\chi}_{k_1}^{(n)+} = \chi_{k_1}^{(n)}$ and $\widetilde{\Theta}_{k_1}^{(n)+} = \Theta_{k_1}^{(n)}$, thus, the amplitude expressions $\mathcal{I}_{k_1}^{+}$ and $\mathcal{I}_{k_1}$ in Eq.~\eqref{AM-Ik} satisfy the relation $\mathcal{I}_{k_1}^{+} = \mathcal{I}_{k_1}$. This indicates that, in mixed soliton solutions, the amplitude of the type-I dark soliton ${\bf S}_{k_1}$ is equal to that in the pure type-I dark soliton solutions, which further confirms that the influence of the type-II dark solitons on the amplitudes of the type-I dark solitons vanishes after their collisions.

Using a similar analytical approach, we conclude that the influence of type-II dark solitons on the shapes of type-I dark solitons also vanishes after their collisions. In particular, the shape of the type-I dark soliton ${\bf S}_{k_1}$ before and after the collisions can be characterized by the factor $\widetilde{\Delta}_{k_1}^{\pm}$, where
\begin{equation}\label{Delta-TI-M}
\begin{aligned}
\widetilde{\Delta}_{k_1}^{\pm} =
\left.\left[\partial_{\zeta_{k_1}\zeta_{k_1}}
\big|u_{k_1}^{\pm}(\zeta_{k_1})\big|\right]\right|_{\zeta_{k_1}=-\widetilde{\Omega}_{k_1}^{\pm}},
\qquad {k_1}=1,2,\dots,N_1.
\end{aligned}
\end{equation}
It follows that $\widetilde{\Delta}_{k_1}^{-} = \widetilde{\Delta}_{k_1}^{-}(p_1, p_2, \dots, p_{N_1}, p_{2N_1+1}, p_{2N_1+2}, \dots, p_{2N_1+N_2})$, whereas $\widetilde{\Delta}_{k_1}^{+} = \widetilde{\Delta}_{k_1}^{+}(p_1, p_2, \dots, p_{N_1})$. Hence, the parameters $p_{2N_1+1}, p_{2N_1+2}, \dots, p_{2N_1+N_2}$ can affect the sign of $\widetilde{\Delta}_{k_1}^{-}$ but have no influence on the sign of $\widetilde{\Delta}_{k_1}^{+}$. This reveals that the type-II dark solitons can modify the shapes of the type-I dark solitons only before the collision, but not after it. In fact, the expression for $\widetilde{\Delta}_{k_1}^{+}$ coincides with the shape factor $\widetilde{\Delta}_{k_1}$ of the pure type-I dark solitons given in Eq.~\eqref{Delta-TI}. Consequently, the shapes of the type-I dark solitons in the mixed soliton solutions are restored to those of the pure type-I dark soliton solutions after the collisions.

\begin{figure}[!htbp]
\centering
\subfigure{\includegraphics[height=5.5cm,width=5.91cm]{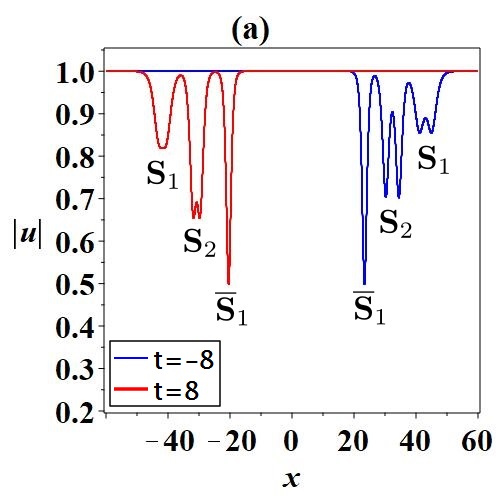}}
\subfigure{\includegraphics[height=5.5cm,width=5.91cm]{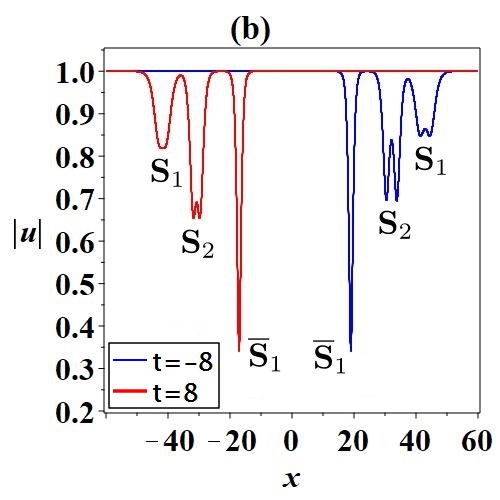}}
\subfigure{\includegraphics[height=5.5cm,width=5.91cm]{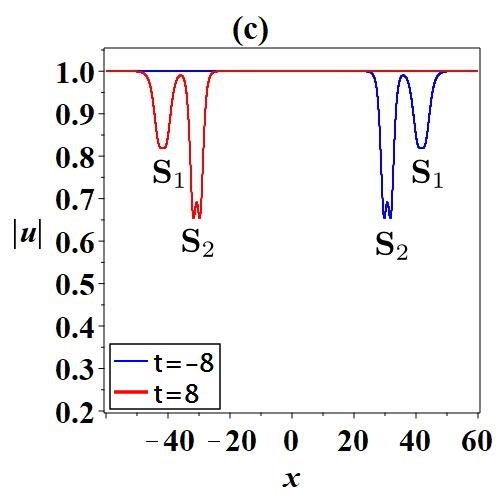}}
\caption{(a)The mixed solutions \eqref{D-SO} composed of two type-I dark solitons and a single type-II dark soliton with $(N_1,N_2)=(2,1)$, $(p_1,p_2,p_5)=(e^{\frac{1}{3}\mathrm{i}\pi},e^{\frac{1}{4}\mathrm{i}\pi},e^{\frac{1}{6}\mathrm{i}\pi})$ and $(\zeta_{1,0},\zeta_{2,0},\zeta_{5,0})=(0,0,0)$;
(b)The mixed solutions \eqref{D-SO} composed of two type-I dark solitons and a single type-II dark soliton with $(N_1,N_2)=(2,1)$, $(p_1,p_2,p_5)=(e^{\frac{1}{3}\mathrm{i}\pi},e^{\frac{1}{4}\mathrm{i}\pi},e^{\frac{1}{9}\mathrm{i}\pi})$ and $(\zeta_{1,0},\zeta_{2,0},\zeta_{5,0})=(0,0,0)$;
(c)The pure two type-I dark soliton solutions \eqref{D-SO} with $(N_1,N_2)=(2,0)$, $(p_1,p_2)=(e^{\frac{1}{3}\mathrm{i}\pi},e^{\frac{1}{4}\mathrm{i}\pi})$ and $(\zeta_{1,0},\zeta_{2,0})=(0,0)$. ~}\label{Fig12}
\end{figure}

To provide a clearer visual understanding of the influence of type-II dark solitons on type-I dark solitons, we present two distinct cases of collisions between two type-I dark solitons and a single type-II dark soliton, together with a comparison to the collision of two pure type-I dark solitons, as shown in Fig.~\ref{Fig12}. In the three panels of this figure, even though the parameters $p_1,p_2$ that determine the identities of the two type-I dark solitons ${\bf S}_1,{\bf S}_2$ are identical, the two solitons nevertheless exhibit clearly distinct shapes and amplitudes before the collision, as indicated by the blue solid lines at $t=-8$. Specifically, changing the value of $p_3$, which corresponds to modifying the properties of the type-II soliton $\mathbf{\overline{S}}_1$, also alters the shapes of the two type-I solitons $\mathbf{S}_1$ and $\mathbf{S}_2$. Both cases indicate that the inclusion of the type-II dark soliton ${\bf \overline{S}}_1$ indeed causes alterations in the shapes and amplitudes of the two type-I dark solitons ${\bf S}_1,{\bf S}_2$. However, after the collision, these two type-I dark solitons share the same shapes and amplitudes as the pure two type-I dark solitons in panel (c), even though they collide with different single type-II dark solitons in panels (a) and (b), as shown by the red solid lines at $t=8$ in all three panels. This indicates that the influence of the single type-II dark soliton on the two type-I dark solitons vanishes, even though the features of the single type-II dark soliton in panels (a) and (b) are distinct.

\section{Conclusion}\label{conclusion}
In this paper, we have derived general dark soliton solutions for the complex modified Korteweg--de Vries equation, expressed in terms of block determinants. The obtained general dark soliton solutions are classified into two distinct types, namely type-I and type-II dark solitons by assigning specific values to the  sub-block orders in these determinant solutions. Strikingly, the single soliton of these two types differ not only in their mathematical structures but also in their three dynamical properties:
(i) The single type-I dark soliton experiences no phase shift as \( x \) varies from \( -\infty \) to \( +\infty \), whereas the phase of the single type-II dark soliton shifts by \( (\pi + 2\vartheta_1) \) over the same range. Here $\vartheta_1$ is correlated to the soliton velocity through Eq.~\eqref{1-v}. (ii) The type-I dark soliton displays three distinct valley profiles: double-valley, single-valley, and flat-bottom (see Fig.~\ref{Fig2}), while the type-II dark soliton only exhibits a single-valley profile.
(iii) The velocity-amplitude relations in these two types of dark solitons are distinct, as shown in Fig.~\ref{Fig3}. Notably, the valley profiles of a single type-I dark soliton are correlated with its velocity, as illustrated in detail in Fig.~\ref{Fig1}(c).

Further, collisions of multiple pure type-I dark solitons and collisions of multiple pure type-II dark solitons are shape-preserving, which do not change the soliton identities except for certain phase shifts. A rarely reported soliton phenomenon has been discovered in the collisions of multiple pure type-I dark solitons, where these multiple dark solitons mutually influence soliton properties of each other. Specifically, changing the velocity or other characteristics of one soliton can result in changes to the properties of all other type-I dark solitons, including their shapes and amplitudes. For instance, as shown in Fig.~\ref{Fig5}, in the collision scenario of two type-I dark solitons, denoted as ${\bf S}_1$ and ${\bf S}_2$, we keep the parameters related to the properties of one soliton unchanged. However, its amplitude still exhibits noticeable changes as the velocity of the other soliton varies, even though the two solitons consistently undergo shape-preserving collisions. This unique characteristic is more intuitively observed in Fig.~\ref{Fig7}, where the valley profiles of soliton ${\bf S}_2$ clearly alter as the velocity of soliton ${\bf S}_1$ varies, even though the parameters related to the velocity of soliton ${\bf S}_2$ remain fixed.
In contrast, multiple pure type-II dark solitons display completely different properties. They behave independently and only undergo phase shifts, without any further influence. Specifically, varying the velocity of one soliton does not affect the shapes or amplitudes of the other type-II solitons.

The coherent structures of type-I dark solitons with type-II dark solitons exhibit shape-altering collisions, resulting in significant changes to the properties of the type-I dark solitons, including their shapes and amplitudes. In contrast, the type-II dark solitons experience only phase shifts without any other changes in their shapes and amplitudes. More uniquely, the influence of type-II dark solitons on the shape alteration of type-I dark solitons exists only before the interaction. After the collision, this shape-altering effect disappears, and the shapes of the type-I dark solitons return to the form observed in the pure type-I dark soliton scenario, as if the type-II dark solitons were absent.  Two specific examples demonstrating this shape-altering collision are shown in Fig.~\ref{Fig9} and Fig.~\ref{Fig12}, where it can be visually observed that, prior to the collision, the shapes of the type-I dark solitons indeed vary with changes in the properties of the type-II dark solitons (see the plots at $t=-6$ in Fig.~\ref{Fig9} and $t=-8$ in Fig.~\ref{Fig12}). However, after the collisions, this shape-altering effect disappears, and the shapes of the type-I dark solitons remain consistent (see the plots at $t=6$ in Fig.~\ref{Fig9} and $t=8$ in Fig.~\ref{Fig12}).
This unique type of shape-altering behavior in dark soliton collisions has rarely been reported before. This study will give impetus into the understanding of non-trivial multiple soliton collision behaviour in spinor condnesates and in multi-mode optical fibers with intricate four-wave mixing. 

\appendix
\section{Derivations of the general multiple dark soliton solutions }\label{App-1}
This appendix provides the detailed derivations of the multiple dark soliton solutions for the cmKdV equation as presented in Theorem \ref{definition of solution}.

Using Eq.~\eqref{D-SO}, the cmKdV equation \eqref{CmKdV} can be transformed into the following bilinear form\cite{Hirota}:
\begin{equation}\label{MKdV-Bi}
\begin{aligned}
&\left(D_t+D_{x}^3-6D_x\right)g \cdot f =0,\\
&\left(D_{x}^2-2\right)f \cdot f =-2|g|^2,
\end{aligned}
\end{equation}
where $f\in\mathbb{R},g\in\mathbb{C}$. As confirmed in \cite{Ohta-SS,Qin-AML},
the bilinear equations:
 \begin{equation}\label{KP-Eq}
\begin{aligned}
\left(D_{x_1}D_{x_{-1}}-2\right)\tau_n\cdot\tau_n&=-2\tau_{n+1}\tau_{n-1},\\
\left(D_{x_{-1}}(D_{x_1}^2-D_{x_2})-4D_{x_{1}}\right)\tau_{n+1}\cdot\tau_{n}&=0,\\
\left(D_{x_1}^3-4D_{x_3}+3D_{x_1}D_{x_2}\right)\tau_{n+1}\cdot\tau_{n}&=0,
\end{aligned}
\end{equation}
admit tau functions of the form
\begin{equation}\label{KP-tau}
\begin{aligned}
\tau_n=\det\limits_{1{\leq}i,j\leq{N}}\left(m_{i,j}^{(n)}\right),
\end{aligned}
\end{equation}
where $m_{i,j}^{(n)}$ is defined through a set of differential relations involving the functions $\Phi_i^{(n)}$ and $\overline{\Phi}_j^{(n)}$ as follows:
\begin{equation}\label{KP-OD1}
\begin{aligned}
\left\{
\begin{aligned}
\partial_{x_1}m_{i,j}^{(n)}=&\Phi_i^{(n)}\overline{\Phi}_j^{(n)},\partial_{x_{-1}}m_{i,j}^{(n)}=-\Phi_i^{(n-1)}\overline{\Phi}_j^{(n+1)},\\
\partial_{x_2}m_{i,j}^{(n)}=&\Phi_i^{(n+1)}\overline{\Phi}_j^{(n)}+\Phi_i^{(n)}\overline{\Phi}_j^{(n-1)},\\
\partial_{x_3}m_{i,j}^{(n)}=&\Phi_i^{(n+2)}\overline{\Phi}_j^{(n)}+\Phi_i^{(n+1)}\overline{\Phi}_j^{(n-1)}+\Phi_i^{(n)}\overline{\Phi}_j^{(n-2)},\\
m_{i,j}^{(n+1)}=&m_{i,j}^{(n)} + \Phi_i^{(n)}\overline{\Phi}_j^{(n+1)},
\end{aligned}
\right.
\end{aligned}
\end{equation}
with $\Phi_i^{(n)}$ and $\overline{\Phi}_j^{(n)}$ being arbitrary functions that satisfy:
 \begin{equation}\label{KP-OD2}
\begin{aligned}
\partial_{x_\ell}\Phi_i^{(n)}=&\Phi_i^{(n+\ell)},
\partial_{x_\ell}\overline{\Phi}_j^{(n)}=-\overline{\Phi}_j^{(n-\ell)}, \ell=-1,1,2,3.
\end{aligned}
\end{equation}
 A simple form for $\Phi_i^{(n)}$ and $\overline{\Phi}_j^{(n)}$, as satisfying Eq.~\eqref{KP-OD2},
  is as follows:
\begin{equation}\label{KP-OD-S1}
\begin{aligned}
\Phi_i^{(n)}=&p_i^ne^{\xi_i},\xi_i=\frac{1}{p_i}x_{-1}+p_ix_1+p_i^2x_2+p_i^3x_3+\xi_{i,0},\\
\overline{\Phi}_j^{(n)}=&\left(-\overline{p}_j\right)^{-n}e^{\overline{\xi}_j},
\overline{\xi}_j=\frac{1}{\overline{p}_j}x_{-1}+\overline{p}_jx_1-\overline{p}_j^2x_2+\overline{p}_j^3x_3+\overline{\xi}_{j,0}.
\end{aligned}
\end{equation}
Using such forms of $\Phi_i^{(n)}$ and $\overline{\Phi}_j^{(n)}$, we can express
 $m_{i,j}^{(n)}$ satisfying Eq.~\eqref{KP-OD1} as:
 \begin{equation}\label{KP-OD-S}
\begin{aligned}
m_{i,j}^{(n)}=&\delta_{i,j}+\frac{1}{p_i+\overline{p}_j}\left(-\frac{p_i}{\overline{p}_j}\right)^ne^{\xi_i+\overline{\xi}_j},
\end{aligned}
\end{equation}
where $\delta_{i,j}=1$ when $i=j$ and $\delta_{i,j}=0$ when $i\neq j$.

To reduce Eq.~\eqref{KP-Eq} to Eq.~\eqref{MKdV-Bi}, we have to first impose the constraint
 \begin{equation}
\begin{aligned}
p_i\overline{p}_i=1,
\end{aligned}
\end{equation}
so that $\tau_n$ can satisfy the dimensional reduction condition:
 \begin{equation}
\begin{aligned}
(\partial_{x_1}-\partial_{x_{-1}})\tau_n=0.
\end{aligned}
\end{equation}
With this condition, \(x_{-1}\) in Eq.~\eqref{KP-Eq} can be replaced by \(x_1\). After performing a simple linear substitution, the terms involving \(x_{-1}\) and \(x_2\) in Eq.~\eqref{KP-Eq} disappear, reducing Eq.~\eqref{KP-Eq} to the following form:
\begin{equation}\label{KP-Eq-1}
\begin{aligned}
\left(D_{x_1}^2-2\right)\tau_n\cdot\tau_n&=-2\tau_{n+1}\tau_{n-1},\\
\left(D_{x_1}^3-3D_{x_1}-D_{x_3}\right)\tau_{n+1}\cdot\tau_{n}&=0.
\end{aligned}
\end{equation}
Since this bilinear system is independent of \(x_{-1}\) and \(x_2\), we can set \(x_{-1} = 0\) and \(x_2 = 0\) in \(\tau_n\).
Next, we impose $x_1,x_3\in\mathbb{R}$ and the following conjugate constraint:
 \begin{equation}
\begin{aligned}
\overline{p}_i = p_i^*, \quad \overline{\xi}_{i,0} = \xi_{i,0}^*,
\end{aligned}
\end{equation}
so that \(\tau_n\) satisfies the condition:
 \begin{equation}
\begin{aligned}
\tau_n^*=\tau_{-n}.
\end{aligned}
\end{equation}
If we further set \(x_1 = x + 3t\), \(x_3 = -t\), and \(\tau_0 = f\), \(\tau_1 = g\), \(\tau_{-1} = g^*\), then Eq.~\eqref{KP-Eq-1} reduces to Eq.~\eqref{MKdV-Bi}. Consequently, the solutions of the complex modified KdV (cmKdV) equation can be expressed through the tau functions as \(u = \frac{\tau_1}{\tau_0}\).

Finally, by taking \(N = 2N_1 + N_2\), \(\zeta_i = \xi_i + \xi_i^*\), and imposing the parameter constraints in Eq.~\eqref{pa-1}, the functions \(\tau_n\) and \(\overline{\delta}_n\) of Eq.~\eqref{so-fg} will admit the following correlation:
 \begin{equation}
\begin{aligned}
\tau_n = \mathcal{C} \overline{\delta}_n,
\end{aligned}
\end{equation}
where $\mathcal{C} = \exp\left( \sum_{i=1}^{2N_1+N_2} \zeta_i \right)$. Since the factor $\mathcal{C}$ can be canceled out without affecting the solution $u = \frac{\tau_1}{\tau_0} = \frac{\mathcal{C} \overline{\delta}_1}{\mathcal{C} \overline{\delta}_0}$, the solutions of the cmKdV equation in the form of Theorem \ref{definition of solution} are derived.

\section{The long-time asymptotic analysis to function $\overline{\delta}_n$}\label{App-2}
In this appendix, we will present the proofs of Lemma \ref{asy-ntype1-delta} and Lemma \ref{asy-ntype2-delta}. The proof of Lemma \ref{asy-ntype1-2-delta} combines the results of Lemma \ref{asy-ntype1-delta} and Lemma \ref{asy-ntype2-delta} and is omitted for brevity. To begin with, we introduce the explicit form of the well-known Cauchy determinant through the following lemma.
\begin{lemma}\label{Cauchy determinant}
    Let $D_N = \operatorname{det}_{1\leq s,j \leq N} \left(\frac{1}{a_s+b_j}\right)$ represent a Cauchy determinant with entries given by $\frac{1}{a_s+b_j}$. Then, it follows that $D_N = \frac{\prod_{1 \leq s < j\leq N}(a_j -a_s)(b_j - b_s)}{\prod_{s,j=1}^N(a_s+b_j)}$.
\end{lemma}
Lemma \ref{Cauchy determinant} can be confirmed through fundamental determinant operations.

\subsection{Proof of Lemma\ref{asy-ntype1-delta}}
In order to prove Lemma \ref{asy-ntype1-delta}, we start by reviewing the necessary preliminary conditions and assumptions associated with it. In this case, we examine the pure $N_1$ type-I dark soliton. The velocities of each soliton are defined as $\nu_{k_1}=2\cos(2\vartheta_{k_1})-4, \quad k_1=1,2,\cdots, N_1$. Here, $\vartheta_{k_1} \in (-\frac{\pi}{2}, \frac{\pi}{2})$, ensuring that $\nu_{k_1} \in (-6,-2]$. Without loss of generality, we assume
\begin{equation}
    \nu_1<\nu_2<\cdots<\nu_{k_1}<\cdots<\nu_{N_1}<0,
\end{equation}
Given the symmetry condition $\zeta_{N_1+s} = -\zeta_{s}+i\pi$, we additionally obtain
\begin{equation}
    \nu_{N_1+1}<\nu_{N_1+2}<\cdots<\nu_{N_1+k_1}<\cdots<\nu_{2N_1}<0.
\end{equation}
Considering these conditions, we can explore the asymptotic behaviors of $\overline{\delta}_n$ as $t \to \infty$ and $t \to -\infty$, respectively.
The discussion is conducted around the neighbourhood of $\zeta_{k_1} \approx 0$, which implies $x \approx \nu_{k_1} t$.\\
(a)\underline{ Before collision $t\rightarrow-\infty$}\\
As $t\rightarrow-\infty$, the following results hold for $e^{\zeta_j}$ and $e^{-\zeta_j}$:
\begin{equation}
    \left\{
    \begin{aligned}
    & e^{\zeta_j} \rightarrow \infty, \quad j>k_1,\\
    & e^{\zeta_j} \rightarrow 0, \quad j<k_1,\\
    \end{aligned}
    \right. \quad \text{and} \quad
    \left\{
    \begin{aligned}
    & e^{-\zeta_j} \rightarrow 0, \quad j>k_1,\\
    & e^{-\zeta_j} \rightarrow \infty, \quad j<k_1,\\
    \end{aligned}
    \right.
\end{equation}
Next, we undertake the asymptotic analysis of the function $\overline{\delta}_n$. By examining the asymptotic properties of $e^{\zeta_j}$ and $e^{-\zeta_j}$ and applying the symmetry condition $\zeta_{N_1+s} = -\zeta_{s}+i\pi$, we derive the following asymptotic expression.

\begin{equation}
\hspace{-1cm}
\begin{aligned}
\overline{\delta}_n \simeq &\overline{\delta}_n^{[k_1]-} \\
= &(-1)^{N_1-k_1}e^{-\sum_{s=1}^{k_1-1} \zeta_{s}+\sum_{s=k_1+1}^{N_1} \zeta_{s}}\times \\
&{\tiny \left|
\begin{array}{cccccccccccccc}
1 & \cdots & 0 & 0 & 0 & \cdots & 0 & 0 &\cdots & 0 & 0 & 0 & \cdots & 0\\
\vdots & \ddots & \vdots& \vdots& \vdots& \ddots & \vdots& \vdots &\ddots &\vdots &\vdots & \vdots & \ddots & \vdots\\
0 & \cdots & 1 & 0 & 0 & \cdots & 0 & 0 &\cdots & 0 &0 & 0 & \cdots & 0\\
0 & \cdots & 0 & \overline{\mathrm{m}}_{k_1,k_1}^{(n)}+e^{-\zeta_{k_1}} & \overline{\mathrm{m}}_{k_1,k_1+1}^{(n)} & \cdots & \overline{\mathrm{m}}_{k_1,N_1}^{(n)} & \overline{\mathrm{m}}_{k_1,N_1+1}^{(n)} &\cdots & \overline{\mathrm{m}}_{k_1,N_1+k_1-1}^{(n)} & \overline{\mathrm{m}}_{k_1,N_1+k_1}^{(n)}& 0 & \cdots & 0\\
0 & \cdots & 0 & \overline{\mathrm{m}}_{k_1+1,k_1}^{(n)} & \overline{\mathrm{m}}_{k_1+1,k_1+1}^{(n)} & \cdots & \overline{\mathrm{m}}_{k_1+1,N_1}^{(n)} & \overline{\mathrm{m}}_{k_1+1,N_1+1}^{(n)} &\cdots & \overline{\mathrm{m}}_{k_1+1,N_1+k_1-1}^{(n)} & \overline{\mathrm{m}}_{k_1+1,N_1+k_1}^{(n)} & 0 & \cdots & 0\\
\vdots & \ddots & \vdots& \vdots& \vdots& \ddots & \vdots& \vdots &\ddots &\vdots &\vdots & \vdots & \ddots & \vdots\\
0 & \cdots & 0 & \overline{\mathrm{m}}_{N_1,k_1}^{(n)} & \overline{\mathrm{m}}_{N_1,k_1+1}^{(n)} & \cdots & \overline{\mathrm{m}}_{N_1,N_1}^{(n)} & \overline{\mathrm{m}}_{N_1,N_1+1}^{(n)} &\cdots & \overline{\mathrm{m}}_{N_1,N_1+k_1-1}^{(n)} & \overline{\mathrm{m}}_{N_1,N_1+k_1}^{(n)} & 0 & \cdots & 0\\
0 & \cdots & 0 & \overline{\mathrm{m}}_{N_1+1,k_1}^{(n)} & \overline{\mathrm{m}}_{N_1+1,k_1+1}^{(n)} & \cdots & \overline{\mathrm{m}}_{N_1+1,N_1}^{(n)} & \overline{\mathrm{m}}_{N_1+1,N_1+1}^{(n)} &\cdots & \overline{\mathrm{m}}_{N_1+1,N_1+k_1-1}^{(n)} & \overline{\mathrm{m}}_{N_1+1,N_1+k_1}^{(n)} &  0& \cdots &0 \\
\vdots & \ddots & \vdots& \vdots& \vdots& \ddots & \vdots& \vdots &\ddots &\vdots &\vdots & \vdots & \ddots & \vdots\\
0 & \cdots & 0 & \overline{\mathrm{m}}_{N_1+k_1-1,k_1}^{(n)} & \overline{\mathrm{m}}_{N_1+k_1-1,k_1+1}^{(n)} & \cdots & \overline{\mathrm{m}}_{N_1+k_1-1,N_1}^{(n)} & \overline{\mathrm{m}}_{N_1+k_1-1,N_1+1}^{(n)} &\cdots & \overline{\mathrm{m}}_{N_1+k_1-1,N_1+k_1-1}^{(n)} & \overline{\mathrm{m}}_{N_1+k_1-1,N_1+k_1}^{(n)} & 0 & \cdots & 0\\
0 & \cdots & 0 & \overline{\mathrm{m}}_{N_1+k_1,k_1}^{(n)} & \overline{\mathrm{m}}_{N_1+k_1,k_1+1}^{(n)} & \cdots & \overline{\mathrm{m}}_{N_1+k_1,N_1}^{(n)} & \overline{\mathrm{m}}_{N_1+k_1,N_1+1}^{(n)} &\cdots & \overline{\mathrm{m}}_{N_1+k_1,N_1+k_1-1}^{(n)} & \overline{\mathrm{m}}_{N_1+k_1,N_1+k_1}^{(n)}+e^{-\zeta_{N_1+k_1}} & 0 & \cdots & 0\\
0 & \cdots & 0 & 0& 0 & \cdots & 0 & 0 &\cdots & 0 & 0& 1 & \cdots & 0\\
\vdots & \ddots & \vdots& \vdots& \vdots& \ddots & \vdots& \vdots &\ddots &\vdots &\vdots & \vdots & \ddots & \vdots\\
0 & \cdots & 0 & 0& 0 & \cdots & 0 & 0 &\cdots & 0 & 0& 0 & \cdots & 1\\
\end{array}
\right|}\\
= &(-1)^{N_1-k_1}e^{-\sum_{s=1}^{k_1-1} \zeta_{s}+\sum_{s=k_1+1}^{N_1} \zeta_{s}}\times \mathcal{A}
\end{aligned}
\end{equation}

with
\begin{equation}
    \begin{aligned}
    \mathcal{A} &=\\
    &{\tiny \left|
\begin{array}{cccccccc}
 \overline{\mathrm{m}}_{k_1,k_1}^{(n)}+e^{-\zeta_{k_1}} & \overline{\mathrm{m}}_{k_1,k_1+1}^{(n)} & \cdots & \overline{\mathrm{m}}_{k_1,N_1}^{(n)} & \overline{\mathrm{m}}_{k_1,N_1+1}^{(n)} &\cdots & \overline{\mathrm{m}}_{k_1,N_1+k_1-1}^{(n)} & \overline{\mathrm{m}}_{k_1,N_1+k_1}^{(n)}\\
 \overline{\mathrm{m}}_{k_1+1,k_1}^{(n)} & \overline{\mathrm{m}}_{k_1+1,k_1+1}^{(n)} & \cdots & \overline{\mathrm{m}}_{k_1+1,N_1}^{(n)} & \overline{\mathrm{m}}_{k_1+1,N_1+1}^{(n)} &\cdots & \overline{\mathrm{m}}_{k_1+1,N_1+k_1-1}^{(n)} & \overline{\mathrm{m}}_{k_1+1,N_1+k_1}^{(n)} \\
 \vdots& \vdots& \ddots & \vdots& \vdots &\ddots &\vdots &\vdots \\
 \overline{\mathrm{m}}_{N_1,k_1}^{(n)} & \overline{\mathrm{m}}_{N_1,k_1+1}^{(n)} & \cdots & \overline{\mathrm{m}}_{N_1,N_1}^{(n)} & \overline{\mathrm{m}}_{N_1,N_1+1}^{(n)} &\cdots & \overline{\mathrm{m}}_{N_1,N_1+k_1-1}^{(n)} & \overline{\mathrm{m}}_{N_1,N_1+k_1}^{(n)} \\
\overline{\mathrm{m}}_{N_1+1,k_1}^{(n)} & \overline{\mathrm{m}}_{N_1+1,k_1+1}^{(n)} & \cdots & \overline{\mathrm{m}}_{N_1+1,N_1}^{(n)} & \overline{\mathrm{m}}_{N_1+1,N_1+1}^{(n)} &\cdots & \overline{\mathrm{m}}_{N_1+1,N_1+k_1-1}^{(n)} & \overline{\mathrm{m}}_{N_1+1,N_1+k_1}^{(n)} \\
\vdots& \vdots& \ddots & \vdots& \vdots &\ddots &\vdots &\vdots\\
\overline{\mathrm{m}}_{N_1+k_1-1,k_1}^{(n)} & \overline{\mathrm{m}}_{N_1+k_1-1,k_1+1}^{(n)} & \cdots & \overline{\mathrm{m}}_{N_1+k_1-1,N_1}^{(n)} & \overline{\mathrm{m}}_{N_1+k_1-1,N_1+1}^{(n)} &\cdots & \overline{\mathrm{m}}_{N_1+k_1-1,N_1+k_1-1}^{(n)} & \overline{\mathrm{m}}_{N_1+k_1-1,N_1+k_1}^{(n)} \\
\overline{\mathrm{m}}_{N_1+k_1,k_1}^{(n)} & \overline{\mathrm{m}}_{N_1+k_1,k_1+1}^{(n)} & \cdots & \overline{\mathrm{m}}_{N_1+k_1,N_1}^{(n)} & \overline{\mathrm{m}}_{N_1+k_1,N_1+1}^{(n)} &\cdots & \overline{\mathrm{m}}_{N_1+k_1,N_1+k_1-1}^{(n)} & \overline{\mathrm{m}}_{N_1+k_1,N_1+k_1}^{(n)}+e^{-\zeta_{N_1+k_1}}\\
\end{array}
\right|}.
  \end{aligned}
\end{equation}
Using the linearity of the determinant, we can express \( \mathcal{A} \) as the sum of four separate determinants:
\begin{equation}
    \begin{aligned}
        \mathcal{A} = \mathcal{A}_1 + \mathcal{A}_2 + \mathcal{A}_3 + \mathcal{A}_4.
    \end{aligned}
\end{equation}
Here,
\begin{equation}
    \begin{aligned}
        \mathcal{A}_1 =& \left|
\begin{array}{cccccccc}
 \overline{\mathrm{m}}_{k_1,k_1}^{(n)} & \overline{\mathrm{m}}_{k_1,k_1+1}^{(n)} & \cdots & \overline{\mathrm{m}}_{k_1,N_1}^{(n)} & \overline{\mathrm{m}}_{k_1,N_1+1}^{(n)} &\cdots & \overline{\mathrm{m}}_{k_1,N_1+k_1-1}^{(n)} & \overline{\mathrm{m}}_{k_1,N_1+k_1}^{(n)}\\
 \overline{\mathrm{m}}_{k_1+1,k_1}^{(n)} & \overline{\mathrm{m}}_{k_1+1,k_1+1}^{(n)} & \cdots & \overline{\mathrm{m}}_{k_1+1,N_1}^{(n)} & \overline{\mathrm{m}}_{k_1+1,N_1+1}^{(n)} &\cdots & \overline{\mathrm{m}}_{k_1+1,N_1+k_1-1}^{(n)} & \overline{\mathrm{m}}_{k_1+1,N_1+k_1}^{(n)} \\
 \vdots& \vdots& \ddots & \vdots& \vdots &\ddots &\vdots &\vdots \\
 \overline{\mathrm{m}}_{N_1,k_1}^{(n)} & \overline{\mathrm{m}}_{N_1,k_1+1}^{(n)} & \cdots & \overline{\mathrm{m}}_{N_1,N_1}^{(n)} & \overline{\mathrm{m}}_{N_1,N_1+1}^{(n)} &\cdots & \overline{\mathrm{m}}_{N_1,N_1+k_1-1}^{(n)} & \overline{\mathrm{m}}_{N_1,N_1+k_1}^{(n)} \\
\overline{\mathrm{m}}_{N_1+1,k_1}^{(n)} & \overline{\mathrm{m}}_{N_1+1,k_1+1}^{(n)} & \cdots & \overline{\mathrm{m}}_{N_1+1,N_1}^{(n)} & \overline{\mathrm{m}}_{N_1+1,N_1+1}^{(n)} &\cdots & \overline{\mathrm{m}}_{N_1+1,N_1+k_1-1}^{(n)} & \overline{\mathrm{m}}_{N_1+1,N_1+k_1}^{(n)} \\
\vdots& \vdots& \ddots & \vdots& \vdots &\ddots &\vdots &\vdots\\
\overline{\mathrm{m}}_{N_1+k_1-1,k_1}^{(n)} & \overline{\mathrm{m}}_{N_1+k_1-1,k_1+1}^{(n)} & \cdots & \overline{\mathrm{m}}_{N_1+k_1-1,N_1}^{(n)} & \overline{\mathrm{m}}_{N_1+k_1-1,N_1+1}^{(n)} &\cdots & \overline{\mathrm{m}}_{N_1+k_1-1,N_1+k_1-1}^{(n)} & \overline{\mathrm{m}}_{N_1+k_1-1,N_1+k_1}^{(n)} \\
\overline{\mathrm{m}}_{N_1+k_1,k_1}^{(n)} & \overline{\mathrm{m}}_{N_1+k_1,k_1+1}^{(n)} & \cdots & \overline{\mathrm{m}}_{N_1+k_1,N_1}^{(n)} & \overline{\mathrm{m}}_{N_1+k_1,N_1+1}^{(n)} &\cdots & \overline{\mathrm{m}}_{N_1+k_1,N_1+k_1-1}^{(n)} & \overline{\mathrm{m}}_{N_1+k_1,N_1+k_1}^{(n)}\\
\end{array}
\right|,\\
\mathcal{A}_2 =& \left|
\begin{array}{cccccccc}
 e^{-\zeta_{k_1}} & 0 & \cdots & 0 & 0 &\cdots & 0 & 0\\
 \overline{\mathrm{m}}_{k_1+1,k_1}^{(n)} & \overline{\mathrm{m}}_{k_1+1,k_1+1}^{(n)} & \cdots & \overline{\mathrm{m}}_{k_1+1,N_1}^{(n)} & \overline{\mathrm{m}}_{k_1+1,N_1+1}^{(n)} &\cdots & \overline{\mathrm{m}}_{k_1+1,N_1+k_1-1}^{(n)} & \overline{\mathrm{m}}_{k_1+1,N_1+k_1}^{(n)} \\
 \vdots& \vdots& \ddots & \vdots& \vdots &\ddots &\vdots &\vdots \\
 \overline{\mathrm{m}}_{N_1,k_1}^{(n)} & \overline{\mathrm{m}}_{N_1,k_1+1}^{(n)} & \cdots & \overline{\mathrm{m}}_{N_1,N_1}^{(n)} & \overline{\mathrm{m}}_{N_1,N_1+1}^{(n)} &\cdots & \overline{\mathrm{m}}_{N_1,N_1+k_1-1}^{(n)} & \overline{\mathrm{m}}_{N_1,N_1+k_1}^{(n)} \\
\overline{\mathrm{m}}_{N_1+1,k_1}^{(n)} & \overline{\mathrm{m}}_{N_1+1,k_1+1}^{(n)} & \cdots & \overline{\mathrm{m}}_{N_1+1,N_1}^{(n)} & \overline{\mathrm{m}}_{N_1+1,N_1+1}^{(n)} &\cdots & \overline{\mathrm{m}}_{N_1+1,N_1+k_1-1}^{(n)} & \overline{\mathrm{m}}_{N_1+1,N_1+k_1}^{(n)} \\
\vdots& \vdots& \ddots & \vdots& \vdots &\ddots &\vdots &\vdots\\
\overline{\mathrm{m}}_{N_1+k_1-1,k_1}^{(n)} & \overline{\mathrm{m}}_{N_1+k_1-1,k_1+1}^{(n)} & \cdots & \overline{\mathrm{m}}_{N_1+k_1-1,N_1}^{(n)} & \overline{\mathrm{m}}_{N_1+k_1-1,N_1+1}^{(n)} &\cdots & \overline{\mathrm{m}}_{N_1+k_1-1,N_1+k_1-1}^{(n)} & \overline{\mathrm{m}}_{N_1+k_1-1,N_1+k_1}^{(n)} \\
\overline{\mathrm{m}}_{N_1+k_1,k_1}^{(n)} & \overline{\mathrm{m}}_{N_1+k_1,k_1+1}^{(n)} & \cdots & \overline{\mathrm{m}}_{N_1+k_1,N_1}^{(n)} & \overline{\mathrm{m}}_{N_1+k_1,N_1+1}^{(n)} &\cdots & \overline{\mathrm{m}}_{N_1+k_1,N_1+k_1-1}^{(n)} & \overline{\mathrm{m}}_{N_1+k_1,N_1+k_1}^{(n)}\\
\end{array}
\right|,\\
\mathcal{A}_3 =& \left|
\begin{array}{cccccccc}
 \overline{\mathrm{m}}_{k_1,k_1}^{(n)} & \overline{\mathrm{m}}_{k_1,k_1+1}^{(n)} & \cdots & \overline{\mathrm{m}}_{k_1,N_1}^{(n)} & \overline{\mathrm{m}}_{k_1,N_1+1}^{(n)} &\cdots & \overline{\mathrm{m}}_{k_1,N_1+k_1-1}^{(n)} & \overline{\mathrm{m}}_{k_1,N_1+k_1}^{(n)}\\
 \overline{\mathrm{m}}_{k_1+1,k_1}^{(n)} & \overline{\mathrm{m}}_{k_1+1,k_1+1}^{(n)} & \cdots & \overline{\mathrm{m}}_{k_1+1,N_1}^{(n)} & \overline{\mathrm{m}}_{k_1+1,N_1+1}^{(n)} &\cdots & \overline{\mathrm{m}}_{k_1+1,N_1+k_1-1}^{(n)} & \overline{\mathrm{m}}_{k_1+1,N_1+k_1}^{(n)} \\
 \vdots& \vdots& \ddots & \vdots& \vdots &\ddots &\vdots &\vdots \\
 \overline{\mathrm{m}}_{N_1,k_1}^{(n)} & \overline{\mathrm{m}}_{N_1,k_1+1}^{(n)} & \cdots & \overline{\mathrm{m}}_{N_1,N_1}^{(n)} & \overline{\mathrm{m}}_{N_1,N_1+1}^{(n)} &\cdots & \overline{\mathrm{m}}_{N_1,N_1+k_1-1}^{(n)} & \overline{\mathrm{m}}_{N_1,N_1+k_1}^{(n)} \\
\overline{\mathrm{m}}_{N_1+1,k_1}^{(n)} & \overline{\mathrm{m}}_{N_1+1,k_1+1}^{(n)} & \cdots & \overline{\mathrm{m}}_{N_1+1,N_1}^{(n)} & \overline{\mathrm{m}}_{N_1+1,N_1+1}^{(n)} &\cdots & \overline{\mathrm{m}}_{N_1+1,N_1+k_1-1}^{(n)} & \overline{\mathrm{m}}_{N_1+1,N_1+k_1}^{(n)} \\
\vdots& \vdots& \ddots & \vdots& \vdots &\ddots &\vdots &\vdots\\
\overline{\mathrm{m}}_{N_1+k_1-1,k_1}^{(n)} & \overline{\mathrm{m}}_{N_1+k_1-1,k_1+1}^{(n)} & \cdots & \overline{\mathrm{m}}_{N_1+k_1-1,N_1}^{(n)} & \overline{\mathrm{m}}_{N_1+k_1-1,N_1+1}^{(n)} &\cdots & \overline{\mathrm{m}}_{N_1+k_1-1,N_1+k_1-1}^{(n)} & \overline{\mathrm{m}}_{N_1+k_1-1,N_1+k_1}^{(n)} \\
0 & 0 & \cdots & 0 & 0 &\cdots & 0 & e^{-\zeta_{N_1+k_1}}\\
\end{array}
\right|,\\
\mathcal{A}_4 =& \left|
\begin{array}{cccccccc}
  e^{-\zeta_{k_1}} & 0 & \cdots & 0 & 0 &\cdots & 0 & 0\\
 \overline{\mathrm{m}}_{k_1+1,k_1}^{(n)} & \overline{\mathrm{m}}_{k_1+1,k_1+1}^{(n)} & \cdots & \overline{\mathrm{m}}_{k_1+1,N_1}^{(n)} & \overline{\mathrm{m}}_{k_1+1,N_1+1}^{(n)} &\cdots & \overline{\mathrm{m}}_{k_1+1,N_1+k_1-1}^{(n)} & \overline{\mathrm{m}}_{k_1+1,N_1+k_1}^{(n)} \\
 \vdots& \vdots& \ddots & \vdots& \vdots &\ddots &\vdots &\vdots \\
 \overline{\mathrm{m}}_{N_1,k_1}^{(n)} & \overline{\mathrm{m}}_{N_1,k_1+1}^{(n)} & \cdots & \overline{\mathrm{m}}_{N_1,N_1}^{(n)} & \overline{\mathrm{m}}_{N_1,N_1+1}^{(n)} &\cdots & \overline{\mathrm{m}}_{N_1,N_1+k_1-1}^{(n)} & \overline{\mathrm{m}}_{N_1,N_1+k_1}^{(n)} \\
\overline{\mathrm{m}}_{N_1+1,k_1}^{(n)} & \overline{\mathrm{m}}_{N_1+1,k_1+1}^{(n)} & \cdots & \overline{\mathrm{m}}_{N_1+1,N_1}^{(n)} & \overline{\mathrm{m}}_{N_1+1,N_1+1}^{(n)} &\cdots & \overline{\mathrm{m}}_{N_1+1,N_1+k_1-1}^{(n)} & \overline{\mathrm{m}}_{N_1+1,N_1+k_1}^{(n)} \\
\vdots& \vdots& \ddots & \vdots& \vdots &\ddots &\vdots &\vdots\\
\overline{\mathrm{m}}_{N_1+k_1-1,k_1}^{(n)} & \overline{\mathrm{m}}_{N_1+k_1-1,k_1+1}^{(n)} & \cdots & \overline{\mathrm{m}}_{N_1+k_1-1,N_1}^{(n)} & \overline{\mathrm{m}}_{N_1+k_1-1,N_1+1}^{(n)} &\cdots & \overline{\mathrm{m}}_{N_1+k_1-1,N_1+k_1-1}^{(n)} & \overline{\mathrm{m}}_{N_1+k_1-1,N_1+k_1}^{(n)} \\
0 & 0 & \cdots & 0 & 0 &\cdots & 0 & e^{-\zeta_{N_1+k_1}}\\
\end{array}
\right|.
    \end{aligned}
\end{equation}
With Eq.~\eqref{mx-sj} and Lemma \ref{Cauchy determinant}, one can find the following results.
\begin{equation}
    \begin{aligned}
        \mathcal{A}_1 =& \prod_{s=k_1}^{N_1+k_1}\left(-\frac{p_s}{p_s^*}\right)^n \mathcal{D}_{[k_1,N_1+k_1]}, \quad
        &\mathcal{A}_2 = \prod_{s=k_1+1}^{N_1+k_1}\left(-\frac{p_s}{p_s^*}\right)^n e^{-\zeta_{k_1}}\mathcal{D}_{[k_1+1,N_1+k_1]},\\
        \mathcal{A}_3 =& \prod_{s=k_1}^{N_1+k_1-1}\left(-\frac{p_s}{p_s^*}\right)^n e^{-\zeta_{N_1+k_1}}\mathcal{D}_{[k_1,N_1+k_1-1]}, \quad
        &\mathcal{A}_4 = -\prod_{s=k_1+1}^{N_1+k_1-1}\left(-\frac{p_s}{p_s^*}\right)^n \mathcal{D}_{[k_1+1,N_1+k_1-1]}.
    \end{aligned}
\end{equation}
The explicit expression of $\mathcal{A}$ can be display as:

\begin{equation}
\begin{aligned}
\mathcal{A}
=&\mathcal{D}_{[k_1+1,N_1+k_1-1]}\prod_{s = k_1+1}^{N_1+k_1-1}\left(-\frac{p_s}{p_s^*}\right)^n\\
&\times  \left[\left(-\frac{p_{k_1}}{p_{k_1}^*}\right)^{2n}\frac{4}{|p_{k_1}-p_{k_1}^*|^2}\frac{(-1)}{|p_{k_1}+p_{k_1}^*|^2} \left(\prod_{s=1}^{k_1-1} \prod_{k_1+1} ^{N_1}\right) \frac{|p_{k_1}+p_{s}|^2}{|p_{s}-p_{k_1}^*|^2}\frac{|p_{k_1}-p_{s}|^2}{|p_{s}+p_{k_1}^*|^2}\right. \\
&\left. \quad + \left(-\frac{p_{k_1}}{p_{k_1}^*}\right)^ne^{-\zeta_{k_1}}\frac{-1}{p_{k_1}+p_{k_1}^*}\prod_{s=1}^{k_1-1}\frac{|p_{k_1}-p_{s}|^2}{|p_{s}+p_{k_1}^*|^2}\prod_{s=k_1+1}^{N_1}\frac{|p_{k_1}+p_{s}|^2}{|p_{s}-p_{k_1}^*|^2}\right.\\
&\quad \left.-\left(-\frac{p_{k_1}}{p_{k_1}^*}\right)^ne^{+\zeta_{k_1}}\frac{1}{p_{k_1}+p_{k_1}^*}\prod_{s=1}^{k_1-1}\frac{|p_{k_1}+p_{s}|^2}{|p_{s}-p_{k_1}^*|^2} \prod_{s=k_1+1}^{N_1}\frac{|p_{k_1}-p_{s}|^2}{|p_{s}+p_{k_1}^*|^2}-
1\right]\\
=&\mathcal{D}_{[k_1+1,N_1+k_1-1]}\left[\chi_{k_1}^{(n)}+\Theta_{k_1}^{(n)}\cosh(\zeta_{k_1}+\Omega_{k_1})\right].
\end{aligned}
\end{equation}

(b)\underline{ After collision $t\rightarrow\infty$}\\
As $t\rightarrow+\infty$, the following results hold for $e^{\zeta_j}$ and $e^{-\zeta_j}$:
\begin{equation}
    \left\{
    \begin{aligned}
    & e^{\zeta_j} \rightarrow 0, \quad j>k_1,\\
    & e^{\zeta_j} \rightarrow \infty, \quad j<k_1.\\
    \end{aligned}
    \right. \quad \text{and} \quad
    \left\{
    \begin{aligned}
    & e^{-\zeta_j} \rightarrow \infty, \quad j>k_1,\\
    & e^{-\zeta_j} \rightarrow 0, \quad j<k_1.\\
    \end{aligned}
    \right.
\end{equation}\\
By examining the asymptotic properties of $e^{\zeta_j}$ and $e^{-\zeta_j}$ and applying the symmetry condition $\zeta_{N_1+s} = -\zeta_{s}+i\pi$, we derive the following asymptotic expression of $\overline{\delta}_n$.

\begin{equation}
\hspace{-1cm}
\begin{aligned}
\overline{\delta}_n \simeq &\overline{\delta}^{[k_1]+}_n \\
= &(-1)^{k_1-1}e^{\sum_{s=1}^{k_1-1} \zeta_{s}- \sum_{s=k_1+1}^{N_1} \zeta_{s}}\times \\
&{\tiny\left|
\begin{array}{cccccccccccccc}
\overline{\mathrm{m}}_{1,1}^{(n)} & \cdots & \overline{\mathrm{m}}_{1,k_1-1}^{(n)} & \overline{\mathrm{m}}_{1,k_1}^{(n)} & 0 & \cdots & 0 & 0 &\cdots & 0 & \overline{\mathrm{m}}_{1,N_1+k_1}^{(n)} & \overline{\mathrm{m}}_{1,N_1+k_1+1}^{(n)} & \cdots & \overline{\mathrm{m}}_{1,2N_1}^{(n)}\\
\vdots & \ddots & \vdots& \vdots& \vdots& \ddots & \vdots& \vdots &\ddots &\vdots &\vdots & \vdots & \ddots & \vdots\\
\overline{\mathrm{m}}_{k_1-1,1}^{(n)} & \cdots & \overline{\mathrm{m}}_{k_1-1,k_1-1}^{(n)}+e^{-\zeta_{k_1-1}} & \overline{\mathrm{m}}_{k_1-1,k_1}^{(n)} & 0 & \cdots & 0 & 0 &\cdots & 0 & \overline{\mathrm{m}}_{k_1,N_1+k_1}^{(n)} & \overline{\mathrm{m}}_{k_1-1,N_1+k_1+1}^{(n)} & \cdots & \overline{\mathrm{m}}_{k_1-1,2N_1}^{(n)}\\
\overline{\mathrm{m}}_{k_1,1}^{(n)} & \cdots & \overline{\mathrm{m}}_{k_1,k_1-1}^{(n)} & \overline{\mathrm{m}}_{k_1,k_1}^{(n)}+e^{-\zeta_{k_1}} & 0 & \cdots & 0 & 0 &\cdots & 0 & \overline{\mathrm{m}}_{k_1,N_1+k_1}^{(n)}&\overline{\mathrm{m}}_{k_1,N_1+k_1+1}^{(n)} & \cdots & \overline{\mathrm{m}}_{k_1,2N_1}^{(n)}\\
0 & \cdots & 0 & 0 & 1 & \cdots & 0 & 0 &\cdots & 0 & 0 & 0 & \cdots & 0\\
\vdots & \ddots & \vdots& \vdots& \vdots& \ddots & \vdots& \vdots &\ddots &\vdots &\vdots & \vdots & \ddots & \vdots\\
0 & \cdots & 0 & 0 & 0 & \cdots & 1 & 0 &\cdots & 0 & 0 & 0 & \cdots & 0\\
0 & \cdots & 0 & 0 & 0 & \cdots & 0 & 1 &\cdots & 0 & 0 &0 & \cdots & 0\\
\vdots & \ddots & \vdots& \vdots& \vdots& \ddots & \vdots& \vdots &\ddots &\vdots &\vdots & \vdots & \ddots & \vdots\\
0 & \cdots & 0 & 0 & 0 & \cdots & 0 & 0 &\cdots & 1 & 0 & 0 & \cdots & 0\\
\overline{\mathrm{m}}_{N_1+k_1,1}^{(n)} & \cdots & \overline{\mathrm{m}}_{N_1+k_1,k_1-1}^{(n)} & \overline{\mathrm{m}}_{N_1+k_1,k_1}^{(n)}  & 0& \cdots & 0 & 0&\cdots & 0 & \overline{\mathrm{m}}_{N_1+k_1,N_1+k_1}^{(n)}+e^{-\zeta_{N_1+k_1}}& \overline{\mathrm{m}}_{N_1+k_1,N_1+k_1+1}^{(n)} & \cdots & \overline{\mathrm{m}}_{N_1+k_1,2N_1}^{(n)}\\
\overline{\mathrm{m}}_{N_1+k_1+1,1}^{(n)} & \cdots & \overline{\mathrm{m}}_{N_1+k_1+1,k_1-1}^{(n)} & \overline{\mathrm{m}}_{N_1+k_1+1,k_1}^{(n)} & 0 & \cdots & 0 & 0 &\cdots & 0 & \overline{\mathrm{m}}_{N_1+k_1+1,N_1+k_1}^{(n)} & \overline{\mathrm{m}}_{N_1+k_1+1,N_1+k_1+1}^{(n)} & \cdots & \overline{\mathrm{m}}_{N_1+k_1+1,2N_1}^{(n)}\\
\vdots & \ddots & \vdots& \vdots& \vdots& \ddots & \vdots& \vdots &\ddots &\vdots &\vdots & \vdots & \ddots & \vdots\\
\overline{\mathrm{m}}_{2N_1,1}^{(n)} & \cdots & \overline{\mathrm{m}}_{2N_1,k_1-1}^{(n)} & \overline{\mathrm{m}}_{2N_1,k_1}^{(n)} & 0 & \cdots & 0 & 0 &\cdots & 0 & \overline{\mathrm{m}}_{2N_1,N_1+k_1}^{(n)} & \overline{\mathrm{m}}_{2N_1,N_1+k_1+1}^{(n)} & \cdots & \overline{\mathrm{m}}_{2N_1,2N_1}^{(n)}\\
\end{array}
\right|}\\
= &(-1)^{k_1-1}e^{\sum_{s=1}^{k_1-1} \zeta_{s}- \sum_{s=k_1+1}^{N_1} \zeta_{s}}\times \\
&{\tiny \left|
\begin{array}{cccccccc}
\overline{\mathrm{m}}_{1,1}^{(n)} & \cdots & \overline{\mathrm{m}}_{1,k_1-1}^{(n)} & \overline{\mathrm{m}}_{1,k_1}^{(n)}  & \overline{\mathrm{m}}_{1,N_1+k_1}^{(n)} & \overline{\mathrm{m}}_{1,N_1+k_1+1}^{(n)} & \cdots & \overline{\mathrm{m}}_{1,2N_1}^{(n)}\\
\vdots & \ddots & \vdots& \vdots& \vdots & \vdots & \ddots & \vdots\\
\overline{\mathrm{m}}_{k_1-1,1}^{(n)} & \cdots & \overline{\mathrm{m}}_{k_1-1,k_1-1}^{(n)}+e^{-\zeta_{k_1-1}} & \overline{\mathrm{m}}_{k_1-1,k_1}^{(n)} & \overline{\mathrm{m}}_{k_1,N_1+k_1}^{(n)} & \overline{\mathrm{m}}_{k_1-1,N_1+k_1+1}^{(n)} & \cdots & \overline{\mathrm{m}}_{k_1-1,2N_1}^{(n)}\\
\overline{\mathrm{m}}_{k_1,1}^{(n)} & \cdots & \overline{\mathrm{m}}_{k_1,k_1-1}^{(n)} & \overline{\mathrm{m}}_{k_1,k_1}^{(n)}+e^{-\zeta_{k_1}} & \overline{\mathrm{m}}_{k_1,N_1+k_1}^{(n)}&\overline{\mathrm{m}}_{k_1,N_1+k_1+1}^{(n)} & \cdots & \overline{\mathrm{m}}_{k_1,2N_1}^{(n)}\\
\overline{\mathrm{m}}_{N_1+k_1,1}^{(n)} & \cdots & \overline{\mathrm{m}}_{N_1+k_1,k_1-1}^{(n)} & \overline{\mathrm{m}}_{N_1+k_1,k_1}^{(n)} & \overline{\mathrm{m}}_{N_1+k_1,N_1+k_1}^{(n)}+e^{-\zeta_{N_1+k_1}}& \overline{\mathrm{m}}_{N_1+k_1,N_1+k_1+1}^{(n)} & \cdots & \overline{\mathrm{m}}_{N_1+k_1,2N_1}^{(n)}\\
\overline{\mathrm{m}}_{N_1+k_1+1,1}^{(n)} & \cdots & \overline{\mathrm{m}}_{N_1+k_1+1,k_1-1}^{(n)} & \overline{\mathrm{m}}_{N_1+k_1+1,k_1}^{(n)} & \overline{\mathrm{m}}_{N_1+k_1+1,N_1+k_1}^{(n)} & \overline{\mathrm{m}}_{N_1+k_1+1,N_1+k_1+1}^{(n)} & \cdots & \overline{\mathrm{m}}_{N_1+k_1+1,2N_1}^{(n)}\\
\vdots & \ddots & \vdots& \vdots &\vdots & \vdots & \ddots & \vdots\\
\overline{\mathrm{m}}_{2N_1,1}^{(n)} & \cdots & \overline{\mathrm{m}}_{2N_1,k_1-1}^{(n)} & \overline{\mathrm{m}}_{2N_1,k_1}^{(n)} & \overline{\mathrm{m}}_{2N_1,N_1+k_1}^{(n)} & \overline{\mathrm{m}}_{2N_1,N_1+k_1+1}^{(n)} & \cdots & \overline{\mathrm{m}}_{2N_1,2N_1}^{(n)}\\
\end{array}
\right|}, t\rightarrow -\infty.\\
\end{aligned}
\end{equation}
Using a similar approach as presented above for the case where $t \rightarrow -\infty$, we can obtain the explicit asymptotic expression for $\overline{\delta}_n$ as $t \rightarrow \infty$, thus completing the proof of Lemma \ref{asy-ntype1-delta}.
\subsection{Proof of Lemma \ref{asy-ntype2-delta}}
In Lemma \ref{asy-ntype2-delta}, we discuss the pure $N_2$ type-II dark soliton and assume that the velocities satisfy the following constraint:
\begin{equation}
    \overline{\nu}_1 < \overline{\nu}_2 < \cdots<\overline{\nu}_{k_2} < \cdots<\overline{\nu}_{N_2} < 0.
\end{equation}
The discussion is conducted around the neighbourhood of $\zeta_{k_2} \approx 0$, which implies $x \approx \overline{\nu}_{k_2} t$.\\
(a)\underline{ Before collision $t\rightarrow-\infty$}\\
We have
\begin{equation}
    \left\{
    \begin{aligned}
    & e^{-\zeta_j} \rightarrow 0, \quad j>k_2,\\
    & e^{-\zeta_j} \rightarrow \infty, \quad j<k_2.\\
    \end{aligned}
    \right.
\end{equation}
The asymptotic expression of $\overline{\delta}_n$ is derived as
\begin{equation*}
\hspace{-1cm}
\begin{aligned}
\overline{\delta}_n
= &\left|
\begin{array}{cccccccc}
\overline{\mathrm{m}}_{1,1}^{(n)}+e^{-\zeta_1} & \cdots & \overline{\mathrm{m}}_{1,k_2-1}^{(n)} & \overline{\mathrm{m}}_{1,k_2}^{(n)} & \overline{\mathrm{m}}_{1,k_2+1}^{(n)} & \cdots & \overline{\mathrm{m}}_{1,N_2}^{(n)} \\
\vdots & \ddots & \vdots& \vdots& \vdots& \ddots & \vdots\\
\overline{\mathrm{m}}_{k_2-1,1}^{(n)} & \cdots & \overline{\mathrm{m}}_{k_2-1,k_2-1}^{(n)}+e^{-\zeta_{k_2-1}} & \overline{\mathrm{m}}_{k_2-1,k_2}^{(n)} & \overline{\mathrm{m}}_{k_2-1,k_2+1}^{(n)} & \cdots & \overline{\mathrm{m}}_{k_2-1,N_2}^{(n)} \\
\overline{\mathrm{m}}_{k_2,1}^{(n)} & \cdots & \overline{\mathrm{m}}_{k_2,k_2-1}^{(n)} & \overline{\mathrm{m}}_{k_2,k_2}^{(n)}+e^{-\zeta_{k_2}} & \overline{\mathrm{m}}_{k_2,k_2+1}^{(n)} & \cdots & \overline{\mathrm{m}}_{k_2,N_2}^{(n)} \\
\overline{\mathrm{m}}_{k_2+1,1}^{(n)} & \cdots & \overline{\mathrm{m}}_{k_2+1,k_2-1}^{(n)} & \overline{\mathrm{m}}_{k_2+1,k_2}^{(n)} & \overline{\mathrm{m}}_{k_2+1,k_2+1}^{(n)}+e^{-\zeta_{k_2+1}} & \cdots & \overline{\mathrm{m}}_{k_2+1,N_2}^{(n)} \\
\vdots & \ddots & \vdots& \vdots& \vdots& \ddots & \vdots\\
\overline{\mathrm{m}}_{N_2,1}^{(n)} & \cdots & \overline{\mathrm{m}}_{N_2,k_2-1}^{(n)} & \overline{\mathrm{m}}_{N_2,k_2}^{(n)} & \overline{\mathrm{m}}_{N_2,k_2+1}^{(n)} & \cdots & \overline{\mathrm{m}}_{N_2,N_2}^{(n)}+e^{-\zeta_{N_2}} \\
\end{array}
\right|\\
\simeq &\overline{\delta}_n^{[k_2]-} \\
=&e^{-\sum_{s=1}^{k_2-1}\zeta_s}\times
\left|
\begin{array}{cccccccc}
1 & \cdots & 0 & 0 & 0 & \cdots & 0 \\
\vdots & \ddots & \vdots& \vdots& \vdots& \ddots & \vdots\\
0 & \cdots & 1  & 0 & 0 & \cdots & 0 \\
0 & \cdots & 0 & \overline{\mathrm{m}}_{k_2,k_2}^{(n)}+e^{-\zeta_{k_2}} & \overline{\mathrm{m}}_{k_2,k_2+1}^{(n)} & \cdots & \overline{\mathrm{m}}_{k_2,N_2}^{(n)} \\
0 & \cdots & 0 & \overline{\mathrm{m}}_{k_2+1,k_2}^{(n)} & \overline{\mathrm{m}}_{k_2+1,k_2+1}^{(n)} & \cdots & \overline{\mathrm{m}}_{k_2+1,N_2}^{(n)} \\
\vdots & \ddots & \vdots& \vdots& \vdots& \ddots & \vdots\\
0 & \cdots & 0 & \overline{\mathrm{m}}_{N_2,k_2}^{(n)} & \overline{\mathrm{m}}_{N_2,k_2+1}^{(n)} & \cdots & \overline{\mathrm{m}}_{N_2,N_2}^{(n)} \\
\end{array}
\right|\\
=&e^{-\sum_{s=1}^{k_2-1}\zeta_s}\times \left(
\left|
\begin{array}{ccccc}
e^{-\zeta_{k_2}} & 0 & \cdots & 0 \\
\overline{\mathrm{m}}_{k_2+1,k_2}^{(n)} & \overline{\mathrm{m}}_{k_2+1,k_2+1}^{(n)} & \cdots & \overline{\mathrm{m}}_{k_2+1,N_2}^{(n)} \\
 \vdots& \vdots& \ddots & \vdots\\
\overline{\mathrm{m}}_{N_2,k_2}^{(n)} & \overline{\mathrm{m}}_{N_2,k_2+1}^{(n)} & \cdots & \overline{\mathrm{m}}_{N_2,N_2}^{(n)} \\
\end{array}
\right|+
\left|
\begin{array}{ccccc}
\overline{\mathrm{m}}_{k_2,k_2}^{(n)} & \overline{\mathrm{m}}_{k_2,k_2+1}^{(n)} & \cdots & \overline{\mathrm{m}}_{k_2,N_2}^{(n)} \\
\overline{\mathrm{m}}_{k_2+1,k_2}^{(n)} & \overline{\mathrm{m}}_{k_2+1,k_2+1}^{(n)} & \cdots & \overline{\mathrm{m}}_{k_2+1,N_2}^{(n)} \\
 \vdots& \vdots& \ddots & \vdots\\
\overline{\mathrm{m}}_{N_2,k_2}^{(n)} & \overline{\mathrm{m}}_{N_2,k_2+1}^{(n)} & \cdots & \overline{\mathrm{m}}_{N_2,N_2}^{(n)} \\
\end{array}
\right|
\right), t\rightarrow -\infty.
\end{aligned}
\end{equation*}

(b)\underline{ After collision $t\rightarrow\infty$}\\
We have
\begin{equation}
    \left\{
    \begin{aligned}
    & e^{-\zeta_j} \rightarrow \infty, \quad j>k_2,\\
    & e^{-\zeta_j} \rightarrow 0, \quad j<k_2.\\
    \end{aligned}
    \right.
\end{equation}\\

The asymptotic expression of $\overline{\delta}_n$ is derived as
\begin{equation*}
\hspace{-1cm}
\begin{aligned}
\overline{\delta}_n
= &\left|
\begin{array}{cccccccc}
\overline{\mathrm{m}}_{1,1}^{(n)}+e^{-\zeta_1} & \cdots & \overline{\mathrm{m}}_{1,k_2-1}^{(n)} & \overline{\mathrm{m}}_{1,k_2}^{(n)} & \overline{\mathrm{m}}_{1,k_2+1}^{(n)} & \cdots & \overline{\mathrm{m}}_{1,N_2}^{(n)} \\
\vdots & \ddots & \vdots& \vdots& \vdots& \ddots & \vdots\\
\overline{\mathrm{m}}_{k_2-1,1}^{(n)} & \cdots & \overline{\mathrm{m}}_{k_2-1,k_2-1}^{(n)}+e^{-\zeta_{k_2-1}} & \overline{\mathrm{m}}_{k_2-1,k_2}^{(n)} & \overline{\mathrm{m}}_{k_2-1,k_2+1}^{(n)} & \cdots & \overline{\mathrm{m}}_{k_2-1,N_2}^{(n)} \\
\overline{\mathrm{m}}_{k_2,1}^{(n)} & \cdots & \overline{\mathrm{m}}_{k_2,k_2-1}^{(n)} & \overline{\mathrm{m}}_{k_2,k_2}^{(n)}+e^{-\zeta_{k_2}} & \overline{\mathrm{m}}_{k_2,k_2+1}^{(n)} & \cdots & \overline{\mathrm{m}}_{k_2,N_2}^{(n)} \\
\overline{\mathrm{m}}_{k_2+1,1}^{(n)} & \cdots & \overline{\mathrm{m}}_{k_2+1,k_2-1}^{(n)} & \overline{\mathrm{m}}_{k_2+1,k_2}^{(n)} & \overline{\mathrm{m}}_{k_2+1,k_2+1}^{(n)}+e^{-\zeta_{k_2+1}} & \cdots & \overline{\mathrm{m}}_{k_2+1,N_2}^{(n)} \\
\vdots & \ddots & \vdots& \vdots& \vdots& \ddots & \vdots\\
\overline{\mathrm{m}}_{N_2,1}^{(n)} & \cdots & \overline{\mathrm{m}}_{N_2,k_2-1}^{(n)} & \overline{\mathrm{m}}_{N_2,k_2}^{(n)} & \overline{\mathrm{m}}_{N_2,k_2+1}^{(n)} & \cdots & \overline{\mathrm{m}}_{N_2,N_2}^{(n)}+e^{-\zeta_{N_2}} \\
\end{array}
\right|\\
\simeq &\overline{\delta}_n^{[k_2]+} \\
=&e^{-\sum_{s=k_2+1}^{N_2}\zeta_s}\times \left|
\begin{array}{cccccccc}
\overline{\mathrm{m}}_{1,1}^{(n)} & \cdots & \overline{\mathrm{m}}_{1,k_2-1}^{(n)} & \overline{\mathrm{m}}_{1,k_2}^{(n)} & 0 & \cdots & 0 \\
\vdots & \ddots & \vdots& \vdots& \vdots& \ddots & \vdots\\
\overline{\mathrm{m}}_{k_2-1,1}^{(n)} & \cdots & \overline{\mathrm{m}}_{k_2-1,k_2-1}^{(n)} & \overline{\mathrm{m}}_{k_2-1,k_2}^{(n)} & 0 & \cdots & 0 \\
\overline{\mathrm{m}}_{k_2,1}^{(n)} & \cdots & \overline{\mathrm{m}}_{k_2,k_2-1}^{(n)} & \overline{\mathrm{m}}_{k_2,k_2}^{(n)}+e^{-\zeta_{k_2}} &0 & \cdots & 0 \\
0 & \cdots & 0 & 0 &1 & \cdots & 0 \\
\vdots & \ddots & \vdots& \vdots& \vdots& \ddots & \vdots\\
0 & \cdots & 0 & 0 & 0 & \cdots & 1 \\
\end{array}
\right|\\
=&e^{-\sum_{s=k_2+1}^{N_2}\zeta_s}\times \left(
\left|
\begin{array}{ccccc}
\overline{\mathrm{m}}_{1,1}^{(n)} & \cdots & \overline{\mathrm{m}}_{1,k_2-1}^{(n)} & \overline{\mathrm{m}}_{1,k_2}^{(n)}\\
\vdots & \ddots & \vdots& \vdots\\
\overline{\mathrm{m}}_{k_2-1,1}^{(n)} & \cdots & \overline{\mathrm{m}}_{k_2-1,k_2-1}^{(n)} & \overline{\mathrm{m}}_{k_2-1,k_2}^{(n)} \\
0& \cdots &0  & e^{-\zeta_{k_2}}
\end{array}
\right|+
\left|
\begin{array}{ccccc}
\overline{\mathrm{m}}_{1,1}^{(n)} & \cdots & \overline{\mathrm{m}}_{1,k_2-1}^{(n)} & \overline{\mathrm{m}}_{1,k_2}^{(n)}\\
\vdots & \ddots & \vdots& \vdots\\
\overline{\mathrm{m}}_{k_2-1,1}^{(n)} & \cdots & \overline{\mathrm{m}}_{k_2-1,k_2-1}^{(n)} & \overline{\mathrm{m}}_{k_2-1,k_2}^{(n)} \\
\overline{\mathrm{m}}_{k_2,1}^{(n)} & \cdots & \overline{\mathrm{m}}_{k_2,k_2-1}^{(n)} & \overline{\mathrm{m}}_{k_2,k_2}^{(n)}
\end{array}
\right|
\right), t\rightarrow\infty.
\end{aligned}
\end{equation*}
With Eq.~\eqref{mx-sj} and Lemma \ref{Cauchy determinant}, one can obtain the result presented in Lemma \ref{asy-ntype2-delta}.

\textbf{Acknowledgments.}
The research is supported by the National Natural Science Foundation of China (Grant No. 12471239 ), the Guangdong Basic and Applied Basic Research Foundation (Grant No. 2024A1515013106) and Student Research Cultivation Program of the Institute for Advanced Study, Shenzhen University. The work of TK was supported
by DST-SERB, Government of India, in the form of a major
research project (File No. EMR/2015/001408). TK also thanks the
Principal and Management of Bishop Heber College, Tiruchirappalli, for constant support
and encouragement.
\bibliographystyle{unsrt}
\bibliography{references}

\end{document}